\documentclass[12pt,hidelinks]{article}
\pdfoutput=1
\usepackage{xr-hyper}
\makeatletter
\newcommand*{\addFileDependency}[1]{
  \typeout{(#1)}
  \@addtofilelist{#1}
  \IfFileExists{#1}{}{\typeout{No file #1.}}
}
\makeatother

\usepackage{mathrsfs,amsmath,epsfig,amssymb,color,graphicx,bm}
\usepackage{algorithm,algorithmic,natbib,caption,hyperref,lineno}
\usepackage[margin=1in]{geometry}
\usepackage[symbol]{footmisc}
\usepackage[dvipsnames]{xcolor}
\graphicspath{{figures/}}
\allowdisplaybreaks

\newtheorem{theorem}{Theorem}
\newtheorem{lemma}{Lemma}
\newtheorem{remark}{Remark}
\newtheorem{assumption}{Assumption}
\newtheorem{proposition}{Proposition}

\newcommand{\bSigma}{\boldsymbol \Sigma}
\newcommand{\bbeta}{\boldsymbol \beta}
\newcommand{\hbeta}{\hat{\boldsymbol \beta}_{\mbox{\tiny \rm  MPL}}}
\newcommand{\tbeta}{\tilde{\boldsymbol \beta}}

\newcommand{\black}{\color{black}}

\newcommand{\bUpsilon}{\boldsymbol \Upsilon}
\newcommand{\bpi}{\boldsymbol \pi}

\newcommand{\X}{\mathbf{X}}
\newcommand{\sumn}{\sum_{i=1}^n}
\newcommand{\sumr}{\sum_{i=1}^r}
\newcommand{\onen}{\frac{1}{n}}
\newcommand{\oner}{\frac{1}{r}}
\newcommand{\app}{\pi_{i}^{\text{app}}}
\newcommand{\appm}{\pi_{\delta i}^{\text{app}}}
\newcommand{\appsi}{\pi_{\delta i}^{\text{app}*}}
\newcommand{\appsj}{\pi_{\delta j}^{\text{app}*}}

\begin{document}
\centerline {\Large\bf Approximating Partial
  Likelihood Estimators }
\centerline {\Large\bf via Optimal Subsampling }
\vspace*{0.2in}
\centerline{ {  Haixiang Zhang$^{1*}$, Lulu Zuo$^{1}$,  HaiYing Wang$^{2}$ and Liuquan Sun$^{3}$ }}
 \vspace*{0.2in}

\centerline{\small \it $^{1}$Center for Applied Mathematics, Tianjin University, Tianjin 300072, China}
\centerline{\small \it $^{2}$Department of Statistics, University of Connecticut, Storrs,
Mansfield, CT 06269, USA}
\centerline{\small \it $^{3}$Academy of Mathematics and Systems Science, Chinese Academy of Sciences, Beijing 100190, China}
\footnotetext[1] {Corresponding author. Email: haixiang.zhang@tju.edu.cn (Haixiang Zhang)}
\vspace{1cm}

\begin{abstract}
  With the growing availability of large-scale biomedical data, it is often
  time-consuming or infeasible to directly perform traditional statistical
  analysis with relatively limited computing resources at hand. We propose a
  fast  subsampling method to effectively approximate the full data
  maximum partial likelihood estimator in Cox's model, which largely reduces the
  computational burden when analyzing massive survival data. We establish
  consistency and asymptotic normality of a general subsample-based
  estimator. The optimal subsampling probabilities with explicit expressions are
  determined via minimizing the trace of the asymptotic variance-covariance
  matrix for a linearly transformed parameter estimator. We propose a two-step
  subsampling algorithm for practical implementation, which has a significant
  reduction in computing time compared to the full data method.  The asymptotic
  properties of the resulting two-step subsample-based estimator is also
  established.  Extensive numerical experiments and a real-world example are provided to assess our subsampling strategy. {\black Supplemental materials for this article are
available online.}

\end{abstract}

\section{Introduction}

With the development of science and technology, the amounts of available data
are rapidly increasing in recent years. A major bottleneck to analyze huge
datasets is that the data volume exceeds the capacity of available
computational resources. It is not always possible to meet the demands for
computational speed and storage memory if we directly perform traditional
analysis for large datasets with a single computer at hand.  To cope with big data,  there are many statistical methods in the literature dealing with the
heavy calculation and storage burden. Basically, we
could classify these methods into three categories.  (i) {\it divide-and-conquer
  approach} (\citeauthor{Zhao2016-AOS}, \citeyear{Zhao2016-AOS}; \citeauthor{Battey2018-AOS}, \citeyear{Battey2018-AOS}; \citeauthor{Shi2018-JASA}, \citeyear{Shi2018-JASA}; \citeauthor{Jordan2019-JASA}, \citeyear{Jordan2019-JASA}; \citeauthor{Volgushev2019-AOS}, \citeyear{Volgushev2019-AOS}; \citeauthor{CLZ-JASA-N2021}, \citeyear{CLZ-JASA-N2021}; \citeauthor{Fan-JASA-2021}, \citeyear{Fan-JASA-2021}).
(ii) {\it online updating approach} (\citeauthor{SWWC-2016}, \citeyear{SWWC-2016}; \citeauthor{LS-2020-JRSSB}, \citeyear{LS-2020-JRSSB}; \citeauthor{RWS-2021-Non}, \citeyear{RWS-2021-Non}; \citeauthor{Luo-2021-JASA}, \citeyear{Luo-2021-JASA}; \citeauthor{wangKBS-2022}, \citeyear{wangKBS-2022}). (iii) {\it subsampling-based approach}. {\black The subsampling is an emerging field for
big data.  Many  papers have been published during recent years.}  For example,
\cite{Wang2018-JASA} and \cite{wang-JLMR-2019} studied the optimal subsampling
for massive logistic regression. \cite{Wang2019-JASA} presented an
information-based subdata selection approach for linear regression with big
datasets. \cite{IHS-2020} studied an effective sketching method for massive
datasets via A-optimal subsampling.  \cite{yao-wang-soft-2019},
\cite{HTYZ-2020-AOS} and \cite{softmax-2021} proposed several subsampling
methods for large-scale multiclass logistic regression.
\cite{Poisson-JASA-2021} considered optimal Poisson subsampling for maximum
quasi-likelihood estimator with massive data. \cite{ZNR-2021-JCGS} proposed a
response-free optimal sampling procedure for generalized linear models under
measurement constraints. \cite{Wang-MA2020} studied the optimal subsampling for
quantile regression in big data.  \cite{Functional-2021} proposed an optimal
subsampling method for the functional linear model via L-optimality criterion.
\cite{Zhang-Wang2021} and \cite{Zuo-2021-CS} considered optimal distributed
subsampling methods for big data in the context of linear and logistic models,
respectively. \cite{Sinica-2021} studied the optimal subsampling method for
generalized linear models under the A-optimality criterion. \cite{Wangtzhang-2022}
proposed an optimal subsampling procedure for multiplicative regression with massive data.
For more related
results on massive data analysis, we refer to several review papers by
\cite{wang-SII-2016}, \cite{Lee-NG-2020}, \cite{OSP-review-2021},
\cite{DC-review}, \cite{Li-Meng-2021} and \cite{SP-review}.

The aforementioned investigations focused on developing statistical methods
for large datasets with uncensored observations. In recent years, huge biomedical datasets become increasingly common, and they
are often subject to censoring
\cite[]{Kleinbaum-2005}. There have been several recent papers on statistical
analysis of massive censored survival data. For example, \cite{Xue2019} and
\cite{JCGS-2021-cox} studied the online updating approach for streams of
survival data. \cite{rare-Cox-2020} presented an optimal Cox regression
subsampling procedure with rare events. \cite{Cox-SGD} and \cite{Xu2020} used
the stochastic gradient descent algorithms to analyse large-scale survival
datasets with Cox's model and the accelerated failure time models,
respectively.  \cite{fastlasso-2020} proposed a batch screening iterative Lasso
method for large-scale and ultrahigh-dimensional Cox
model. \cite{zuo2021-sim}
proposed a sampling-based method for massive survival data with additive hazards
model.  \cite{DC-cox} studied an efficient divide-and-conquer algorithm to fit
high-dimensional Cox regression for massive datasets. \cite{SIM-AFT-sampling} studied the optimal subsampling algorithms for parametric accelerate failure time models with massive survival data. In spite of the
aforementioned papers, existing research on massive survival data
is relatively limited, and it is meaningful to further investigate the
statistical theories in the area of large-scale survival analysis.

It is worthy mentioning that subsampling is an emerging area of research,
which has attracted great attentions in both statistics and computer
science \citep{PingMa2014-JMLR,GNN-2021}. Subsampling methods
focus on selecting a small proportion of the full data as a surrogate to perform statistical
computations. A key to success of subsampling is to design nonuniform sampling
probabilities so that those influential or informative data points are sampled
with high probabilities. Although significant progress has been made towards
developing optimal subsampling theory for uncensored observations, to the best
of our knowledge, the research on optimal subsampling for large-scale
survival data lags behind.  In consideration of the important role of
Cox's model in the field of survival analysis \cite[]{Cox1972,
  Fleming-1991}, it is desirable to develop effective subsampling
methods in the context of Cox's model for massive survival data. This paper aims
to close this gap by developing a subsample-based estimator to fast and
effectively approximate the full data maximum partial likelihood estimator.
Our aim is to design an efficient subsampling and estimation strategy to
better balance the trade-off between computational efficiency and statistical
efficiency.  Here are some key differences between our proposed
  subsampling approach and some recently developed approaches on Cox's model
  with large-scale data: (i) \cite{rare-Cox-2020} proposed a subsampling-based estimation for Cox's model with rare events by including all observed failures, while our
 optimal subsampling method is developed for Cox's model under the regular
 setting that observed failure times are not rare compared with the
   observed censoring times. (ii) \cite{Cox-SGD} presented a stochastic gradient descent (SGD) procedure
 for Cox's model. This method primarily intends to resolve the problems
   that the whole dataset cannot be easily loaded in memory; the main aim is to
 deal with the out-of-memory issue rather than speeding up the calculation. (iii) \cite{fastlasso-2020} and  \cite{DC-cox} studied the
 variable selection problem for ultrahigh-dimensional Cox's  model, which
   is different from the focus of our paper on dealing with very large sample sizes.

The main contributions of
our proposed subsampling method include three aspects: First, the computation
of our subsample-based estimator is much faster than that of the full data
estimator calculated by the standard R function {\tt coxph}.  Therefore, it effectively reduces the computational burden when analysing massive
survival data with Cox's model. Second, we provide an explicit expression for
the optimal subsampling distribution, which has much better performance than the
uniform subsampling distribution in terms of statistical efficiency. Third, we
establish consistency and asymptotic normality of the proposed subsample
estimator, which is useful for performing statistical
inference (e.g. constructing confidence intervals and
testing hypotheses). 

The remainder of this paper is organized as follows.  In Section \ref{sec2}, we review
the setup and notations for Cox's model. A general subsample-based estimator is
proposed to approximate the full data maximum partial likelihood
estimator. In Section \ref{sec3}, we establish consistency and
asymptotic normality of a general subsample-based estimator. The optimal subsampling probabilities are
explicitly specified in the context of L-optimality criterion. In Section \ref{sec4}, we
give a two-step subsampling algorithm together with the asymptotic properties of
the resulting estimator.  In Section \ref{sec5}, extensive
simulations together with an application are conducted to verify the validity of the proposed subsampling
procedure. Some concluding remarks are
presented in Section \ref{sec6}.  Technical proofs  are given in the supplement.

\section{Model and Subsample-Based Estimation}\label{sec2}

In many biomedical applications, the outcome of interest is measured as a
``time-to-event'', such as death and onset of cancer. The time to
occurrence of an event is referred to as a failure time
\cite[]{Kalbfleisch-2002}, and its typical characteristic is subject to possible
right censoring. For $i=1,\cdots,n$, let $T_i$ be the failure time, $C_i$ be the
censoring time, and  {\black $\X_i$ be the $p$-dimensional vector of  time-independent covariates}
(e.g., treatment indicator, blood pressure, age, and gender). We assume that
$T_i$ and $C_i$ are conditionally independent given $\X_i$. The observed
failure time is $Y_i = \min(T_i, C_i)$, and the failure indicator is
$\Delta_i = I(T_i \leq C_i)$, where $I(\cdot)$ is the indicator function. For
convenience, we denote the full data of independent and identically
distributed observations from the population as
$\mathcal{D}_n = \{(\X_i, \Delta_i, Y_i), i=1,\cdots,n\}$. The Cox's proportional
hazard regression model \cite[]{Cox1972} {\black is commonly used} to describe the
relationship between covariates of an individual and the risk of experiencing an
event. This model assumes that the conditional hazard rate function of $T_i$
given $\X_i$ is
\begin{align}\label{01}
  \lambda(t|\X_i)= \lambda_0(t) \exp(\bbeta^\prime \X_i),
\end{align}
where $\lambda_0(t)$ is an unknown baseline hazard function,
$\bbeta = (\beta_1,\cdots,\beta_p)^\prime$ is a $p$-dimensional vector of
regression parameters, and its true value belongs to a compact set
$\Theta \subset \mathbb{R}^p$.  To estimate $\bbeta$, \cite{COX1975} proposed a
novel {\it partial likelihood} method.
The negative log-partial likelihood function is
\begin{align}\label{PLF}
  \ell(\bbeta)
  &= -\onen\sumn  \int_0^\tau \left[\bbeta^\prime \X_i -  \log\left\{\sum_{j=1}^n I(Y_j \geq t)\exp(\bbeta^\prime \X_j)\right\}  \right] dN_i(t),
\end{align}
where $N_i(t) = I(\Delta_i =1, Y_i \leq t)$ is a counting process and
$\tau$ is a prespecified positive constant. One advantage of Cox's {\it partial
  likelihood} method is that the criterion function given in (\ref{PLF}) does
not involve the nonparametric baseline hazard function $\lambda_0(t)$,
and the resulting estimator of $\bbeta$ is asymptotically equivalent to the
parameter estimator obtained by maximizing the full likelihood function
\cite[]{COX1975}.

For convenience, we introduce the following notations to ease the presentation:
\begin{align*}
  S^{(k)} (t,\bbeta)= \onen \sumn I(Y_i \geq t) \X_i^{\otimes k}\exp(\bbeta^\prime \X_i), ~k=0, 1~ {\rm and}~ 2,
\end{align*}
where the notation $\mathbf{u}^{\otimes k}$ means $\mathbf{u}^{\otimes 0} = 1$, $\mathbf{u}^{\otimes 1} = \mathbf{u}$ and
$\mathbf{u}^{\otimes 2} = \mathbf{u}\mathbf{u}^\prime$ for a vector
$\mathbf{u}$.  Throughout this paper,
 $\|\mathbf{A}\| = (\sum_{1\leq i,j \leq p} a_{ij}^2)^{1/2}$ for a matrix
$\mathbf{A}=(a_{ij})$. 

The gradient of $\ell(\bbeta)$ is
\begin{align}\label{SCF1}
  \dot{\ell}(\bbeta)
  &=-\onen \sumn \int_0^\tau \{\X_i - \bar{\X}(t,\bbeta)\}dN_i(t)\nonumber\\
  &=-\onen \sumn \int_0^\tau \{\X_i - \bar{\X}(t,\bbeta)\}dM_i(t,\bbeta),
\end{align}
where
$M_i(t,\bbeta) = N_i(t) - \int_0^t I(Y_i \geq u)\exp(\bbeta^\prime
\X_i)\lambda_0(u)du$, and
\begin{align}\label{X-bar-24}
  \bar{\X}(t,\bbeta) = \frac{S^{(1)} (t,\bbeta)}{S^{(0)} (t,\bbeta)}.
\end{align}
The Hessian matrix of $\ell(\bbeta)$ is given by
\begin{align}\label{Hess1}
  \ddot{\ell}(\bbeta)=\onen \sumn \int_0^\tau \left[\frac{S^{(2)} (t,\bbeta)}{S^{(0)} (t,\bbeta)} -\left\{\frac{S^{(1)} (t,\bbeta)}{S^{(0)} (t,\bbeta)}\right\}^{\otimes 2} \right]dN_i(t).
\end{align}

According to \cite{COX1975}, the full data maximum partial likelihood (MPL) estimator
$\hbeta$ is the solution to $\dot{\ell}(\bbeta) = 0$, and the asymptotic
properties of $\hbeta$ have been investigated by \cite{Andersen1982}.
There is no closed-form to $\hbeta$, and it is numerically calculated by
Newton's method through iteratively applying
\begin{align}\label{newton-1}
  \bbeta^{(m+1)} =  \bbeta^{(m)} - \{\ddot{\ell}(\bbeta^{(m)})\}^{-1}\dot{\ell}(\bbeta^{(m)}).
\end{align}
For small datasets with hundreds of observations or even fewer, the iterative
algorithm given in (\ref{newton-1})  is able to converge in a reasonable time. For
moderate datasets, it is common to use the gold standard {\tt coxph} function in the R package
of \cite{survival-package},  where a smart updating procedure is adopted
to speed up the computation \cite[]{Simon-JSS}.  The computational efficiency of {\tt coxph} will be presented in the simulation section.

It is desirable to develop an effective and computationally stable method when handling massive survival datasets with Cox's model.  Recently, \cite{rare-Cox-2020} introduced a novel subsampling procedure for Cox regression with rare events, while our aim is to propose a subsampling procedure for large-scale Cox model under non-rare-events setting. To be specific, we assign subsampling probabilities $\{\pi_i\}_{i=1}^n$ to the full data $\mathcal{D}_n$, where $\sumn \pi_i = 1$ and $\pi_i > 0$ for
$i=1,\cdots,n$. Draw a random subsample of size $r$ with replacement based on
$\{\pi_i\}_{i=1}^n$ from the full data $\mathcal{D}_{n}$, where $r$ is typically
much smaller than $n$.
Let $\mathcal{D}_{r}^* = \{ (\X_i^*, \Delta_i^*, Y_i^*,
\pi_i^*)\}_{i=1}^r$ be a selected subsample with size $r$ from the full data
$\mathcal{D}_n$, where $\X_i^*$, $\Delta_i^*$, $Y_i^*$, and $\pi_i^*$ are the
covariate, the failure indicator, the observed failure times, and the
subsampling probability, respectively, in the subsample.
We propose a weighted pseudo log partial
likelihood using the subsample $\mathcal{D}_{r}^*$: 
\begin{align}\label{subLF}
  \ell^*(\bbeta)
  &= -\oner\sumr  \frac{1}{n\pi_i^*}\int_0^\tau \left\{\bbeta^\prime \X^*_i -  \log \left[r^{-1}\sum_{j=1}^r \pi_j^{*-1} I(Y_j^* \geq t)\exp(\bbeta^\prime \X_j^*)\right]  \right\} dN_i^*(t),
\end{align}
where $N_i^*(t) = I(\Delta_i^* =1, Y^*_i \leq t)$.
The inverse probability weighting in (\ref{subLF}) is to ensure the
consistency of the resulting subsample estimator towards $\hbeta$, which will
be carefully investigated in Section \ref{sec3}. 
The corresponding weighted subsample score function is
\begin{align}\label{SS24}
  \dot{\ell}^*(\bbeta) &=  -\oner \sumr \frac{1}{n\pi_i^*} \int_0^\tau \{ \X_i^* - \bar{\X}^{*}(t, \bbeta)\}dN_i^*(t)\\
                       &=  -\oner \sumr \frac{1}{n\pi_i^*} \int_0^\tau \{ \X_i^* - \bar{\X}^{*}(t, \bbeta)\}dM_i^*(t, \bbeta),\nonumber
\end{align}
where
$M^*_i(t,\bbeta) = N^*_i(t) - \int_0^t I(Y_i^* \geq u)\exp(\bbeta^\prime
\X_i^*)\lambda_0(u)du$, and
\begin{eqnarray}\label{X-bar-S}
  \bar{\X}^{*}(t, \bbeta) = \frac{S^{*(1)}(t, \bbeta)}{S^{*(0)}(t, \bbeta)}
\end{eqnarray}
with
\begin{align} {S}^{*(k)}(t,\bbeta) =& \frac{1}{nr} \sum_{i=1}^{r}
  \frac{1}{\pi_i^*}I(Y_i^{*} \geq t) (\X_i^{*})^{\otimes k} \exp
  (\bbeta^\prime \X_i^{*}),~~k=0,1~{\rm and}~2.
\end{align}

The subsample-based estimator $\tbeta$ is the solution to
$ \dot{\ell}^*(\bbeta) = 0$, which is computationally easier to solved by
Newton's method due to the smaller subsample size. Here $\tbeta$ can be
viewed as a subsample approximation to the full data $\hbeta$.
A natural question is how to select the subsample so that $\tbeta$ and $\hbeta$
are close. We will derive the asymptotic distribution of $\tbeta-\hbeta$ and
then find the probabilities that minimize a function of the asymptotic
variance.


\section{Asymptotic Properties and Subsampling Strategy}\label{sec3}
In this section, we establish the asymptotic properties of a general
subsample-based estimator $\tbeta$ obtained via solving
$ \dot{\ell}^*(\tbeta) = 0$, where $\dot{\ell}^*({\bbeta})$ is given in
(\ref{SS24}).  A strategy on how to specify optimal subsampling probabilities
for our method is presented. We need the following assumptions for
theoretical derivation. Throughout this paper we allow $\pi_i$'s to depend on
the data, so they may be random.
\begin{assumption}\label{assu1}
The baseline hazard satisfies that $\int_0^\tau \lambda_0(t)dt <
  \infty$, and $P(T_i \geq \tau)>0$. 
\end{assumption}
\begin{assumption}\label{assu2}
The quantity
  $\onen\sumn\int_0^\tau\Big[\frac{S^{(2)}(t,{\bbeta})}{S^{(0)}(t,
    \bbeta)} - \left\{\frac{S^{(1)}(t, \bbeta)}{S^{(0)}(t,
      \bbeta)}\right\}^{\otimes 2} \Big]dN_i(t)$
  converges in probability to a
  positive definite matrix  for all $\bbeta\in\Theta$, where $\Theta$ is a compact set containing the true value of $\bbeta$.
\end{assumption}
\begin{assumption}\label{assu3}
 The time-independent covariates $\mathbf{X}_i$'s are bounded.
\end{assumption}
\begin{assumption}\label{assu4}
The subsampling probabilities satisfy $\max_{1\leq i \leq n} (n\pi_i)^{-1} = O_P(1)$.
\end{assumption}

Assumptions \ref{assu1} and \ref{assu2} are two classical regularity
conditions for Cox's model \cite[]{Andersen1982}; Assumption \ref{assu3} is a bounded
condition, which was commonly imposed in the literature about Cox's model, e.g.,
\cite{Huang-J2013-AOS} and \cite{Fang2016-JRSSB}. This assumption is reasonable
in most practical applications, because for biomedical survival data the
covariates of an individual are often treatment indicator, blood pressure, age,
and gender, etc. These biomedical related features are usually bounded
\cite[]{rare-Cox-2020}. Assumption \ref{assu4} is required to protect the weighted
subsample pseudo-score function given in (\ref{SS24}) from being dominated by
those data points with extremely small subsampling
probabilities. i.e.,   Assumption 4 requires that the minimum subsampling probability is at the same order of $1/n$ in probability.  This assumption  was also imposed by \cite{IEEE-Poisson}.

We establish the consistency and asymptotic normality of the subsample-based
estimator $\tbeta$ conditional on the full data $\mathcal{D}_n$ in the
following. This result plays an important role in performing statistical
inference. In addition, the asymptotic distribution is a key foundation to
design optimal subsampling probabilities for our method. Throughout this paper, the notation
$b = O_{P|\mathcal{D}_{n}}(1)$ denotes that $b$ is bounded in conditional
probability, i.e., for any $\epsilon > 0$, there exists a finite
$b_\epsilon > 0$ such that
$P\{P(|b| \geq b_\epsilon | \mathcal{D}_n) < \epsilon\}\rightarrow 1$.

\begin{theorem}\label{Th1}
 Under assumptions~\ref{assu1}-\ref{assu4}, if $r=o(n)$ as $n\rightarrow \infty$ and
 $r\rightarrow \infty$, then the subsample-based
  estimator $\tbeta$ is consistent to $\hbeta$ with a convergence rate
  $O_{P|\mathcal{D}_n}(r^{-1/2})$. In addition, conditional on
  $\mathcal{D}_n$ in probability, we have
  \begin{align}\label{Eq4}
    \bSigma^{-1/2}(\tbeta - \hbeta) \stackrel{d}{\longrightarrow} N(\mathbf{0},\mathbf{I}),
  \end{align}
  where $\stackrel{d}{\longrightarrow}$ denotes convergence in distribution,
  $\bSigma =\mathbf{\Psi}^{-1}\mathbf{\Gamma}\mathbf{\Psi}^{-1}$ with
  \begin{align}\label{Eq5}
    \mathbf{\Psi} = \onen\sumn\int_0^\tau\left[\frac{S^{(2)}(t,\hat{\bbeta}_{\mbox{\tiny \rm  MPL}})}{S^{(0)}(t, \hbeta)} -  \left\{\frac{S^{(1)}(t, \hbeta)}{S^{(0)}(t, \hbeta)}\right\}^{\otimes 2}  \right]dN_i(t),
  \end{align}
  and
  \begin{align}\label{Eq6}
    \mathbf{\Gamma} &=\frac{1}{n^2r} \sumn \frac{1}{\pi_i} \left[\int_0^\tau\left\{\X_i - \bar{\X}(t, \hbeta)\right\}dM_i(t,\hbeta)\right]^{\otimes 2}.
  \end{align}
\end{theorem}


\begin{remark} 
The convergence rate indicates that
  $\|\tbeta - \hbeta\| = O_{P|\mathcal{D}_n}(r^{-1/2})$.
  Since a random sequence that converges to zero in conditional probability also
  converges to zero in unconditional probability \citep{XLletter-2008}, we know that $\|\tbeta - \hbeta\| = o_{P|\mathcal{D}_n}(1) = o_{P}(1)$ as
  $r\rightarrow \infty$.
  Therefore, the subsample-based estimator $\tbeta$ is close to $\hbeta$ as long as $r$ is large enough. It is reasonable to use $\tbeta$ as a surrogate for  $\hbeta$ in order to reduce computational burden when handling large-scale Cox's model in practice.
\end{remark}

\begin{remark}\label{remark3}
  The asymptotic normality in condition distribution indicates that the
  distribution of the error term $\tbeta - \hbeta$ conditional on
  $\mathcal{D}_n$ can be approximated by that of a normal random variable, say
  $\mathbf{Z}$, with conditional distribution $N(\mathbf{0},\bSigma)$. This
  means that for any $\epsilon>0$, the probability
  $P(\|\tbeta - \hbeta\|\geq \epsilon|\mathcal{D}_n)$ is accurately approximated
  by $P(\|\mathbf{Z}\|\geq \epsilon|\mathcal{D}_n)$. Hence, a smaller variance
  ensures a smaller excess error bound. 
  This sheds light on how to design optimal subsampling probabilities for our
  proposed sampling method.
\end{remark}

For practical application of the proposed sampling strategy, we need to specify
the subsampling distribution $\bpi = \{\pi_i\}_{i=1}^n$.  A simple choice is the
uniform subsampling with $\{\pi_i = n^{-1}\}_{i=1}^n$.  However, this is not
optimal, because it does not distinguish the importances among different data
points. It is desirable to design nonuniform subsampling probabilities such that
more informative data points are more likely to be selected into a subsample
(\citeauthor{Wang2018-JASA}, \citeyear{Wang2018-JASA}; \citeauthor{Wang-MA2020}, \citeyear{Wang-MA2020}). In view of Remark~\ref{remark3}, we
propose to determine nonuniform subsampling probabilities by minimizing the
asymptotic variance-covariance matrix $\bSigma$ given in (\ref{Eq4}). However,
the meaning of ``minimizing'' a matrix needs to be carefully defined. Here we
adopt the idea from design of experiments \cite[]{Kiefer1959}, and determine the
optimal subsampling probabilities by minimizing a convex function of
$\bSigma$. 
We follow the idea of \cite{Wang2018-JASA} and focus on minimizing
tr($\mathbf{\Psi}\mathbf{\Sigma}\mathbf{\Psi}$)=tr($\mathbf{\Gamma}$),
where
\begin{eqnarray*}
  {\rm tr}(\mathbf\Gamma)&=&{\rm tr}\bigg(\frac{1}{n^2r} \sumn \frac{1}{\pi_i} \left[\int_0^\tau\left\{\X_i - \bar{\X}(t, \hbeta)\right\}dM_i(t, \hbeta)\right]^{\otimes 2}\bigg)\\
                   &=& \frac{1}{rn^{2}}\sumn \frac{1}{\pi_i} \left\|\int_0^\tau\left\{\X_i - \bar{\X}(t, \hbeta)\right\}dM_i(t,\hbeta)\right\|^2.
\end{eqnarray*}

As a matter of fact, this optimality criterion of minimizing
tr($\mathbf{\Gamma}$) is a version of L-optimality criterion \cite[]{Atkinson-2007}, because tr($\mathbf{\Gamma}$) is trace of  the asymptotic variance-covariance matrix of $\mathbf{\Psi}\tbeta$, which is a linearly transformed subsample estimator. The following theorem  provides an explicit expression for the optimal subsampling distribution $\bpi^{\rm Lopt} =\{\pi_{i}^{\rm Lopt}\}_{i=1}^n$ in the context of L-optimality criterion.

\begin{theorem}\label{Th2}
 If the subsampling probabilities are chosen as
  \begin{align}\label{002}
    \pi_{i}^{\rm Lopt} =  \frac{\|\int_0^\tau\{\X_i - \bar{\X}(t, \hbeta)\}dM_i(t,\hbeta)\|}{\sum_{j=1}^n \|\int_0^\tau\{\X_j - \bar{\X}(t, \hbeta)\}dM_j(t,\hbeta)\|},~  \ \textrm{$i=1,\cdots,n$},
  \end{align}
  then $tr(\mathbf\Gamma)$ attains its minimum.
\end{theorem}

\begin{remark} 
  The numerator of $\pi_{i}^{\rm Lopt}$ has a term
  $\bar{\X}(t, \hbeta)$, which contains all individuals of the full data
  $\mathcal{D}_n$.
  This is different from existing results on parametric models without
  censoring, for which numerators of optimal subsampling probabilities involve
  only individual observations' information (except the dependency of the full
  data estimator). Practical adjustments are required to implement the optimal
  subsampling probabilities to tackle the additional computational challenge due to
  censored survival data with Cox's model.
  We will discuss this in Section~\ref{sec4}.
\end{remark}

\begin{remark}\label{remark-N4}
{\black With the $A$-optimality criterion \citep{Wang2018-JASA}, we can derive
  the corresponding optimal subsampling probabilities that minimize
tr($\mathbf{\Sigma}$). They are
  \begin{align}\label{Eq-Aopt}
    \pi_{i}^{\rm Aopt} =  \frac{\|\mathbf{\Psi}^{-1}\int_0^\tau\{\X_i - \bar{\X}(t, \hbeta)\}dM_i(t,\hbeta)\|}{\sum_{j=1}^n \|\mathbf{\Psi}^{-1}\int_0^\tau\{\X_j - \bar{\X}(t, \hbeta)\}dM_j(t,\hbeta)\|},~  \ \textrm{$i=1,\cdots,n$},
  \end{align}
where $\mathbf{\Psi}$ is given in (\ref{Eq5}). Due to the term $\mathbf{\Psi}$
in (\ref{Eq-Aopt}), the computational burden of $\pi_{i}^{\rm Aopt}$ is much
heavier than that of $\pi_{i}^{\rm Lopt}$. Therefore, we focus on $\pi_{i}^{\rm
  Lopt}$ in the presentation of our subsampling procedure. We provide numerical comparisons between the A-optimality criterion and the L-optimality criterion in Section~\ref{sec:simulation}.}
\end{remark}

We provide more insights on the optimal subsampling probabilities
$\{\pi_{i}^{\rm Lopt}\}_{i=1}^n$ from two aspects: First, the numerator
$\int_0^\tau\{\X_i - \bar{\X}(t, \hbeta)\}dM_i(t,\hbeta)$ is
actually the $i$th sample's score given in (\ref{SCF1}), which is also referred
to as the residual \citep{Martingale-1990-Biom}. {\black The subsampling probabilities in
\cite{rare-Cox-2020} for censored individuals also share a similar spirit, but
the subsampling probabilities for observed events are one in
\cite{rare-Cox-2020}'s approach.}
Second, since the failure times are observed for uncensored
observations, they contain more information than censored observations. The
optimal subsampling probabilities give higher preferences to uncensored
observations compared with censored observations. This will be demonstrated
numerically in Section~\ref{sec:simulation}.

\section{Practical Implementation}\label{sec4}
\subsection{Two-Step Subsampling Algorithm}
In this section, we discuss some issues on practical implementation and provide
strategies to resolve them.

First, the optimal subsampling probabilities $\{\pi_{i}^{\rm Lopt}\}_{i=1}^n$
contain the full data estimator $\hbeta$. We take a pilot subsample from $\mathcal{D}_n$ by uniform
subsampling with replacement, say $\mathcal{D}_{r_0}^* = \{(\X^{0*}_i, \Delta^{0*}_i, Y^{0*}_i),
i=1,\cdots,r_0\}$, obtain a pilot estimator $\tbeta_0$ using
$\mathcal{D}_{r_0}^*$, and use $\tbeta_0$ to replace the $\hbeta$ in
(\ref{002}) for practical implementation.

Second, the resultant probabilities
still involve a term $\bar{\X}(t, \tbeta_0)$ after replacing $\hbeta$ with
$\tbeta_0$. This term involves the full data $\mathcal{D}_n$ so it requires heavy computation
burden. To tackle this problem, we recommend replacing
$\bar{\X}(t, \tbeta_0)$ with
$\bar{\X}^{0*}(t,\tbeta_0)$,  
where
\begin{align}\label{X0}
  \bar{\X}^{0*}(t,\bbeta)= \frac{ \sum_{j=1}^{r_0} I(Y^{0*}_j \geq t)\X_j^{0*}\exp(\bbeta^\prime\X_j^{0*})}{ \sum_{j=1}^{r_0} I(Y^{0*}_j \geq t)\exp(\bbeta^\prime\X_j^{0*})}.
\end{align}
This is a reasonable choice because it can be shown that
$\bar{\X}^{0*}(t, \bbeta) = \bar{\X}(t, \bbeta) +
o_{P}(1)$, for any $t\in [0,\tau]$ and $\bbeta \in \Theta$ (see Eq. (S.6) in
 the Appendix).

Third, the term
$dM_i(t,\hbeta)$ 
involves the unknown baseline hazard function
$\lambda_0(t)$.  
We propose a subsample
Breslow-type estimator for $\Lambda_0(t)=\int_0^t\lambda_0(s)ds$ using $\mathcal{D}_{r_0}^{*}$ as
follows:
\begin{align}\label{pi_Brew}
  \hat{\Lambda}_0^{\mbox{\tiny\rm  UNIF}}(t,\bbeta)
  &= \sum_{i=1}^{r_0} \left\{\frac{\Delta_i^{0*} I(Y_i^{0*} \leq t)}{ \sum_{j=1}^{r_0} I(Y_j^{0*} \geq Y_i^{0*}) \exp(\bbeta^\prime \X_j^{0*})}\right\}.
\end{align}
Taking into account previous discussions, the approximated optimal subsampling
probabilities are
\begin{align}\label{Aosp}
 \pi_{i}^{\text{app}}= \frac{\|\int_0^\tau\{\X_i - \bar{\X}^{0*}(t, {\tbeta_0})\}d\hat{M}_i(t,\tbeta_0)\|}{\sum_{j=1}^n \|\int_0^\tau\{\X_j - \bar{\X}^{0*}(t, {\tbeta_0})\}d\hat{M}_j(t,\tbeta_0)\|},~~i=1,\cdots,n,
\end{align}
where
$d\hat{M}_i(t,\tbeta_0) = dN_i(t) - I(Y_i \geq t)\exp(\tbeta_0^\prime
\X_i)d\hat{\Lambda}_0^{\mbox{\tiny\rm UNIF}}(t,\tbeta_0)$. 

{\black Fourth, we see from (\ref{Aosp}) that $\app$ is proportional to
$\|\int_0^\tau\{\X_i - \bar{\X}^{0*}(t, \tbeta_0)\}d\hat{M}_i(t,\tbeta_0)\|$,
which could be small for some data points. Since $\app$'s are obtained by
inserting the pilot $\tbeta_0$, they are not the real optimal probabilities. The
variation of $\tbeta_0$ may be significantly amplified by data points with much
smaller values of $\app$ than other data points, because the sampling
probabilities appear in the denominator of the asymptotic variance as shown
in~(\ref{Eq6}). From another angle, if some data points with much smaller $\app$
are selected into a subsample, the weighted subsample pseudo-score function
(\ref{SS24}) may be dominated by these data points and thus the variance of the
resulting estimator is inflated by them.  Following the idea of defensive
importance sampling \citep{Hesterberg-1995-TE,Owen-JASA-2000}, we mix the
approximated optimal subsampling distribution with the uniform subsampling
distribution. Specifically, we use $\appm = (1-\delta)\app + \delta/n$ instead
of $\app$ in (\ref{Aosp}) for practical implementation, where $\delta \in (0,1)$
controls the proportion of mixture. A main advantage of this approach is that
$n\appm$ is lower-bounded by $\delta$, so it ensures robustness of the
subsampling estimator.  The same idea was also adopted by other subsampling
methods in the literature, such as \cite{PingMa2014-JMLR, Poisson-JASA-2021,
  IEEE-Poisson}.
We use $\delta=0.1$ in the numerical simulations and real-world application in
Section~\ref{sec5}, and this choice works well.\black

We present a practical two-step subsampling method for Cox's model in Algorithm~\ref{algo1}.


\begin{algorithm}[htp]
  \caption {Two-Step Subsampling Procedure}\label{algo1}
  $\bullet$ {\bf Step 1}. Take a pilot subsample of size $r_0$
  $ \mathcal{D}_{r_0}^* = \{(\X^{0*}_i, \Delta^{0*}_i,
  Y^{0*}_i)\}_{i=1}^{r_0}$ using uniform subsampling with replacement
  from the full data $\mathcal{D}_n$. Here $r_0$ is typically much smaller than
  $r$. Compute a pilot estimator $\tbeta_0$ by solving
  \begin{align*}
    \dot{\ell}^{0*}(\bbeta) = -\frac{1}{r_0} \sum_{i=1}^{r_0} \Delta^{0*}_i
    \{\X^{0*}_i - \bar{\X}^{0*}(Y^{0*}_i,\bbeta)\} = 0,
  \end{align*}
  where $\bar{\X}^{0*}(t,\bbeta)$ is given in (\ref{X0}). 
  Calculate
  \begin{align}\label{PSP1}
    \appm= (1-\delta)\app + \onen\delta,
    \quad, i=1, ..., n,
  \end{align}
  where $\app$'s are given in (\ref{Aosp}) and
  $\delta$ is often a small number. e.g., $\delta = 0.1$.

  $\bullet$ {\bf Step 2}. Draw $r$ data points with replacement from the full
  data $\mathcal{D}_n$ using the subsampling probabilities
  $\{\appm\}_{i=1}^n$ given in (\ref{PSP1}). 
  Let
  $\mathcal{D}_{r}^* = \{ (\X_i^*, \Delta_i^*, Y_i^*,
  \appsi\}_{i=1}^r$ be the selected
  subsample. 
  Obtain the two-step subsample-based estimator ${\breve{\bbeta}}$ by solving
  \begin{align}\label{SC-45}
    \dot{\ell}_{\tbeta_0}^*(\bbeta) =  -\oner \sumr \frac{1}{n\appsi} \Delta_i^*
    \{ \X_i^* - \bar{\X}_{\tbeta_0}^{*}(Y_i^*, \bbeta)\} =0,
  \end{align}
  where $\bar{\X}_{\tbeta_0}^{*}(t, \bbeta)$ has the same expression as
  $\bar{\X}^{*}(t, \bbeta)$ given in (\ref{X-bar-S}) except that
  $\pi_i^*$ is replaced with $\appsi$. 
\end{algorithm}

Note that we do not recommend combining $ \mathcal{D}_{r_0}^*$ and
$ \mathcal{D}_{r}^*$ together for Step 2 of Algorithm~\ref{algo1}.
If we are able to handle the calculation on the combined data from
$\mathcal{D}_{r_0}^*$ and $\mathcal{D}_{r}^*$, then it is more efficient to
increase the second step subsample size to $r_0 + r$. {\black That is to say, the pilot subsample $ \mathcal{D}_{r_0}^*$ does not come into the estimation step for ${\breve{\bbeta}}$. Therefore, we do not need to allocate two subsample sizes $r_0$ and $r$ when implementing our method. Some discussion and guidance on the selection of $r_0$
are provided in {\black Section~\ref{sec:simulation}}.
}

We established the asymptotic normality of the estimator ${\breve{\bbeta}}$ from
the practical Algorithm~\ref{algo1} in the following theorem.

\begin{theorem} \label{Th3}
 Under assumptions~\ref{assu1}-\ref{assu3}, if $r=o(n)$, then
   as $r_0 \rightarrow \infty$, $r \rightarrow \infty$ and
  $n\rightarrow \infty$, conditional on $\mathcal{D}_n$ and $\tbeta_0$,
  the two-step estimator $\breve{\bbeta}$ in Algorithm~\ref{algo1} is consistent
  to $\hbeta$ with convergence rate $r^{-1/2}$. Furthermore, the approximation
  error has an asymptotically normal distribution, that is
  \begin{align}\label{eq:1}
    \bSigma^{-1/2}(\breve{\bbeta}- \hbeta)
    \stackrel{d}{\longrightarrow} N(0,\mathbf{I}),
  \end{align}
  where $\bSigma =\mathbf{\Psi}^{-1}\mathbf{\Gamma}\mathbf{\Psi}^{-1}$ with
  $\mathbf{\Psi}$ defined in (\ref{Eq5}),
  \begin{align}
    \mathbf{\Gamma}
    &=\frac{1}{n^2r} \sumn
      \frac{1}{\pi_{\delta i}^{\rm Lopt}} \left[\int_0^\tau\left\{\X_i -
      \bar{\X}(t, \hbeta)\right\}dM_i(t,\hbeta)\right]^{\otimes 2},
  \end{align}
  and 
  \begin{align}\label{Pd-46}
    \pi_{\delta i}^{\rm Lopt}= (1- \delta)\pi_i^{\rm Lopt} + \frac{\delta}{n}, \quad
    i=1,\cdots,n.
  \end{align}
\end{theorem}

\begin{remark} \label{remark5}
{\black  Note that the full data estimator $\hbeta$ converges to the true parameter at a rate of
  $n^{-1/2}$, so the full data estimator $\hbeta$ in \eqref{eq:1}
  can be replaced by the true parameter since $r=o(n)$. Thus the asymptotic
  result in Theorem \ref{Th3}  can be used for inference on the true parameter.
 Since the subsampling rate is often very small when dealing with large-scale
 datasets, it is reasonable to apply the asymptotic normality in practice.}
\end{remark}

{\black

The following proposition gives the unconditional convergence rate and
asymptotic normality of the two-step subsample estimator towards the true
parameter, which are very useful when we perform inference with respect to the
true parameter.

\begin{proposition} \label{Th4}
 Under assumptions~\ref{assu1}-\ref{assu3}, if $r=o(n)$, then
   as $r_0 \rightarrow \infty$, $r \rightarrow \infty$ and
  $n\rightarrow \infty$,
  the two-step estimator $\breve{\bbeta}$ in Algorithm~\ref{algo1} is consistent
  to the true parameter $\bbeta_0$ with convergence rate $r^{-1/2}$. i.e., we
  have $\|\breve{\bbeta}- \bbeta_0\| = O_P(r^{-1/2})$. Moreover,
  $\breve{\bbeta}$ is asymptotically normal, that is
    \begin{align*}
    \bSigma^{-1/2}(\breve{\bbeta}- \bbeta_0)
    \stackrel{d}{\longrightarrow} N(0,\mathbf{I}),
  \end{align*}
  where $\bSigma$ is given in Theorem~\ref{Th3}.
\end{proposition}

}

In view of Proposition \ref{Th4} and Remark \ref{remark5}, we need to provide an estimate for the variance-covariance matrix of
${\breve{\bbeta}}$ when conducting statistical inference for the true parameter. A
simple method is to replace $\hbeta$ with ${\breve{\bbeta}}$ in the asymptotic
variance-covariance matrix $\bSigma$. However, this requires the calculation
on the full data $\mathcal{D}_n$. To reduce computational cost, we propose to
estimate the variance-covariance matrix of $\breve{\bbeta}$ using the subsample
$\mathcal{D}_r^*$ only with
\begin{align}\label{SE1}
  \breve{\mathbf{\Sigma}}  =\breve{\mathbf{\Psi}}^{-1}\breve{\mathbf{\Gamma}}\breve{\mathbf{\Psi}}^{-1},
\end{align}
where
\begin{align*}
  \breve{\mathbf{\Psi}}
  &= \frac{1}{rn}\sum_{i=1}^{r}  \frac{\Delta_i^*}{\appsi} \left[\frac{S^{*(2)}(Y_i^*,\breve{\bbeta})}{S^{*(0)}(Y_i^*, \breve{\bbeta})} -  \left\{\frac{S^{*(1)}(Y_i^*, \breve{\bbeta})}{S^{*(0)}(Y_i^*, \breve{\bbeta})}\right\}^{\otimes 2}  \right],\\
  \breve{\mathbf{\Gamma}}
  &=\frac{1}{r^2n^{2}}\sum_{i=1}^{r}\frac{1}{\{\appsi\}^2} \left[\int_0^\tau  \{\X^*_i - \bar{\X}^{0*}(t, \breve{\bbeta})\}d \hat{M}_i^*(t,\breve{\bbeta})\right]^{\otimes 2},
\end{align*}
$S^{*(k)} (Y_i^*,\breve{\bbeta}) = {(rn)}^{-1} \sum_{j=1}^{r} {\appsj}^{-1}I(Y_j^{*} \geq Y_i^{*})
\X_j^{*\otimes k}\exp(\breve{\bbeta}^\prime \X^{*}_j)$ for
$k=0,1,2$,
$\bar{\X}^{0*}(t, {\breve{\bbeta}})$ is defined in
(\ref{X0}), and
$d\hat{M}^*_i(t,\breve{\bbeta}) = dN_i^*(t) - I(Y^*_i \geq
t)\exp(\breve{\bbeta}^\prime \X^*_i)d\hat{\Lambda}_0^{\mbox{\tiny\rm
    UNIF}}(t,\breve{\bbeta})$. 
We will assess the performance of formula (\ref{SE1}) by numerical simulations.

Lastly, the cumulative hazard function $\Lambda_0(t)$ plays an important role for
predicting the survival probability of an individual in many biomedical
applications.
It has an expression of
$S(t|\X) = P(T > t|\X)= \exp\{-\exp(\bbeta^\prime \X)\Lambda_0(t)\}$ with Cox's
model.
The Breslow estimator $\hat{\Lambda}_0(t,\bbeta)$  is the
maximum likelihood estimator of $\Lambda_0(t)$, where
\begin{align}\label{Bre1}
  \hat{\Lambda}_0(t,\bbeta) &=\sumn \int_0^t \frac{ d N_i(s)}{ \sum_{j=1}^n I(Y_j \geq s) \exp(\bbeta^\prime \X_j)} \nonumber\\
                            &= \sumn \frac{\Delta_i I(Y_i \leq t)}{ \sum_{j=1}^n I(Y_j \geq Y_i) \exp(\bbeta^\prime \X_j)}.
\end{align}
{\black Based on the subsample estimator $\breve{\bbeta}$, it is easy to obtain a
Breslow type estimator $\hat{\Lambda}_0(t,\breve{\bbeta})$ by replacing $\bbeta$
with $\breve{\bbeta}$ in (\ref{Bre1}), i.e., an estimated cumulative hazard
function is based on the entire dataset but with a subsampling-based estimator of
$\bbeta$. As pointed out by an reviewer, the computation burden of this
Breslow estimator is not heavy if the observed failure times are sorted in an
increasing order, because it has an explicit expression and no optimization
process is required. The Breslow type estimator has a computation complexity
$O(n\log(n)) + O(n)$, where $O(n\log(n))$ is due to the sorting of the full data
and $O(n)$ is from the summation.}

\section{Numerical Studies}\label{sec5}
\subsection{Simulation}
\label{sec:simulation}
In this section, we conduct simulations to evaluate the performance of our proposed subsampling
method. We generate failure times $T_i$'s from Cox's model with
a baseline hazard function $\lambda_0(t) = 0.5 t$ and the true parameter $\bbeta_0
= (-1,-0.5,0,0.5,1)^{\prime}$ with $p=5$. We consider four settings for the
covariate $\X_i = (X_{i1},\cdots,X_{i5})^\prime$.\\
{\black {\it Case} \uppercase \expandafter {\romannumeral 1} : components of $\X_i$ are independent  uniform random variables over $(-1, 1)$.}\\
{\it Case} \uppercase \expandafter {\romannumeral 2}: $\X_i$ follows $0.5N(-\mathbf{1},\bUpsilon)+0.5N(\mathbf{1},\bUpsilon)$, where $\Upsilon_{jk}=0.5^{|j-k|}$, i.e., $\X_i$ follows a mixture of two multivariate normal distributions.\\
{\it Case} \uppercase \expandafter {\romannumeral 3}: components of $\X_i$ are independent exponential random variables with probability density function $f(x)=2e^{-2x}I(x>0)$.\\
{\it Case} \uppercase \expandafter {\romannumeral 4}: $\X_i$ follows a
multivariate $t$ distribution with degree of freedom 10, mean zero and
covariance matrix $\bUpsilon$ where $\Upsilon_{jk}=0.5^{|j-k|}$.

The censoring times $C_i$'s are independently generated from a uniform
distribution over $(0,c_0)$ with $c_0$ being chosen so that {\black the censoring rate
(CR) is about $20\%$ and $60\%$, respectively.} Results were calculated based on
1000 replications of the simulation.  We set the full data sample size to
$n = 10^6$, and consider the subsample sizes of $r= 400, 600, 800$, and
$1000$, respectively.


We evaluate the proposed method using the empirical mean squared error (MSE), defined as
 \begin{align}\label{MSE28}
 {\rm MSE}(\breve{\bbeta}) = \frac{1}{1000} \sum_{b=1}^{1000} \|\breve{\bbeta}^{(b)} - \hbeta\|^2,
 \end{align}
where $\breve{\bbeta}^{(b)}$ is the estimate from the $b$th subsample with $\delta=0.1$.

We studied the effect of the pilot subsample size $r_0$ first.  Table~\ref{tab:1}
presents the MSEs of subsampling-based estimator by varying the pilot subsample
size $r_0$ = 300, 400 and 500.  We see that the influence of $r_0$ on
$\breve{\bbeta}$ is not significant if we choose a reasonably large pilot
subsample. Hence, we suggest to use $r_0$= 300 for settings similar to the
simulation setup. Users may adopt a larger pilot subsample if the dimension of the
problem is higher or if the censoring rate is higher.

\begin{table}[H] 
  \begin{center}
    \caption{The MSE of subsampling-based estimator with different pilot subsample size $r_0$.}
    \label{tab:1}
    \vspace{0.1in} \small
    \begin{tabular}{lccccccccccc}
      \hline
      & &  & \multicolumn{4}{c}{CR=$20\%$} &  & \multicolumn{4}{c}{CR=$60\%$} \\
      \cline{4-7}\cline{9-12}
      & $r_0$ & &$r=400$ &$r=600$ &$r=800$ &$r=1000$ & & $r=400$ &$r=600$ &$r=800$ &$r=1000$\\
      \hline
      Case I
& 300 &&0.0320 & 0.0215 & 0.0159 & 0.0130 & &0.0590 & 0.0392 &0.0279 & 0.0229 \\
& 400 &&0.0321 & 0.0211 & 0.0163 & 0.0123 & &0.0586 &0.0385 &0.0279 & 0.0220 \\
& 500 &&0.0322 & 0.0203 & 0.0161 & 0.0124 & &0.0573 &0.0365 &0.0279 & 0.0219 \\
      \hline
      Case II
& 300 &&0.0340 & 0.0214 & 0.0165 & 0.0127 & &0.0592 &0.0374 &0.0284 & 0.0221 \\
& 400 &&0.0332 & 0.0219 & 0.0164 & 0.0128 & &0.0599 &0.0381 &0.0299 & 0.0227 \\
& 500 &&0.0338 & 0.0222 & 0.0167 & 0.0125 & &0.0593 &0.0381 &0.0281 & 0.0224 \\
      \hline
      Case III
& 300 &&0.0418 & 0.0272 & 0.0206 & 0.0164 & &0.0804 &0.0510 &0.0379 &0.0294 \\
& 400 &&0.0401 & 0.0262 &0.0191 & 0.0155 & &0.0800 & 0.0515 & 0.0367 & 0.0310 \\
& 500 && 0.0392 & 0.0257 &0.0187 &0.0149 & & 0.0769 & 0.0500 & 0.0359 & 0.0290 \\
      \hline
      Case IV
& 300 &&0.0167 &  0.0108 &0.0083 & 0.0065 & &0.0226 &0.0157 &0.0108 &0.0086 \\
& 400 &&0.0152 & 0.0102 &  0.0080 &0.0060 & &0.0241 &0.0150 &0.0112 &0.0088 \\
& 500 &&0.0151 & 0.0100 & 0.0075 & 0.0060 & &0.0224 &0.0144 &0.0107 & 0.0084 \\
      \hline
    \end{tabular}
  \end{center}
\end{table}

Next we investigated how the MSEs behave as a function of $\delta$. From the
expression of ${\boldsymbol\pi}_{\delta}^{\rm app}=\{\appm\}_{i=1}^n$ given in
(\ref{PSP1}), we know the sampling distribution
${\boldsymbol\pi}_{\delta}^{\rm app}$ is close to the optimal subsampling
distribution when $\delta$ is small, while it is close to the uniform
subsampling distribution if $\delta$ is close to 1.   In
  Table~\ref{tab:2}, we present the MSEs of the subsampling estimator for
  different values of $\delta$: 0, 0.1, 0.3, and 0.5. It is
  seen that $\delta$ = 0.1 produce the best result most frequently among.
Hence, we recommend
$\delta$ = 0.1 when implementing our method in practical applications with similar settings.
\begin{table}[H] 
  \begin{center}
    \caption{The MSE of subsampling-based estimator with different mixing rate $\delta$.}
    \label{tab:2}
    \vspace{0.1in} \small
    \begin{tabular}{lcccccccccccccccc}
      \hline
      & &  & \multicolumn{4}{c}{CR=$20\%$} &  & \multicolumn{4}{c}{CR=$60\%$} \\
      \cline{4-7}\cline{9-12}
& $r$ & &$\delta=0$ &$\delta=0.1$  &$\delta=0.3$ &$\delta=0.5$ & & $\delta=0$ &$\delta=0.1$ &$\delta=0.3$ &$\delta=0.5$\\
      \hline
      Case I
& 400 &&0.0326&0.0320  & 0.0322 &0.0346 &  &0.0596&0.0590  & 0.0613 & 0.0670 \\
& 600 &&0.0216&0.0215  & 0.0223 & 0.0230 & &0.0393&0.0392  & 0.0408 & 0.0443 \\
& 800 &&0.0163&0.0159  & 0.0166 & 0.0176 & &0.0290&0.0279  & 0.0290 & 0.0317 \\
&1000 &&0.0131&0.0130  & 0.0132 & 0.0141 & &0.0222&0.0229  & 0.0243 & 0.0263 \\
      \hline
      Case II
& 400 &&0.0347&0.0340  & 0.0357 & 0.0376 & &0.0594&0.0592  & 0.0623 & 0.0691 \\
& 600 &&0.0219&0.0214  & 0.0225 & 0.0242 & &0.0372&0.0374  & 0.0394 &  0.0432 \\
& 800 &&0.0164&0.0165  & 0.0170 & 0.0180 & & 0.0284&0.0284  & 0.0302 & 0.0335 \\
&1000 &&0.0127&0.0127  & 0.0134 & 0.0141 & &0.0222&0.0221  & 0.0231 & 0.0248 \\
      \hline
      Case III
& 400 && 0.0418&0.0418  & 0.0437 & 0.0463& &0.0924&0.0804  & 0.0837 & 0.0891 \\
& 600 &&0.0272&0.0272  & 0.0281 & 0.0300 & &0.0535&0.0510  & 0.0537 & 0.0582 \\
& 800 &&0.0209&0.0206  & 0.0211 & 0.0223 & &0.0359&0.0379  & 0.0404 & 0.0431 \\
&1000 &&0.0164&0.0164  & 0.0169 & 0.0177 & &0.0295&0.0294  & 0.0306 & 0.0336 \\
      \hline
      Case IV
& 400 &&0.0174&0.0167  & 0.0166 & 0.0178 & &0.0241&0.0226  & 0.0232 & 0.0256 \\
& 600 &&0.0106&0.0108  & 0.0106 & 0.0112 & &0.0154&0.0157  & 0.0157 & 0.0169 \\
& 800 &&0.0082&0.0083  & 0.0083 & 0.0084 & &0.0111&0.0108  & 0.0113 & 0.0128 \\
&1000 &&0.0065&0.0065  & 0.0065 & 0.0067 & &0.0091&0.0086  & 0.0089 & 0.0097 \\
      \hline
    \end{tabular}
  \end{center}
\end{table}

\begin{table}[htp] 
  \begin{center}
    \caption{Simulation results of the subsample estimator $\breve{\beta}_1$ with CR = $20\%$$^\ddag$.}
    \label{tab:3}
    \vspace{0.1in} \small
    \begin{tabular}{lccccccccccc}
      \hline
      & &  & \multicolumn{4}{c}{Lopt} &  & \multicolumn{4}{c}{UNIF} \\
      \cline{4-7}\cline{9-12}
      & r & &Bias &ESE &SE &CP & & Bias &ESE &SE &CP\\
      \hline
      Case I
      & 400 & &-0.0021& 0.0860  & 0.0901 & 0.960 & &-0.0069 &0.1121  &0.1148 & 0.947 \\
      & 600 & &0.0008 & 0.0717  & 0.0728 & 0.947 & &-0.0033& 0.0896  &0.0933 & 0.961 \\
      & 800 & &-0.0015 & 0.0603 & 0.0629 & 0.958 & &-0.0041 & 0.0787 &0.0810 & 0.963 \\
      & 1000& &0.0021 & 0.0551  & 0.0559 & 0.948 & &-0.0039 &  0.0672 &0.0722 &0.961 \\
      \hline
      Case II
      & 400 & &-0.0008& 0.0836  & 0.0844 & 0.946 & &-0.0156 & 0.1032 & 0.1039 & 0.952 \\
      & 600 & &-0.0016 & 0.0662 & 0.0681 & 0.949 & &-0.0159 &0.0857  & 0.0847 & 0.950 \\
      & 800 & &-0.0037 &0.0594  & 0.0588 & 0.949 & &-0.0149 & 0.0769  &0.0731 &0.935 \\
      & 1000& & -0.0020 &0.0538 & 0.0523 & 0.944 & &-0.0090 &0.0672 & 0.0649 & 0.939\\
      \hline
      Case III
      & 400 & &-0.0025&0.1047   &0.0997  & 0.937 & &-0.0109 &0.1341  &0.1297 & 0.943 \\
      & 600 & &-0.0002 &0.0827  &0.0808  &0.936 & & -0.0029 &0.1132  &0.1048 & 0.944 \\
      & 800 & &-0.0029 &0.0704  &0.0696  &0.938 & & -0.0037 &0.0969  &0.0906 & 0.939 \\
      & 1000& &-0.0015 &0.0615  &0.0621  & 0.952 & & 0.0008 &0.0872  &0.0808 & 0.938 \\
      \hline
      Case IV
      & 400 & &0.0011 & 0.0602  & 0.0625 & 0.953 & &-0.0093 &0.0763  &0.0758  & 0.942 \\
      & 600 & &-0.0010 &0.0496  & 0.0505 & 0.958 & &-0.0072 &0.0600  &0.0616  & 0.955 \\
      & 800 & &0.0015 & 0.0421  & 0.0433 & 0.953 & &-0.0057 &0.0505  &0.0531  & 0.950 \\
      & 1000& &-0.0003 &0.0369  & 0.0386 & 0.966 & & -0.0058 &0.0461  & 0.0473 & 0.953 \\
      \hline
    \end{tabular}
  \end{center}
{\vspace{0cm} \hspace{-0.3cm} $\ddag$ \footnotesize The (Bias, SE) of full data MPL estimator $\hat{\beta}_1$ is Case I: (Bias, SE) = (0.0001, 0.0021), Case II:  (Bias, SE) = (-0.0036, 0.0020), Case III: (Bias, SE) = (0.0004, 0.0027), Case IV: (Bias, SE) = (-0.0001, 0.0014).}
\end{table}

We considered the proposed subsampling method
with approximated optimal subsampling probabilities in Algorithm~\ref{algo1}
with $\delta = 0.1$ (``Lopt estimator''), and the uniform subsampling method
(``UNIF estimator''). We calculated the empirical biases (Bias),
the mean estimated standard errors (SE) calculated using (\ref{SE1}),
the empirical standard errors (ESE), and the empirical 95\% coverage probability
(CP) towards the  true parameter $\bbeta$. The pilot sample size is $r_0=300$.

\begin{table}[htp] 
  \begin{center}
    \caption{Simulation results of the subsample estimator $\breve{\beta}_1$ with CR = $60\%$$^\ddag$.}
    \label{tab:4}
    \vspace{0.1in} \small
    \begin{tabular}{lccccccccccc}
      \hline
      & &  & \multicolumn{4}{c}{Lopt} &  & \multicolumn{4}{c}{UNIF} \\
      \cline{4-7}\cline{9-12}
      & r & &Bias &ESE &SE &CP & & Bias &ESE &SE &CP\\
      \hline
      Case I
      & 400 & &0.0074 & 0.1119  & 0.1219 & 0.971 & &-0.0094& 0.1509  & 0.1717 & 0.965 \\
      & 600 & &0.0066 & 0.0926  & 0.0984 & 0.966 & &-0.0142 &0.1221  & 0.1404 & 0.979 \\
      & 800 & &0.0036 & 0.0769  & 0.0849 & 0.967 & &-0.0057 &0.1089  & 0.1216 & 0.968 \\
      & 1000& &0.0056 & 0.0708  & 0.0758 & 0.962 & &0.0001  & 0.0927  &0.1083 & 0.979\\
      \hline
      Case II
      & 400 & &-0.0086 &0.1032  & 0.1012 & 0.947 & &-0.0242 &0.1398  & 0.1409 & 0.952 \\
      & 600 & &-0.0033 &0.0783  & 0.0813 & 0.961 & &-0.0141 &0.1164  & 0.1139 & 0.945 \\
      & 800 & &-0.0046 &0.0728  & 0.0699 & 0.937 & &-0.0053 &0.0979  & 0.0984 & 0.948 \\
      & 1000& &-0.0007 &0.0623  & 0.0622 & 0.950  & &-0.0088 &0.0869 &0.0878  & 0.945 \\
      \hline
      Case III
      & 400 & &-0.0033 &0.1562  & 0.1647 & 0.959 & &-0.0282 &0.2186  & 0.2190 & 0.949 \\
      & 600 & &-0.0004 &0.1244  & 0.1322 & 0.963 & &-0.0199 &0.1845  & 0.1777 &  0.942 \\
      & 800 & &-0.0015 &0.1113  & 0.1141 & 0.962 & &-0.0075 &0.1497  & 0.1533 & 0.948 \\
      & 1000& &0.0027  &0.0972  & 0.1015 & 0.958 & &-0.0085 &0.1358  & 0.1364 & 0.954\\
      \hline
      Case IV
      & 400 & &0.0037  & 0.0685  & 0.0672 & 0.939 &&-0.0122&0.0997 &0.0975 & 0.952\\
      & 600 & &-0.0002 & 0.0585  &0.0540 & 0.922  && -0.0026 & 0.0784 &0.0774 &0.944 \\
      & 800 & &0.0022  & 0.0469  & 0.0465 &0.948 & & -0.0061 &0.0701& 0.0663 &0.936 \\
      & 1000& &0.0018  & 0.0422  & 0.0414 &0.944& &-0.0047& 0.0599 &0.0588 & 0.945\\
      \hline
    \end{tabular}
  \end{center}
{\vspace{0cm} \hspace{-0.3cm}$\ddag$ \footnotesize  The (Bias, SE) of full data MPL estimator $\hat{\beta}_1$ is Case I: (Bias, SE) = (0.0026, 0.0029), Case II:  (Bias, SE) = (-0.0013, 0.0028), Case III: (Bias, SE) = (0.0012, 0.0042), Case IV: (Bias, SE) = (0.0021, 0.0019).}
\end{table}

We present the estimation results about $\beta_1$ in Tables~\ref{tab:3} and
\ref{tab:4}, indicating that both Lopt and UNIF estimators are asymptotically
unbiased. The SE and ESE are similar and the coverage
probabilities are close to the nominal level, which support the asymptotic
normality of the proposed estimator and demonstrate that the subsample-based
variance-covariance matrix estimator in (\ref{SE1}) is accurate. Both subsample
estimators get better as the sampling size $r$ increases. In addition, the Bias
and ESE of the Lopt estimator are much smaller than those of UNIF estimator with
the same subsample size $r$. This agrees with the conclusion in Theorem~\ref{Th3}.
Results for other regression coefficients are similar and thus are
omitted. {\black Furthermore, we report the Bias and SE of the full data MPL
estimator towards the true parameter in the footnotes of
Tables~\ref{tab:3} and \ref{tab:4}.}

We calculated the five-number summary statistics of $\pi_{\delta}^{\rm app}$'s for the
censored and uncensored observations separately to demonstrate the impact of
censoring on optimal subsampling probabilities numerically.
Table~\ref{tab:fiveS} reports the results, including the Minimum, Lower-hinge (the first quartile), Median, Upper-hinge (the third
quartile), and the Maximum. Uncensored observations have larger subsampling
probabilities than censored observations in general, i.e., uncensored
observations are more likely to be selected into a subsample by our subsampling method.

\begin{table}[htp] 
 \begin{center}
    \caption{The five-number summary statistics  for $\pi_{\delta}^{\rm app}$ ($\times10^{6}$)$^\ddag$.}
    \label{tab:fiveS}
    \vspace{0.1in} \small
    \begin{tabular}{lllcccccccccccccc}
      \hline
      &&  & &Minimum &Lower-hinge &Median &Upper-hinge &Maximum\\
      \hline
 CR=20\%    & Case I
 & $\pi_{\delta, c}^{\rm app}$ & &0.1000&0.1894 & 0.4307 & 1.0141& 28.3688\\
 && $\pi_{\delta,u}^{\rm app}$ & &0.1194&0.5507 & 0.8835 & 1.2732& 21.5598\\
   & Case II
 & $\pi_{\delta, c}^{\rm app}$ & &0.1000& 0.1621 & 0.3547 & 0.8058&28.1762\\
 && $\pi_{\delta,u}^{\rm app}$ & &0.1182& 0.5451 & 0.8588 & 1.3208&40.5288\\
    & Case III
 & $\pi_{\delta, c}^{\rm app}$ & &0.1000&0.1698 &0.3853 &0.8781& 58.9218\\
 && $\pi_{\delta,u}^{\rm app}$ & &0.1240&0.5132 &0.8090 &1.2228& 93.0236\\
     & Case IV
 & $\pi_{\delta, c}^{\rm app}$ & &0.1000& 0.1854 & 0.4018 &0.8461 & 33.6698\\
 && $\pi_{\delta,u}^{\rm app}$ & &0.1430& 0.5188 & 0.7948 &1.2636& 57.4029\\
       \hline
 CR=60\%    & Case I
 & $\pi_{\delta, c}^{\rm app}$ & &0.1000&0.1759 &0.4512 &0.9822& 21.7717\\
 && $\pi_{\delta,u}^{\rm app}$ & &0.1475&0.7905 &1.2350 &1.7154& 24.5950\\
   & Case II
 & $\pi_{\delta, c}^{\rm app}$ & &0.1000& 0.1801 & 0.3659 & 0.7793& 27.9329\\
 && $\pi_{\delta,u}^{\rm app}$ & &0.1163& 0.8631 & 1.3230 & 1.9279& 43.5112\\
    & Case III
 & $\pi_{\delta, c}^{\rm app}$ & &0.1000& 0.1607 & 0.3527 &  0.7858&54.5352\\
 && $\pi_{\delta,u}^{\rm app}$ & &0.1181& 0.8581 & 1.2927 & 1.8870 &61.6812\\
     & Case IV
 & $\pi_{\delta, c}^{\rm app}$ & &0.1000& 0.1259 & 0.2200 & 0.4839& 54.4077\\
 && $\pi_{\delta,u}^{\rm app}$ & &0.1724& 0.8910 & 1.4241 & 2.1763& 167.3048\\
 \hline
    \end{tabular}
  \end{center}
   {\vspace{0cm} \hspace{-0.3cm} $\ddag$ \footnotesize $\pi_{\delta, c}^{\rm app}$  and $\pi_{\delta,u}^{\rm app}$ denote the mixed  approximated optimal subsampling probabilities for censored and uncensored samples, respectively; $\delta=0.1$.}
\end{table}

Furthermore, we assessed the computational efficiency of our optimal subsampling
method. {\black For comparison, we also considered the UNIF, full data estimator
  and SGD estimator \cite[]{Cox-SGD}, where the full data estimator was
  calculated with the R function {\tt coxph} and the SGD estimator was obtained
  by the R function {\tt bigSurvSGD} (using default settings).}  The
computations were carried out using R \citep{R-cite} on a desktop computer with
64GB memory. We restricted the calculations to access one CPU core and recorded
the average CPU time from 100 repetitions.  Table~\ref{tab:6} reports the
results for Case \uppercase \expandafter {\romannumeral 1}, where the subsample
size is $r=1000$.  The computational speed of the Lopt estimator is much faster
than that of the full data estimator with {\tt coxph}. The computational burden
of the full data method gets heavier as the increase of full data sample size.
In other words, subsampling is desirable in Cox's regression because it reduce
the computational cost significantly.  The UNIF estimator is faster to compute
than the Lopt estimator, because it does not need the step of calculating the
sampling probabilities, but it has a lower estimation efficiency as we have seen
in previous results. {\black Note that the SGD estimator is slower than the full
  data estimator in terms of computation speed. We point out that the main aim
  of the SGD estimator was to deal with large datasets where {\tt coxph} cannot
  be used (due to out-of-memory issues) rather than speeding up the
  calculations.  In Table \ref{Tab:OU7r}, we present more comparisons between
  Lopt and UNIF methods when the CPU computation times are similar. It is seen
  that the Lopt and UNIF may have similar estimation efficiency using similar
  CPU times. However, the UNIF uses larger sample sizes and thus larger
  memory. The optimal subsampling method achieves the same estimation efficiency
  with less computing resources in these scenarios. }

\begin{table}[htp] 
  \begin{center}
    \caption{The CPU time for Case \uppercase \expandafter {\romannumeral 1} with  $r=1000$ (in seconds)$^\dagger$.} \label{tab:6}
    \vspace{0.1in}
    \begin{tabular}{llllllllll}
      \hline
      && & &\multicolumn{3}{c}{$n$}\\
      \cline{5-7}
      &Methods& & &$10^6$ &$5\times10^6$ &$10^7$  \\
      \hline
  CR = 20\%    &UNIF & & & 0.17   &  0.25    &  0.34        \\
               &Lopt  & & & 0.39   &  1.22    &  2.28         \\
          &full data & & & 6.75   &  45.71    & 100.65           \\
          &SGD       & & & 94.19  &  603.81    & 1294.70        \\
      \hline
  CR = 60\%    &UNIF & & & 0.10   &  0.17    &  0.27          \\
               &Lopt  & & & 0.32   &  1.17    & 2.33          \\
          &full data & & & 6.16   &  45.87    & 99.54        \\
          &SGD       & & & 99.50   & 530.84    & 1112.85      \\
      \hline
    \end{tabular}
  \end{center}
  \vspace{-0.3cm}
  {\hspace{2.8cm} \noindent$^\dagger$ \footnotesize ``full data": calculated with  R function {\tt coxph}; ``SGD": calculated \\
  {\vspace{0cm}
  \hspace{2.8cm} with R function {\tt bigSurvSGD}.}}
\end{table}

\begin{table}[htp] 
\begin{center} \caption{Comparisons of CPU times between Lopt and UNIF (in seconds)$^\dagger$.}
  \label{Tab:OU7r}
  {\rule{135mm}{0.16mm}\\
    \begin{tabular}{llcccccccccccccccc}
      &&\multicolumn{3}{c}{CR=20\%}&&\multicolumn{3}{c}{CR=60\%}\\
      \cline{3-5}\cline{7-9}
      &&CPU &$r$&MSE&&CPU &$r$&MSE&\\
      \hline
  Case I   & Lopt  &0.4304   & 1000  &0.01215    &&0.3834   &1100  &0.02047 \\
           & UNIF &0.4823   & 1700  &0.01113    &&0.3631   &2100  &0.01930 \\
          \hline
Case II    & Lopt  &0.3670   & 800   & 0.01515   &&0.4591   &1400  & 0.01602 \\
           & UNIF &0.3415   & 1400  &0.01519    &&0.4568   &2300  &0.01757 \\
            \hline
Case III    & Lopt  &0.3842   &850   &0.01667    &&0.4972   &1400  &0.02158 \\
            & UNIF &0.4465   &1500  &0.01675    &&0.5259   &2300  &0.02424 \\
          \hline
Case IV    & Lopt  &0.4749   &1100  & 0.00575    &&0.4097   &1100  & 0.00819 \\
           & UNIF &0.5260   &1650  &0.00565     &&0.3726   &2000  & 0.00882 \\
    \end{tabular}\\
    \rule{135mm}{0.16mm}}\\
  {\vspace{-0.0cm}  \hspace{-1.0cm}\footnotesize $\dagger$ ``CPU" denotes average CPU time from 100 repetitions; the full data size is  $n=10^6$.}
\end{center}
\end{table}

{\black
Finally, we compared the two subsampling probabilities derived from  the
$L$-optimality criterion (Lopt) and $A$-optimality criterion (Aopt), respectively. By Remark \ref{remark-N4},
the optimal subsampling probabilities under the $A$-optimality criterion are obtained by minimizing
tr($\mathbf{\Sigma}$), where $\mathbf{\Sigma}$ is given in (\ref{Eq4}).  Using a similar
deduction as that of (\ref{Aosp}), we can obtain the approximated optimal subsampling
probabilities under the A-optimality criterion:
\begin{align}\label{Aopt-11}
 \pi_{i}^{\text{Aopt}} = \frac{\|\mathbf{\Psi}^{0*-1}\int_0^\tau\{\X_i - \bar{\X}^{0*}(t, {\tbeta_0})\}d\hat{M}_i(t,\tbeta_0)\|}{\sum_{j=1}^n \|\mathbf{\Psi}^{0*-1}\int_0^\tau\{\X_j - \bar{\X}^{0*}(t, {\tbeta_0})\}d\hat{M}_j(t,\tbeta_0)\|},~~i=1,\cdots,n,
\end{align}
where
\begin{align*}
{\mathbf{\Psi}^{0*}}
  &= \frac{1}{r_0}\sum_{i=1}^{r_0} {\Delta_i^{0*}} \left[\frac{S^{0*(2)}(Y_i^{0*},\tbeta_0)}{S^{0*(0)}(Y_i^{0*}, \tbeta_0)} -  \left\{\frac{S^{0*(1)}(Y_i^{0*}, \tbeta_0)}{S^{0*(0)}(Y_i^{0*}, \tbeta_0)}\right\}^{\otimes 2}  \right],
\end{align*}
and $S^{0*(k)} (Y_i^{0*},\tbeta_0) = {(r_0)}^{-1} \sum_{j=1}^{r_0}I(Y_j^{0*} \geq Y_i^{0*})
\X_j^{0*\otimes k}\exp(\tbeta_0^\prime \X^{0*}_j)$ for
$k=0,1,2$. The corresponding mixed subsampling probabilities with A-optimality criterion are
\begin{align*}
\pi_{\delta i}^{\text{Aopt}} = (1-\delta) \pi_{i}^{\text{Aopt}} + \frac{\delta}{n},~~ i=1,\cdots,n.
\end{align*}
In Figures \ref{fig1} and \ref{fig2}, we report the empirical MSEs of subsample
estimators with Lopt, Aopt and UNIF methods, where $\delta = 0.1$. The results
indicates that Lopt and Aopt have similar performance. In addition, the UNIF has
the largest MSE compared with Lopt and Aopt methods. It is clear that the speed
of Aopt is slower than Lopt, because there is an additional term
${\mathbf{\Psi}^{0*}}$ involved in (\ref{Aopt-11}).  As a summary, we recommend
using the Lopt for our subsampling method in practical applications.

}

 \begin{figure}[htp] 
    \includegraphics[width=3.0in]{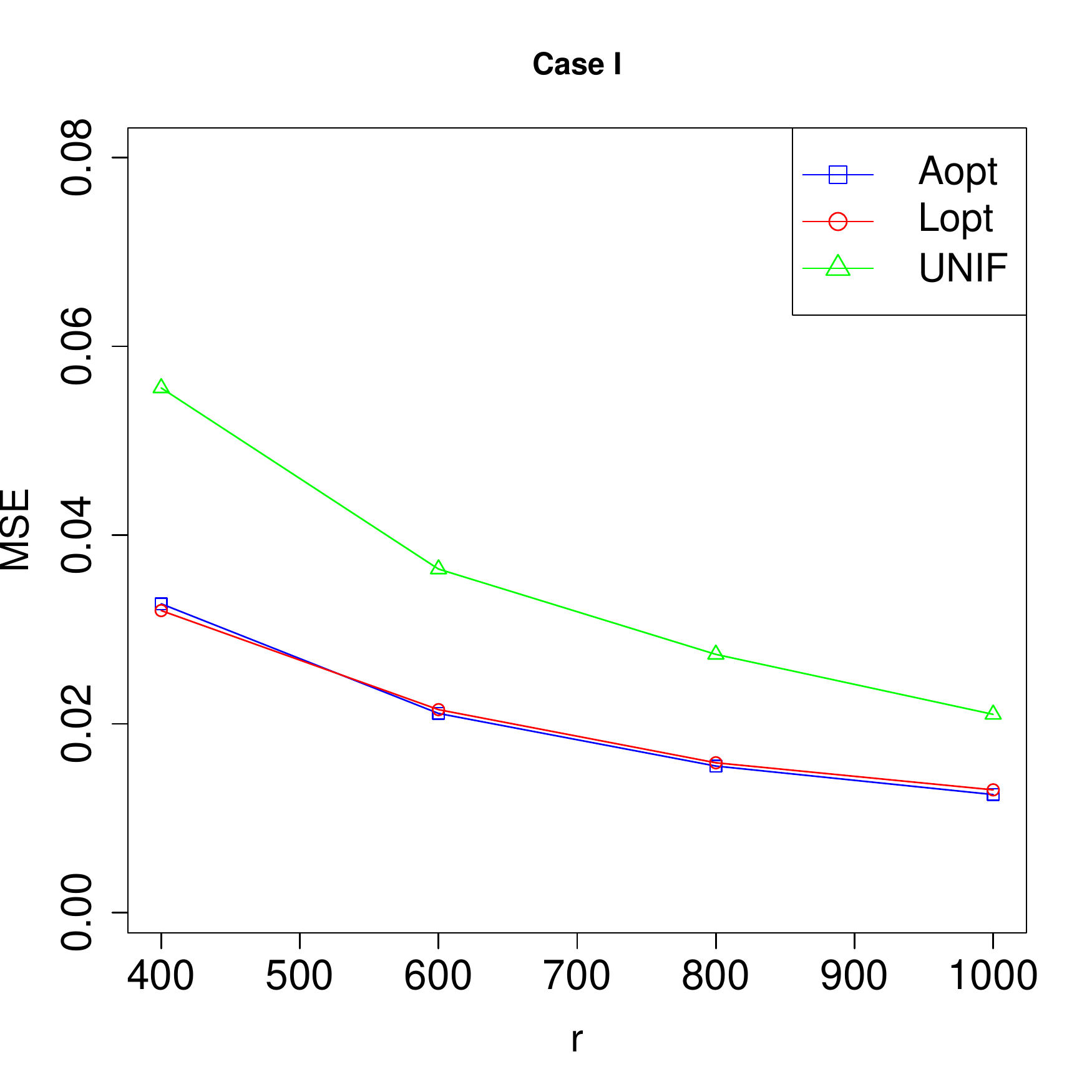}\hfill\includegraphics[width=3.0in]{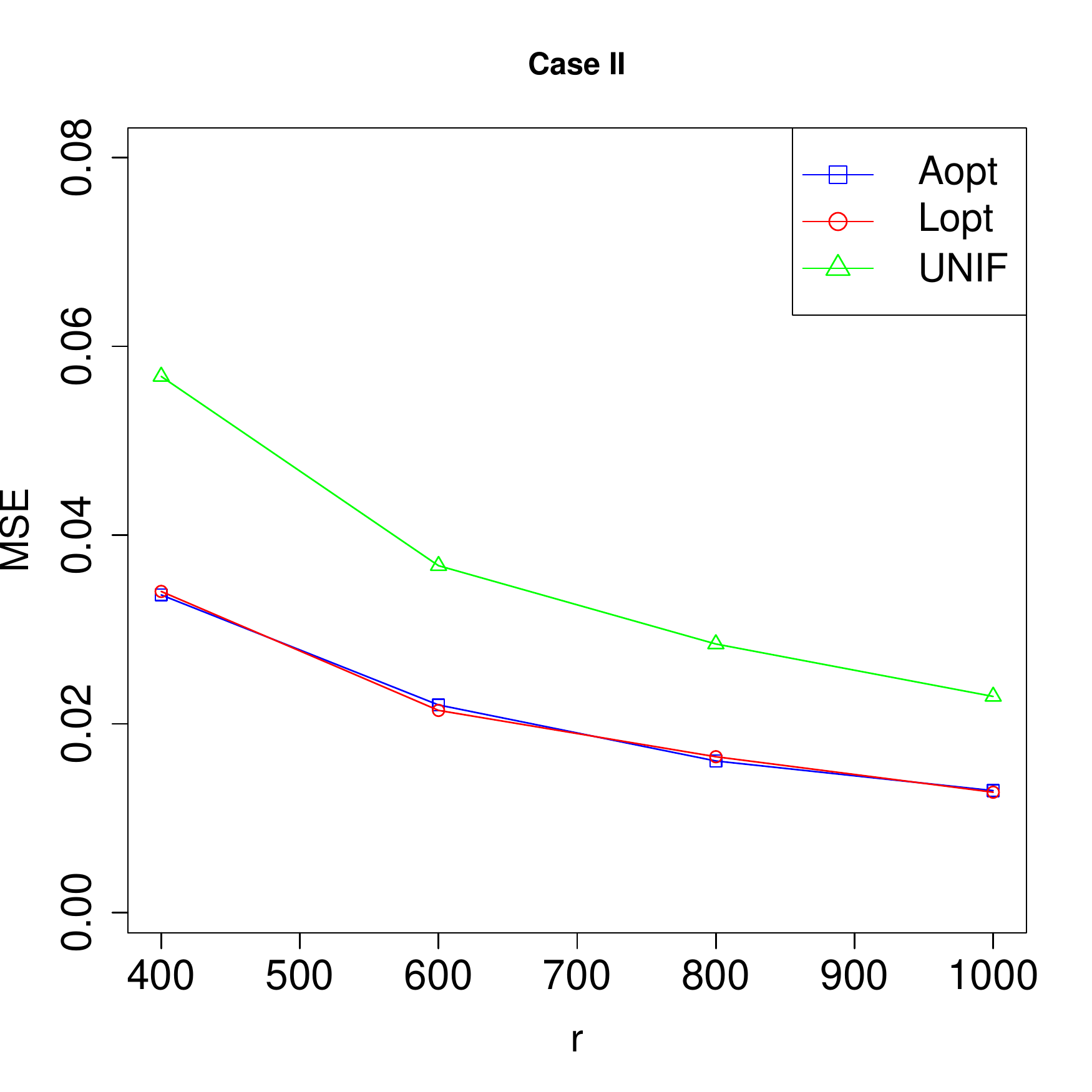}\\
    \includegraphics[width=3.0in]{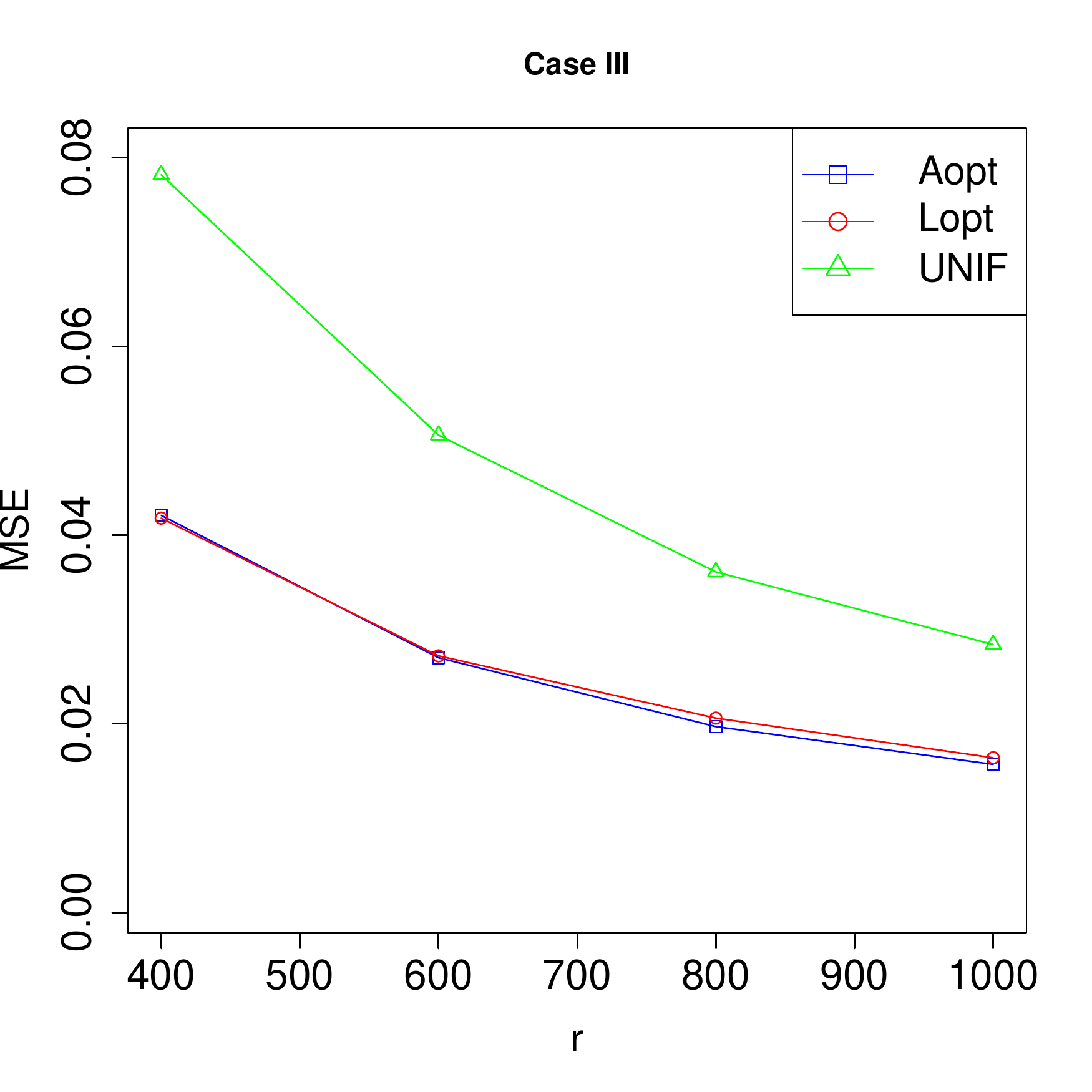}\hfill\includegraphics[width=3.0in]{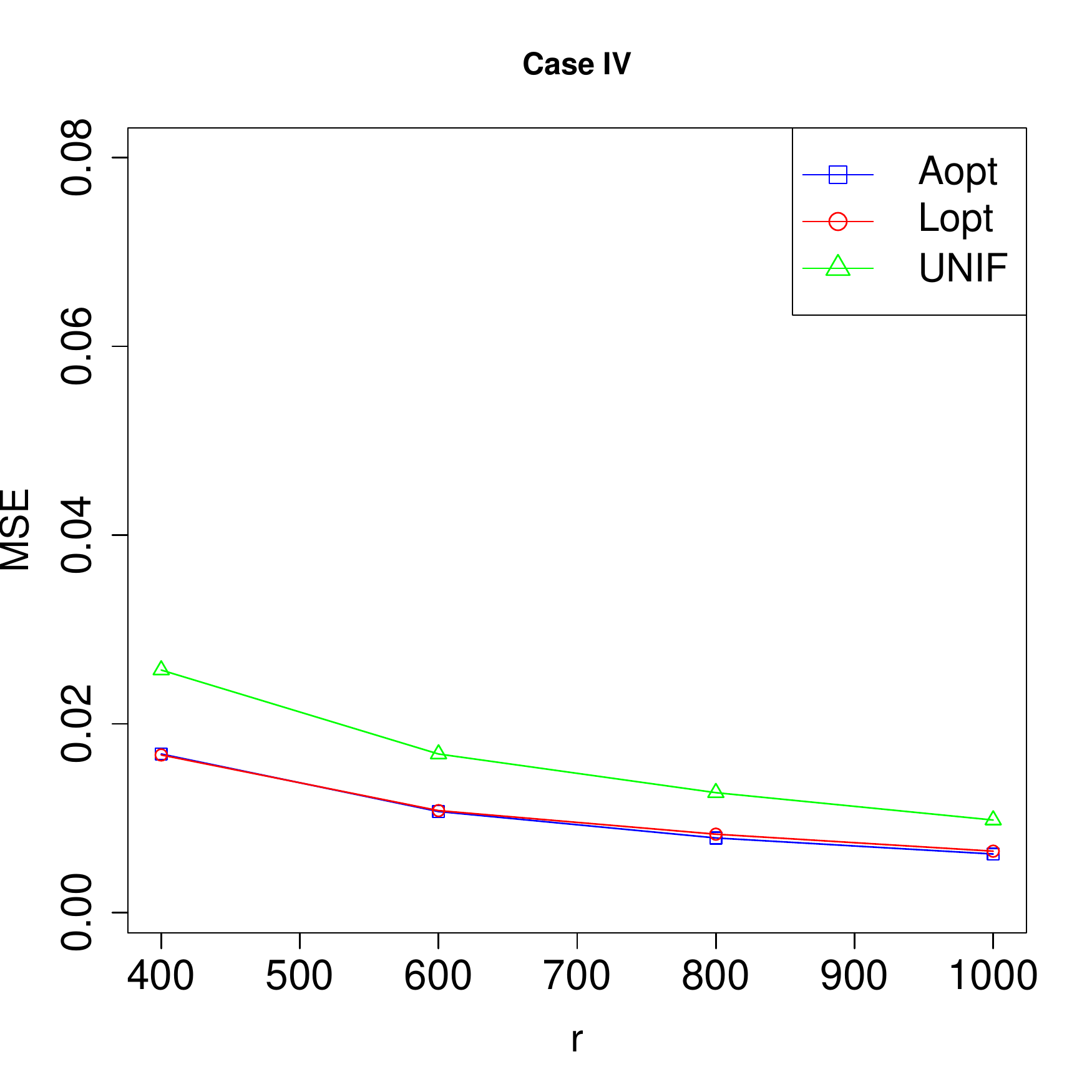}\\
    \begin{center}
     \caption{The MSEs for different subsampling methods
        with CR= 20\%.} \label{fig1}
    \end{center}
  \end{figure}

\begin{figure}[H]
\includegraphics[width=3.0in]{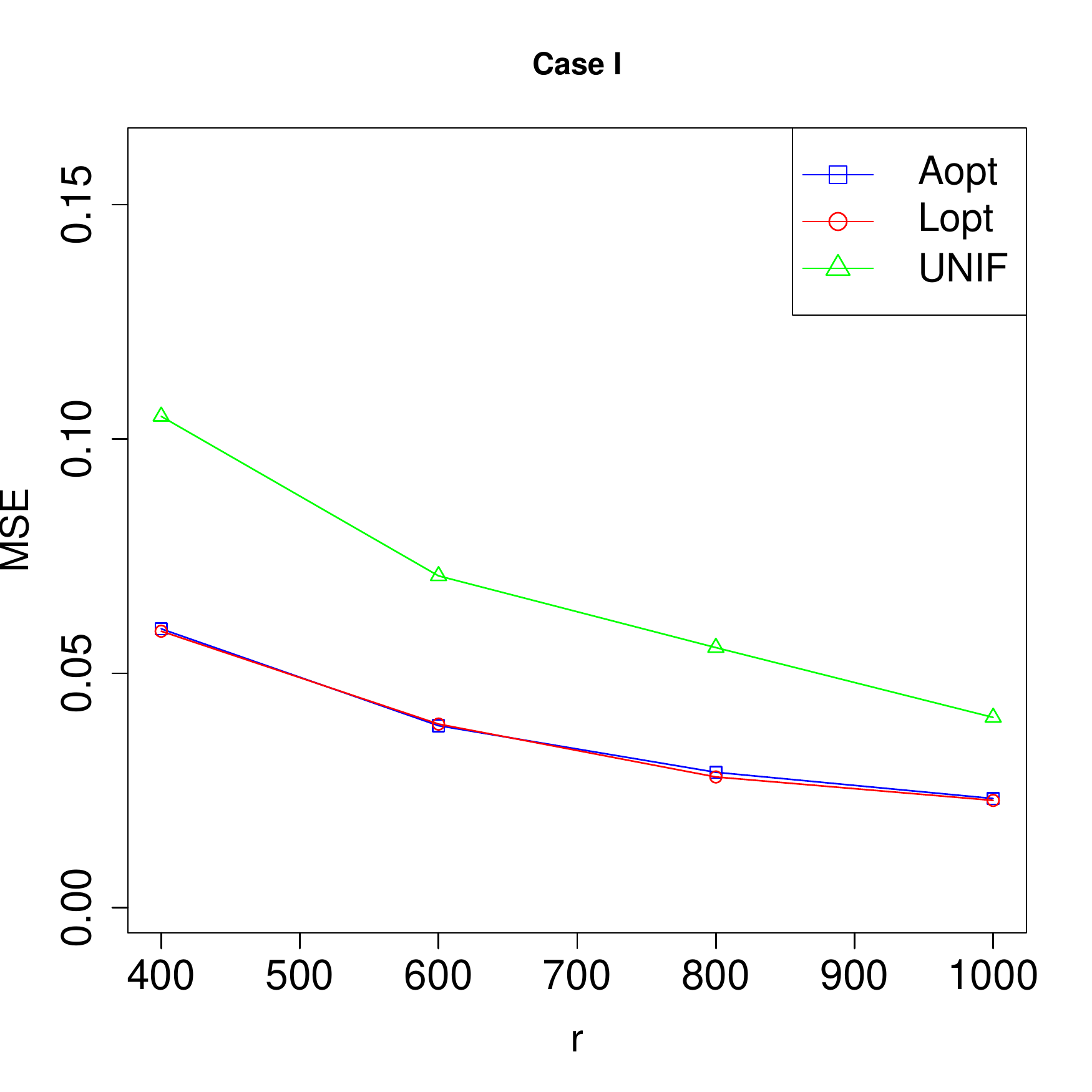}\hfill\includegraphics[width=3.0in]{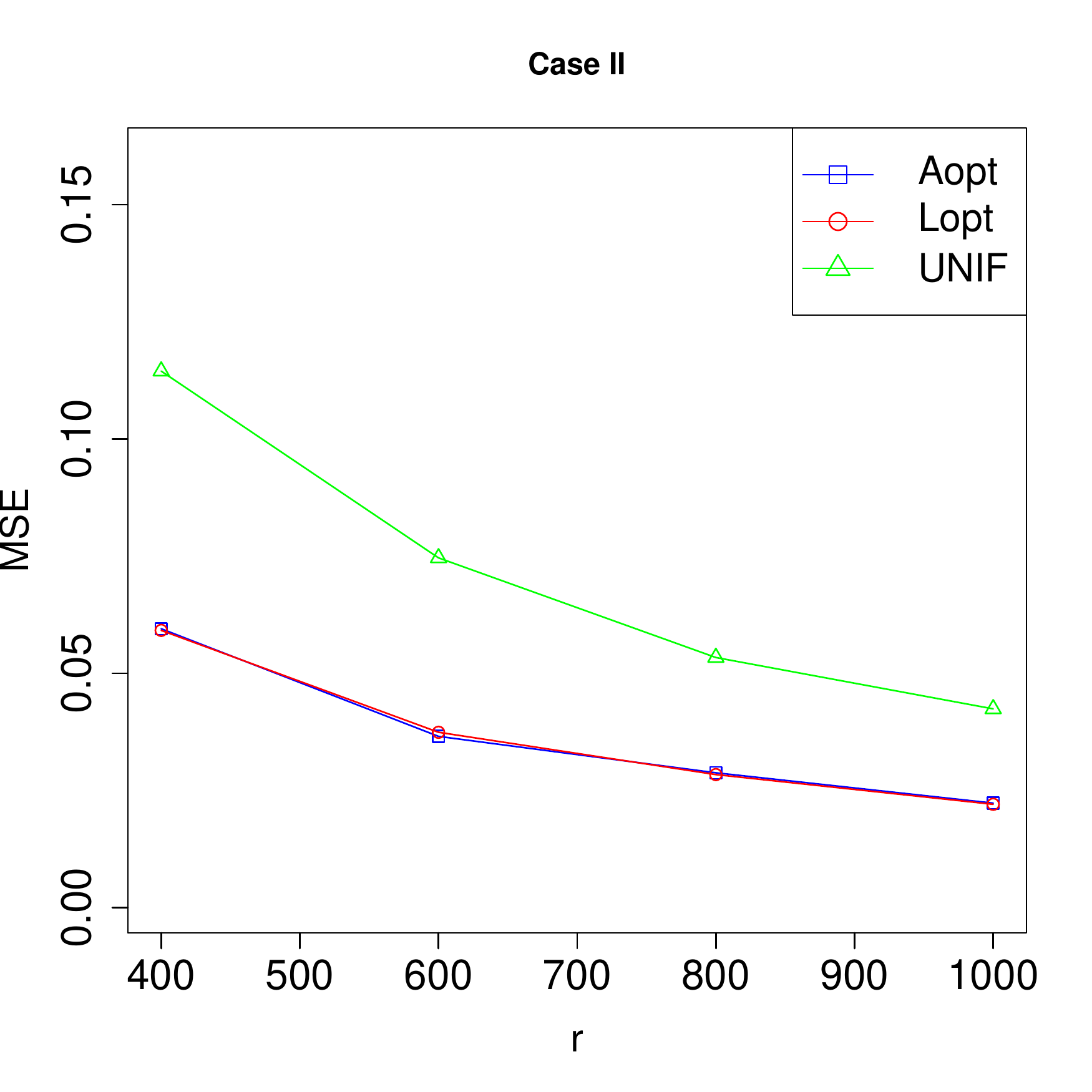}\\
\includegraphics[width=3.0in]{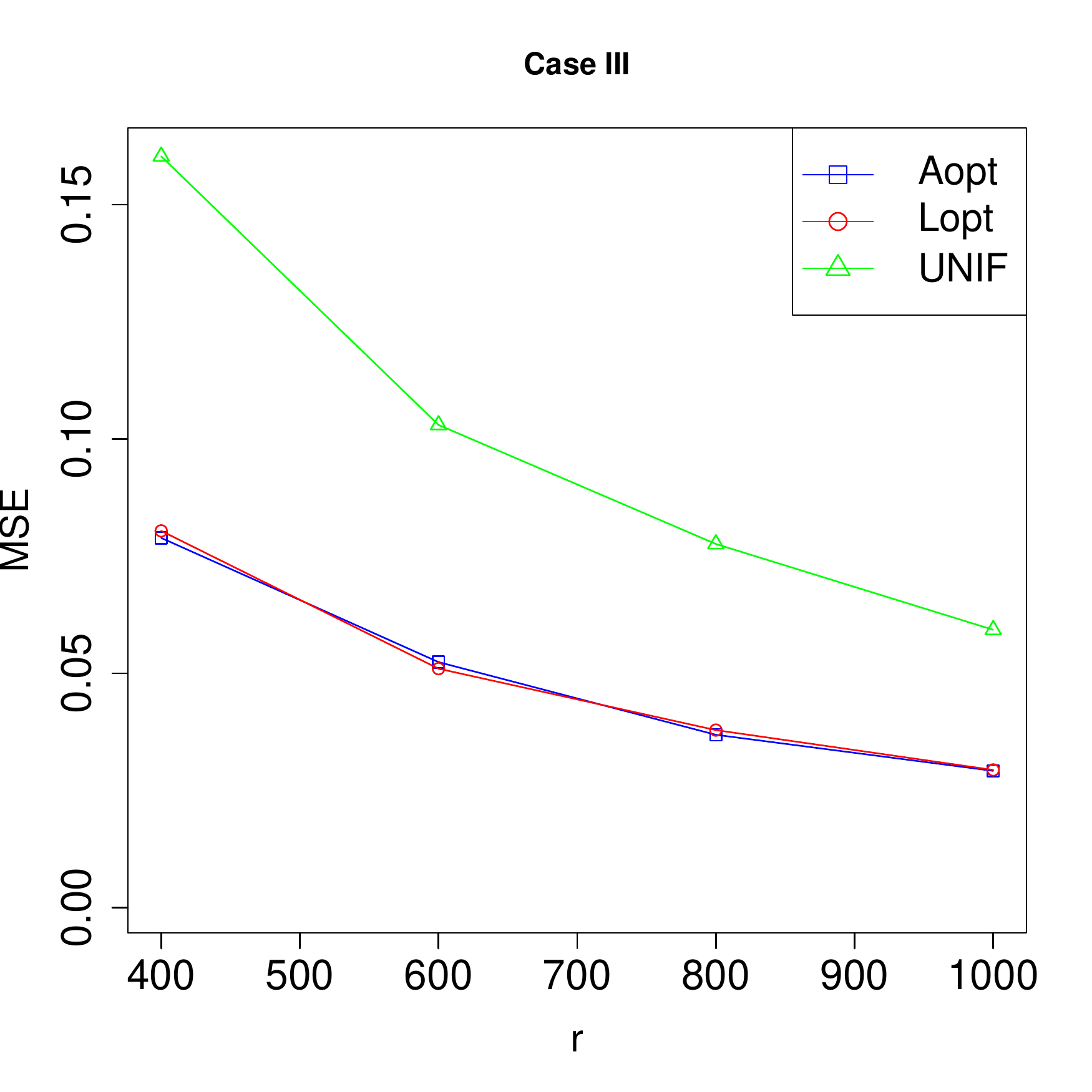}\hfill\includegraphics[width=3.0in]{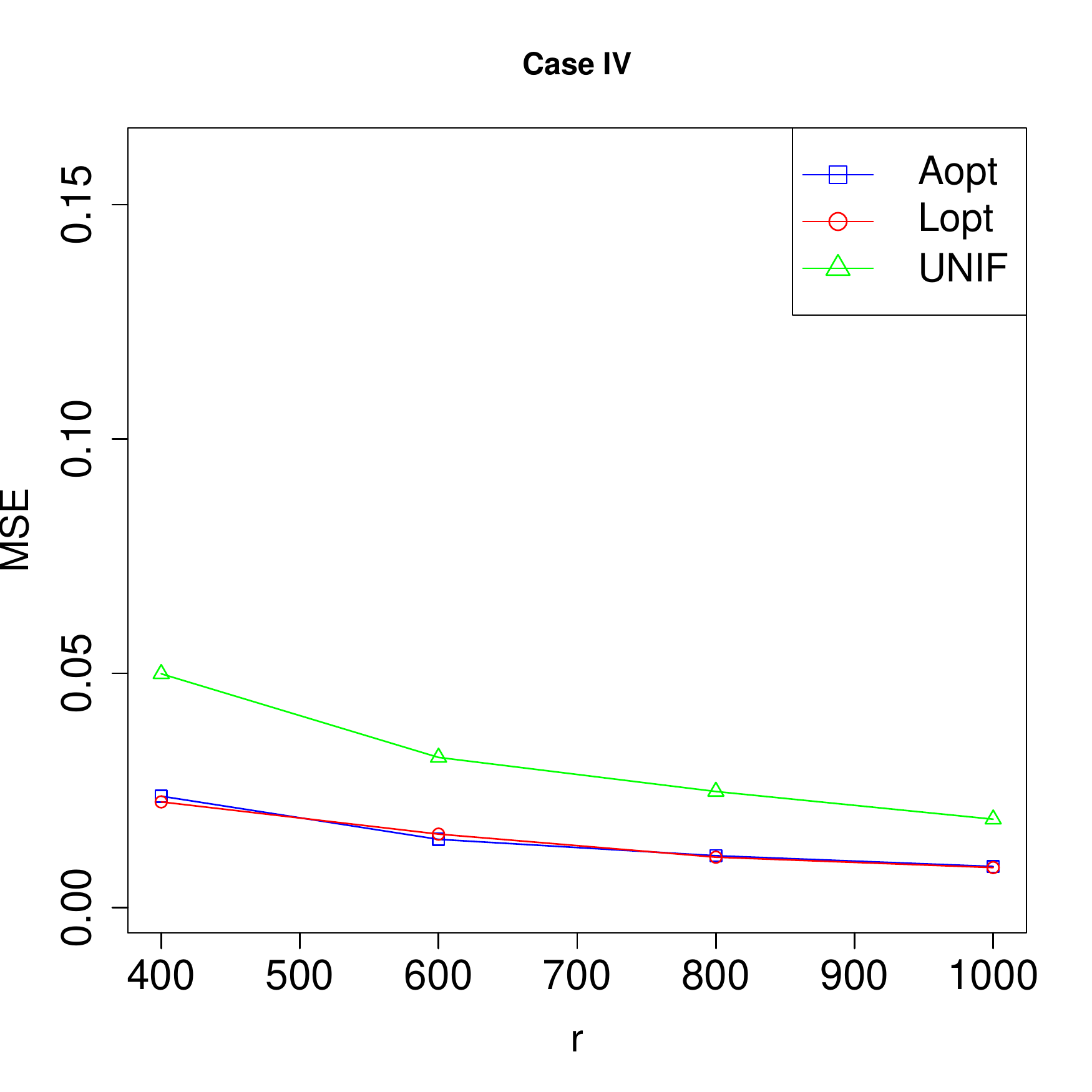}\\
\begin{center}
\caption{The MSEs for different subsampling methods with  CR= 60\%.}\label{fig2}
\end{center}
\end{figure}
\subsection{Application}
\label{realdata52}
{\black
In this section, we apply the proposed subsampling method to a real-world data example about
the USA airline, where the dataset is publicly available from \cite{DVN2008-data}. We are interested in analysing those arrival delayed  airlines, where there were totally 57,729,435 arrival delayed commercial flights within the USA from October 1987 to April 2008. For the $i$th airline,  the failure time $T_i$ is defined as the delayed time from  scheduled arrival time to actual arrival time (in minutes).  Among those 57,729,435 arrival delayed flights, 33,142,872 subjects  experienced an actual arrival within 15 minutes (the delayed arrival time is less than 15 minutes). The censoring rate is about 42.6\%.  For analysis, the risk factors $\X_i = (X_{i1},X_{i2})^\prime$ in Cox's
model are departure status (departure on time or ahead of schedule = 0 and departure delayed = 1) and distance (continuous, in thousands of miles), respectively.}

\begin{table}[htp] 
  \begin{center}
    \caption{Estimation results for the airline arrival delay data with one subsample.}
    \label{tab:8S}
    \vspace{0.1in}
    \begin{tabular}{lccccccccc}
      \hline
      & &  & \multicolumn{3}{c}{Lopt} &  & \multicolumn{3}{c}{UNIF} \\
      \cline{4-6}\cline{8-10}
      & $\beta$ & &Est &SE &CI & & Est &SE &CI\\
      \hline
      $r=400$
& $\beta_1$ &&-1.0009 &0.1303 &(-1.2562, -0.7456) &&-1.2038 & 0.1353&(-1.4691, -0.9385)\\
& $\beta_2$ &&-0.3188 &0.1194 &(-0.5527, -0.0848) &&-0.2471 & 0.1296&(-0.5009, 0.0069) \\
      \hline
      $r=600$
& $\beta_1$ &&-1.0633 &0.1036 &(-1.2663, -0.8602) &&-1.0865 &0.1298 &(-1.3409, -0.8322)\\
& $\beta_2$ &&-0.2341 &0.0869 &(-0.4045, -0.0637) &&-0.3207 &0.1195 &(-0.5549, -0.0865)\\
      \hline
      $r=800$
& $\beta_1$ &&-1.1121 &0.0826 &(-1.2741, -0.9502) &&-1.0575 & 0.1001&(-1.2536, -0.8613) \\
& $\beta_2$ &&-0.3418 &0.0689 &(-0.4769, -0.2067)&&-0.2674 & 0.1031 &(-0.4694, -0.0654) \\
      \hline
      $r=1000$
& $\beta_1$ &&-1.1099 &0.0745 &(-1.2559, -0.9638) &&-1.1147& 0.0857 &(-1.2826, -0.9468) \\
& $\beta_2$ &&-0.2498 &0.0598 &(-0.3669, -0.1326) &&-0.2505& 0.0882 &(-0.4233, -0.0777) \\
      \hline
    \end{tabular}
  \end{center}
\end{table}

{\black
For comparison, we calculated the full data estimator
$\hbeta =(-1.1301, -0.2396)^\prime$ with {\tt coxph}, and the corresponding SEs are  0.00035 and 0.00034, respectively. Hence, departure status and distance have negative effects on the hazard rate of  airline's arrival. That is to say, it is expected that a departure delayed airline with long distance would owns a longer arrival delay. In addition,  we calculated the Lopt
estimator with $\delta=0.1$ and the UNIF estimator. We present the
results on the subsampling-based estimator (Est), the SE and the 95\% confidence
interval (CI) based on one subsample in Table~\ref{tab:8S}. Both Lopt and UNIF
estimators are close to $\hbeta$, especially when the subsample size is large
(e.g. $r=1000$). The SE of the Lopt estimator is much smaller than that of the UNIF
estimator, which supports the theoretical conclusion in Theorem~\ref{Th2}.  To further validate the usefulness of our method, we
report the Bias, SE and ESE of the subsample-based estimates based on 1000
subsamples in Table~\ref{tab:9S}, {\black where the Bias denotes the average
  bias of the subsampling estimator with respect to the full-data MPL estimator.} Both subsample-based estimates are
unbiased, and the SE is close to ESE indicating that the estimated
variance-covariance matrix in (\ref{SE1}) works well. The results in
Table~\ref{tab:9S} again demonstrate that the Lopt estimator is more efficient than the UNIF estimator. Finally, the full data MPL estimator with {\tt coxph} needs 265.58 seconds, where the computer is the same as that used in the simulation. For $r=1000$, the Lopt only requires 9.03 seconds to output the subsample estimators and their SEs (UNIF needs 1.08 seconds). i.e., the Lopt method has a much faster
computation efficiency than the {\tt coxph} when we face with large-scale survival dataset in practice.
}

\begin{table}[htp] 
  \begin{center}
    \caption{The Bias and (ESE, SE) for subsample estimates in real data.} \label{tab:9S}
    \vspace{0.1in}
    \begin{tabular}{lcccc}
      \hline
      & $\beta$ & &Lopt &UNIF   \\
      \hline
      $r=400$
      &$\beta_1$&   &0.0030  (0.1228, 0.1289) &-0.0059 (0.1351, 0.1466) \\
      &$\beta_2$&   &0.0022  (0.1033, 0.1094) &-0.0035 (0.1346, 0.1419) \\

      \hline
      $r=600$
      &$\beta_1$&  &0.0003  (0.1017, 0.1016) &-0.0055 (0.1129, 0.1087) \\
      &$\beta_2$&  &-0.0018 (0.0833, 0.0804) &-0.0019 (0.1103, 0.1061) \\

      \hline
      $r=800$
      &$\beta_1$&   &0.0019  (0.0827, 0.0854) &-0.0026 (0.0982, 0.0932) \\
      &$\beta_2$&   &-0.0002 (0.0680, 0.0692) &-0.0008 (0.0938, 0.0912) \\

      \hline
      $r=1000$
      &$\beta_1$&   &0.0002  (0.0799, 0.0781) &-0.0009 (0.0859, 0.0886) \\
      &$\beta_2$&   &-0.0027 (0.0616, 0.0657) &-0.0030 (0.0849, 0.0859) \\
      \hline
    \end{tabular}
  \end{center}
\end{table}

\section{Concluding Remarks}\label{sec6}

In this paper, we have studied the statistical properties of a general
subsampling algorithm for Cox's model with massive survival data.
We provided the optimal subsampling probabilities, and established
asymptotic properties of the two-step subsample-based parameter estimator
conditional on the full data.  Extensive simulations and a real
data example have been used to validate the practical usefulness of our
method. {\black  Note that the proposed approach is appropriate when the outcome of interest is common and the dataset includes enough observed events, i.e., our subsample method is suitable for the regular time-to-event data. Faced with
massive survival datasets with rare events, \cite{rare-Cox-2020} proposed a
novel
and interesting subsampling procedure to deal with computational challenges in
massive data Cox regression. Their procedure is based on
counting process type score function, while we derive the asymptotic
distribution of subsample-based estimator from the martingale-type
subsample score function. \cite{rare-Cox-2020} avoided
the need
of estimating the cumulative baseline hazard function for the optimal subsampling  probabilities, which was in contrast to our approach.  In addition, our proposed subsample estimator approximates the maximizer of the full data
partial likelihood, and the approximation error is not significantly affected by
the correctness of the Cox model. In other words, if the proportional
hazards assumption is violated, the subsample estimator is still close to the
full data estimator, but the full data estimator may not be the best estimator
any more.

There are four important topics for further research. First, it is desirable to
investigate how the proportional hazards assumption can be adequately checked
based on subsamples. Second, 
the numerator of $\pi_{i}^{\text{app}}$ only involves the
$i$th subject and
the pilot subsample, which sheds light on the feasibility of distributed or
parallel algorithms when calculating the optimal sampling probabilities. For example, by splitting the full data into multiple blocks, it is possible to calculate the terms
$\|\int_0^\tau\{\X_i - \bar{\X}^{0*}(t, {\tbeta_0})\}d\hat{M}_i(t,\tbeta_0)\|$
with distributed computing environments. In this case, the computational speed of optimal subsampling method would be significantly improved.
  Third, in many practical applications, observed data are often corrupted by outliers \cite[]{LowCon-2021}. Therefore,
 it is interesting to study the optimal subsampling method for Cox's model with the presence of outliers.
 Forth, the tuning parameter $\delta$ perform well with a value of 0.1 in our
 numerical results, but there is not theoretically justification to show that
 this value will work in all scenarios. How to select $\delta$ attentively
 requires further investigations.

}

\section*{Supplementary Materials}
{\black
\begin{description}
\item[Supplement] The supplementary PDF file contains proofs of all the
theoretical results in this paper.
\item[R codes] The zip file contains
the R codes used to perform the subsampling methods described in the article, where the readme file describes details about the codes.
\end{description}
}

\section*{Acknowledgement}
The authors would like to thank the Editor, the Associate Editor and two
reviewers for their constructive and insightful comments that greatly
improved the manuscript. We also thank Aliasghar Tarkhan for providing some helpful comments on the usage of
R package ``bigSurvSGD". The work of Wang was supported by National Science Foundation (NSF), USA grant
CCF-2105571. The work of Sun was supported in part by the National Natural
Science Foundation of China (Grant No. 12171463).

\section*{Conflict of Interest}
The authors declare that there are no conflicts of interest.


\clearpage

\begin{center}
  Supplementary Materials for\\
  \large\bf Approximating Partial
  Likelihood Estimators via Optimal Subsampling
\end{center}
\vspace*{0.2in}
\centerline{ {  Haixiang Zhang$^{}$, Lulu Zuo$^{}$,  HaiYing Wang$^{}$ and Liuquan Sun$^{}$ }}

\appendix

\renewcommand{\theequation}{S.\arabic{equation}}
\renewcommand{\thefigure}{S.\arabic{figure}}
\renewcommand{\thelemma}{S.\arabic{lemma}}
\renewcommand{\thetable}{S.\arabic{table}}

\section{Proofs}
\setcounter{equation}{0}

In this section, we give the proof details of Theorems \ref{Th1}-\ref{Th3} and Proposition \ref{Th4}.  For these goals, we first need the following lemmas.\\

\begin{lemma}\label{Lem0}\cite[]{CoxXP-2009-JASA}
Suppose that as $n\rightarrow \infty$,
\begin{align*}
\sup_{t\in[0,\tau]} |h_n(t) - h(t)| \rightarrow 0, ~~~ \sup_{t\in[0,\tau]}|g_n(t) - g(t)|\rightarrow 0,
\end{align*}
where $h$ is continuous on $[0, \tau]$, $g_n(\cdot)$ and $g(\cdot)$ are left-continuous on
$[0,\tau]$,  with their total variations bounded by a constant that is independent of $n$. Then, $n\rightarrow \infty$,
\begin{align*}
\sup_{t\in[0,\tau]} \left|\int_0^t h_n(u)dg_n(u) - \int_0^t h(u)dg(u)\right| \rightarrow 0,
\end{align*}
and
\begin{align*}
\sup_{t\in[0,\tau]} \left|\int_0^t h_n(u)dg_n(u) - \int_0^t h_n(u)dg(u)\right| \rightarrow 0.
\end{align*}
\end{lemma}

\vspace{0.5cm}

\begin{lemma} 
Suppose the assumptions \ref{assu1}-\ref{assu4} hold, then as
  $n \rightarrow \infty$ and $r \rightarrow \infty$, conditional on
  $\mathcal{D}_n$, for any $\bbeta \in \Theta$ we have
  \begin{eqnarray}\label{A1}
    \mathbf{U}^*(\bbeta) = \dot{\ell}(\bbeta) + O_{P|\mathcal{D}_{n}}(r^{-1/2}),
\end{eqnarray}
and
\begin{eqnarray}\label{A2}
  \dot{\ell}^*(\bbeta) = \mathbf{U}^*(\bbeta)  + o_{P|\mathcal{D}_{n}}(r^{-1/2}),
\end{eqnarray}
where $\Theta $ is a compact set containing the true value of $\bbeta$, $\dot{\ell}(\bbeta)$ and $\dot{\ell}^*(\bbeta)$ are given in (\ref{SCF1}) and (\ref{SS24}), respectively. Moreover,
$$\mathbf{U}^*(\bbeta) = -\frac{1}{nr}\sumr \frac{1}{\pi_i^*} \int_0^\tau \{\X_i^* - \bar{\X}(t,\bbeta)\} dM^*_i(t,\bbeta)$$ with $dM^*_i(t,\bbeta) = dN^*_i(t) - I(Y_i^* \geq t)\exp(\bbeta^\prime \X_i^*)\lambda_0(t) dt$, $i=1,\cdots,r$.
\end{lemma}

\noindent{\bf Proof.}  For $i=1,\cdots,r$, denote
\begin{eqnarray*}
  \zeta_i^*(\bbeta) = -\frac{1}{n\pi_i^*} \int_0^\tau \{\X_i^* - \bar{\X}(t,\bbeta)\} dM^*_i(t,\bbeta).
\end{eqnarray*}
Conditional on $\mathcal{D}_n$, $\zeta_1^*(\bbeta)$, $\cdots$,
$\zeta_r^*(\bbeta)$ are independent and identically distributed random vectors,
it is straightforward to derive that
\begin{eqnarray*}
  E\{\zeta_1^*(\bbeta) | \mathcal{D}_n\} &=& -\onen \sumn \int_0^\tau \{\X_i - \bar{\X}(t,\bbeta)\}dM_i(t,\bbeta)\\
                                         &=& \dot{\ell}(\bbeta).
\end{eqnarray*}
Note that $\mathbf{U}^*(\bbeta) = r^{-1}\sumr \zeta_i^*(\bbeta)$, then
$E\{\mathbf{U}^*(\bbeta)|\mathcal{D}_n\} = E\{\zeta_1^*(\bbeta) |
\mathcal{D}_n\} = \dot{\ell}(\bbeta)$.

Let $\mathbf{U}_j^*(\bbeta)$ be the $j$th component of $\mathbf{U}^*(\bbeta)$
for $j=1,\cdots,p$, then we have
\begin{eqnarray*}
  Var\{\mathbf{U}_j^*(\bbeta) | \mathcal{D}_n\}&=& \frac{1}{n^2 r}\sumn \frac{1}{\pi_i} \left[\int_0^\tau\{ \X_{ij} - \bar{\X}_j(t,\bbeta)\}dM_i(t,\bbeta)\right]^2\\
                                               &&- \frac{1}{n^2 r}\left[\sumn\int_0^\tau\{ \X_{ij} - \bar{\X}_j(t,\bbeta)\}dM_i(t,\bbeta)\right]^2\\
                                               &\leq& \frac{1}{n^2 r} \sumn \frac{1}{\pi_i}\left\|\int_0^\tau \{\X_{i} - \bar{\X}(t,\bbeta)\}dM_i(t,\bbeta)\right\|^2\\
                                               &\leq& \max_{1\leq i \leq n}\left\{\frac{1}{n\pi_i}\right\} \frac{1}{n r} \sumn \left\|\int_0^\tau \{\X_{i} - \bar{\X}(t,\bbeta)\}dM_i(t,\bbeta)\right\|^2\\
                                               &=& O_P(r^{-1}).
\end{eqnarray*}
Here the last equality is from the assumption \ref{assu4}, together with
\begin{align*}
  \onen \sumn \left\|\int_0^\tau \{\X_{i} - \bar{\X}(t,\bbeta)\}dM_i(t,\bbeta)\right\|^2 = O_P(1),
\end{align*}
which can be deduced by the boundedness of $\X_i$'s in $\mathcal{D}_n$,
together with the assumptions \ref{assu1} and \ref{assu3}.  The Markov's inequality implies
that $\mathbf{U}_j^*(\bbeta) - \dot{\ell}_j(\bbeta) = O_{P|\mathcal{D}_{n}}(r^{-1/2})$.
Therefore,
$\mathbf{U}^*(\bbeta) = \dot{\ell}(\bbeta) +
O_{P|\mathcal{D}_{n}}(r^{-1/2})$, i.e., the conclusion given in (\ref{A1}) is
established.

For the sake of proving (\ref{A2}), we rewrite the expression of
$\dot{\ell}^*(\bbeta)$ as
\begin{align}\label{LLU}
  \dot{\ell}^*(\bbeta) &= -\frac{1}{rn} \sumr \frac{1}{\pi_i^*} \int_0^\tau \{ \X_i^*  - \bar{\X}(t, \bbeta) +  \bar{\X}(t, \bbeta) - \bar{\X}^{*}(t, \bbeta)\}dM_i^*(t,\bbeta)\nonumber\\
                       &= \mathbf{U}^*(\bbeta)  + \underbrace{\frac{1}{rn} \sumr \frac{1}{\pi_i^*} \int_0^\tau\{\bar{\X}^{*}(t, \bbeta)-\bar{\X}(t, \bbeta) \}dM_i^*(t,\bbeta)}_{\mathbf{R}^*(\bbeta)}.
\end{align}

Recall that
$\bar{\X}^{*}(t, \bbeta) = S^{*(1)}(t, \bbeta)/ S^{*(0)}(t, \bbeta)$,
where
$ {S}^{*(1)}(t,\bbeta) = (nr)^{-1} \sum_{i=1}^{r} \pi_i^{*-1}I(Y_i^{*} \geq t)
\X_i^{*} \exp (\bbeta^\prime \X_i^{*})$ and
$S^{*(0)}(t,\bbeta) = (nr)^{-1} \sum_{i=1}^{r} \pi_i^{*-1}I(Y_i^{*} \geq t)
\exp (\bbeta^\prime \X_i^{*}).  $ Conditional on $\mathcal{D}_n$,
$(\X_i^{*}, Y_i^{*}, \pi_i^{*})$'s are independent and identically
distributed variables. For  a subsample $\mathcal{D}_r^* = \{Z_i^*\}_{i=1}^r$ with $Z_i^* = (\X_i^*, \Delta_i^*, Y_i^*, \pi_i^*)$, we define a subsample empirical measure given the full data $\mathcal{D}_n$,
\begin{eqnarray*}
\mathbb{P}_{r|\mathcal{D}_n} = \oner\sumr \delta_{Z_i^*},
\end{eqnarray*}
where $\delta_Z$ is a measure that assigns mass 1 at $Z$ and 0 elsewhere. For a measurable function $f: \mathcal{D}_n \mapsto \mathbb{R}$, we denote
\begin{eqnarray*}
\mathbb{P}_{r|\mathcal{D}_n}f = \oner\sumr f(Z_i^*).
\end{eqnarray*}
Using the conditional empirical measure $\mathbb{P}_{r|\mathcal{D}_n}$,
we can rewrite the $ {S}^{*(k)}(t,\bbeta)$ as
\begin{eqnarray*}
{S}^{*(k)}(t,\bbeta) = \mathbb{P}_{r|\mathcal{D}_n}\{(n\pi^*)^{-1}I(Y^*\geq t)\X^{*\otimes k}\exp(\bbeta^\prime \X^*)\}, ~~k=0,1, 2.
\end{eqnarray*}
In order to use the technique of  empirical process \cite[]{van-emprical-1996}, we denote $\mathbf{P}_{\mathcal{D}_n}$ as
taking expectation conditional on the full data $\mathcal{D}_n$. e.g.
\begin{eqnarray}\label{PD-726}
\mathbf{P}_{\mathcal{D}_n} f (Z^*) = E\left\{f(Z^*)~\vline ~\mathcal{D}_n\right\} = \sumn \pi_i f(Z_i).
\end{eqnarray}
From (\ref{PD-726}), we have the following expressions:
\begin{eqnarray*}
\mathbf{P}_{\mathcal{D}_n} \{(n\pi^*)^{-1}I(Y^*\geq t) \X^{*\otimes k} \exp(\bbeta^\prime \X^*)\}&=&
E\left[\frac{1}{n\pi^*}I(Y^*\geq t) \X^{*\otimes k} \exp(\bbeta^\prime \X^*)~ \vline ~\mathcal{D}_n\right]\\
&=&\onen\sumn I(Y_i\geq t) \X_i^{\otimes k} \exp(\bbeta^\prime \X_i)\\
&=& {S}^{(k)}(t,\bbeta).
\end{eqnarray*}

By \cite{Kosorok-2008} and the assumptions \ref{assu3} and \ref{assu4}, we know $\{(n\pi)^{-1}I(Y\geq t)\X^{\otimes k}\exp(\bbeta^\prime \X): t\in [0,\tau], \bbeta \in \Theta\}$ and $\{(n\pi)^{-1} N(t): t\in [0,\tau]\}$ are Donsker, where $k=0$, 1 and 2. Therefore, conditional on $\mathcal{D}_n$ we have
\begin{align}\label{SU7.5}
\|S^{*(k)}(t, \bbeta) - S^{(k)}(t, \bbeta)\| \stackrel{P}{\longrightarrow} 0~{\rm uniformly~towards}~t.
\end{align}
Because ${S}^{(0)}(t,\bbeta)$ is  bounded away from zero \cite[]{Andersen1982}, then conditional on $\mathcal{D}_n$,
\begin{eqnarray*}
\sup_{t\in [0,\tau]}\left\|\frac{\mathbb{P}_{r|\mathcal{D}_n}\{(n\pi^*)^{-1}I(Y^*\geq t)\X^{*}\exp(\bbeta^\prime \X^*)\}}{\mathbb{P}_{r|\mathcal{D}_n}\{(n\pi^*)^{-1}I(Y^*\geq t)\exp(\bbeta^\prime \X^*)\}} - \frac{\mathbf{P}_{\mathcal{D}_n} \{(n\pi^*)^{-1}I(Y^*\geq t)\X^{*} \exp(\bbeta^\prime \X^*)\}}{\mathbf{P}_{\mathcal{D}_n} \{(n\pi^*)^{-1}I(Y^*\geq t) \exp(\bbeta^\prime \X^*)\}}\right\| \stackrel{P}{\longrightarrow}0.
\end{eqnarray*}
i.e., as $r\rightarrow \infty$,
\begin{align}\label{S8.5}
\|\bar{\X}^{*}(t, \bbeta) - \bar{\X}(t, \bbeta)\| \stackrel{P}{\longrightarrow} 0~{\rm uniformly~towards}~t.
\end{align}

Combining (\ref{LLU}) and (\ref{S8.5}), as $r\rightarrow \infty$ some calculations lead to
\begin{eqnarray*}
  \mathbf{R}^*(\bbeta)
  &=&\frac{1}{rn} \sum_{i=1}^r \frac{1}{\pi_i^*} \int_0^\tau\{\bar{\mathbf{X}}^{*}(t, \bbeta) - \bar{\mathbf{X}}(t, \bbeta)\}dM_i^*(t)\\
  &=& \underbrace{\int_0^\tau\{\bar{\mathbf{X}}^{*}(t, \bbeta) - \bar{\mathbf{X}}(t, \bbeta)\}d\bar{N}_r^*(t)}_{\mathbf{R}_1^*(\bbeta)} - \underbrace{\int_0^\tau\{\bar{\mathbf{X}}^{*}(t, \bbeta) - \bar{\mathbf{X}}(t, \bbeta)\}d\bar{\Lambda}_r^*(t)}_{\mathbf{R}_2^*(\bbeta)},
\end{eqnarray*}
where $\bar{N}_r^*(t) = \frac{1}{rn} \sum_{i=1}^r \frac{1}{\pi_i^*}N_i^*(t)$ and
$\bar{\Lambda}_r^*(t) = \frac{1}{rn} \sum_{i=1}^r \frac{1}{\pi_i^*}\int_0^t I(Y_i^* \geq u)\exp(\bbeta^\prime
\X_i^*)\lambda_0(u)du $. Note that $\bar{N}_r^*(t)$ and $\bar{\Lambda}_r^*(t)$ are two nondecreasing processes, due to (\ref{S8.5}) we have
\begin{eqnarray*}
\|\mathbf{R}_1^*(\bbeta)\| &=& \left\|\int_0^\tau\{\bar{\mathbf{X}}^{*}(t, \bbeta) - \bar{\mathbf{X}}(t, \bbeta)\}d\bar{N}_r^*(t)\right\|\\
&\leq& \int_0^\tau\|\bar{\mathbf{X}}^{*}(t, \bbeta) - \bar{\mathbf{X}}(t, \bbeta)\|d\bar{N}_r^*(t)\\
&=&\bar{N}_r^*(\tau)o_P(1),
\end{eqnarray*}
and
\begin{eqnarray*}
\|\mathbf{R}_2^*(\bbeta)\| &=& \left\|\int_0^\tau\{\bar{\mathbf{X}}^{*}(t, \bbeta) - \bar{\mathbf{X}}(t, \bbeta)\}d\bar{\Lambda}_r^*(t)\right\|\\
&\leq& \int_0^\tau\|\bar{\mathbf{X}}^{*}(t, \bbeta) - \bar{\mathbf{X}}(t, \bbeta)\|d\bar{\Lambda}_r^*(t)\\
&=&\bar{\Lambda}_r^*(\tau)o_P(1).
\end{eqnarray*}
Therefore,
\begin{eqnarray*}
\mathbf{R}^*(\bbeta) &=& \{\bar{N}_r^*(\tau) - \bar{\Lambda}_r^*(\tau)\}o_P(1)\\
&=&\left\{\frac{1}{rn} \sum_{i=1}^r \frac{1}{\pi_i^*}M_i^*(\tau)\right\} o_P(1).
\end{eqnarray*}

In view of the martingale property $E\{M(\tau)\} = 0$,
conditional on $\mathcal{D}_n$ we observe the following two facts:
\begin{align*}
  E\left\{\frac{1}{rn}\sum_{i=1}^r  \frac{1}{\pi_i^*}M_i^*(\tau) ~\vline~\mathcal{D}_n\right\} = \frac{1}{n}\sum_{i=1}^n M_i(\tau) = o_P(1),
\end{align*}
and
\begin{align*}
  Var\left\{\frac{1}{rn}\sum_{i=1}^r  \frac{1}{\pi_i^*}M_i^*(\tau) ~\vline~\mathcal{D}_n\right\} &= \frac{1}{n^2r}\sum_{i=1}^n\frac{1}{\pi_i}M_i^2(\tau)  - \frac{1}{r}\left\{\frac{1}{n}\sum_{i=1}^n M_i(\tau)\right\}^2\nonumber\\
   &\leq \max_{1\leq i \leq n}\left\{\frac{1}{n\pi_i}\right\}\frac{1}{rn}\sum_{i=1}^nM_i^2(\tau) + o_P(r^{-1})\\
    & = O_{P}(r^{-1}),
\end{align*}
where the last equality is due to the assumptions \ref{assu1}, \ref{assu3} and \ref{assu4}. By the Markov's
inequality, we have
\begin{align}\label{r77}
  \frac{1}{rn}\sum_{i=1}^r  \frac{1}{\pi_i^*}M_i^*(\tau) =  O_{P|\mathcal{D}_n}(r^{-1/2}).
\end{align}
Therefore, we know that
\begin{align}\label{Rstar-79}
  \mathbf{R}^*(\bbeta) = O_{P|\mathcal{D}_n} (r^{-1/2})o_P(1) = o_{P|\mathcal{D}_n} (r^{-1/2}).
\end{align}
Combining (\ref{LLU}) and (\ref{Rstar-79}), we get
$\dot{\ell}^*(\bbeta) = \mathbf{U}^*(\bbeta) + o_{P|\mathcal{D}_{n}}(r^{-1/2})$.
This ends the proof.

\vspace{1.5cm}
\begin{lemma}\label{lem3} 
If the assumptions \ref{assu1}-\ref{assu4} hold, as $n \rightarrow \infty$ and $r \rightarrow \infty$, conditional on $\mathcal{D}_n$, we have
\begin{align}\label{A2.7}
\dot{\ell}^*(\hbeta) = O_{P|\mathcal{D}_n} (r^{-1/2}),
\end{align}
and
\begin{align}\label{A2.8}
\mathbf{\ddot{\ell}}^*(\hbeta) = \mathbf{\Psi} + o_{P} (1),
\end{align}
where $\mathbf{\Psi}$ is given in (\ref{Eq5}), and
\begin{align}\label{LL2}
\mathbf{\ddot{\ell}}^*(\hbeta) = \frac{1}{nr}\sum_{i=1}^{r} \frac{\Delta_i^*}{\pi_i^*} \left[\frac{S^{*(2)}(Y_i^*,\hat{\bbeta}_{\mbox{\tiny \rm  MPL}})}{S^{*(0)}(Y_i^*, \hbeta)} -  \left\{\frac{S^{*(1)}(Y_i^*, \hbeta)}{S^{*(0)}(Y_i^*, \hbeta)}\right\}^{\otimes 2}  \right].
\end{align}
\end{lemma}
\noindent{\bf Proof}. In view of (\ref{A1}) and (\ref{A2}), we know
$\mathbf{\dot{\ell}}^*(\hbeta) = \dot{\ell}(\hbeta) +
O_{P|\mathcal{D}_{n}}(r^{-1/2})$. From \cite{COX1975}, the full data maximum
partial likelihood estimator $\hbeta$ satisfying $\dot{\ell}(\hbeta) = 0$, hence
the conclusion given in (\ref{A2.7}) holds.

Based on the subsample
$\mathcal{D}_{r}^* = \{ (\X_i^*, \Delta_i^*, Y_i^*,
\pi_i^*)\}_{i=1}^r$, we introduce an auxiliary term
\begin{eqnarray}
  \mathbf{V}^*(\bbeta) = -\frac{1}{nr}\sumr \frac{1}{\pi_i^*} \int_0^\tau \{\X_i^* - \bar{\X}(t,\bbeta)\} dN^*_i(t,\bbeta),
\end{eqnarray}
where $N_i^*(t) = I(\Delta_i^* =1, Y^*_i \leq t)$, and
$\bar{\X}(t,\bbeta)$ is given in (\ref{X-bar-24}).  Some calculations
lead to the following expression:
\begin{align}\label{VE-1}
  \mathbf{\dot{V}}^*(\hbeta) =  \frac{1}{nr}\sum_{i=1}^{r} \frac{\Delta_i^*}{\pi_i^*} \left[\frac{S^{(2)}(Y_i^*,\hat{\bbeta}_{\mbox{\tiny \rm  MPL}})}{S^{(0)}(Y_i^*, \hbeta)} -  \left\{\frac{S^{(1)}(Y_i^*, \hbeta)}{S^{(0)}(Y_i^*, \hbeta)}\right\}^{\otimes 2}  \right].
\end{align}
Conditional on $\mathcal{D}_n$, it is straightforward to deduce that
\begin{eqnarray*}
  E\{\mathbf{\dot{V}}^*(\hbeta)|\mathcal{D}_n\} &=& \onen\sumn\Delta_i\left[\frac{S^{(2)}(Y_i,\hat{\bbeta}_{\mbox{\tiny \rm  MPL}})}{S^{(0)}(Y_i, \hbeta)} -  \left\{\frac{S^{(1)}(Y_i, \hbeta)}{S^{(0)}(Y_i, \hbeta)}\right\}^{\otimes 2}  \right]\\
                                                &=&\mathbf{\Psi}.
\end{eqnarray*}
For any $1\leq j_1,j_2\leq p$, denote $\mathbf{\dot{V}}_{j_1j_2}^*(\hbeta)$ and
$\mathbf{\Psi}_{j_1j_2}$ as any components of $\mathbf{\dot{V}}^*(\hbeta)$ and
$\mathbf{\Psi}$, respectively. Then we have
\begin{align*}
  Var\{\mathbf{\dot{V}}_{j_1j_2}^*(\hbeta) | \mathcal{D}_n\} &= \frac{1}{rn^2} \sumn \frac{\Delta_i}{\pi_i} \left[\frac{S^{(2)}_{j_1j_2}(Y_i,\hat{\bbeta}_{\mbox{\tiny \rm  MPL}})}{S^{(0)}(Y_i, \hbeta)} -  \left\{\frac{S^{(1)}(Y_i, \hbeta)}{S^{(0)}(Y_i, \hbeta)}\right\}_{j_1j_2}^{\otimes 2}  \right]^2 - \oner\mathbf{\Psi}_{j_1j_2}^2\\
   &\leq \frac{1}{rn^2} \sumn \frac{\Delta_i}{\pi_i} \left\|\frac{S^{(2)}(Y_i,\hat{\bbeta}_{\mbox{\tiny \rm  MPL}})}{S^{(0)}(Y_i, \hbeta)} -  \left\{\frac{S^{(1)}(Y_i, \hbeta)}{S^{(0)}(Y_i, \hbeta)}\right\}^{\otimes 2} \right\|^2\\
   &\leq \max_{1\leq i \leq n}\left\{\frac{1}{n\pi_i}\right\}\frac{1}{rn} \sumn {\Delta_i} \left\|\frac{S^{(2)}(Y_i,\hat{\bbeta}_{\mbox{\tiny \rm  MPL}})}{S^{(0)}(Y_i, \hbeta)} -  \left\{\frac{S^{(1)}(Y_i, \hbeta)}{S^{(0)}(Y_i, \hbeta)}\right\}^{\otimes 2} \right\|^2\\
                                                             &= O_{P|\mathcal{D}_n} (r^{-1}).
\end{align*}
Here the last equality is from the assumption \ref{assu4}, along with
\begin{eqnarray*}
  \onen \sumn {\Delta_i} \left\|\frac{S^{(2)}(Y_i,\hat{\bbeta}_{\mbox{\tiny \rm  MPL}})}{S^{(0)}(Y_i, \hbeta)} -  \left\{\frac{S^{(1)}(Y_i, \hbeta)}{S^{(0)}(Y_i, \hbeta)}\right\}^{\otimes 2} \right\|^2 = O_{P|\mathcal{D}_n} (1),
\end{eqnarray*}
which is derived from the boundedness of $\X_i$'s in $\mathcal{D}_n$,
the assumption \ref{assu3} and  $S^{(0)}(Y_i, \hbeta)$ is bounded away from zero. By the Markov's inequality, we get
\begin{align}\label{F79}
  \mathbf{\dot{V}}^*(\hbeta) = \mathbf{\Psi} + O_{P|\mathcal{D}_n} (r^{-1/2}).
\end{align}

Conditional on $\mathcal{D}_n$,  some calculations lead
to
\begin{eqnarray}\label{LB-710}
  \|\mathbf{\ddot{\ell}}^*(\hbeta) - \mathbf{\dot{V}}^*(\hbeta)\|&\leq& \frac{1}{rn} \sumr \frac{\Delta_i^*}{\pi_i^*}\left\|\frac{S^{*(2)}(Y_i^*,\hat{\bbeta}_{\mbox{\tiny \rm  MPL}})}{S^{*(0)}(Y_i^*, \hbeta)} - \frac{S^{(2)}(Y_i^*,\hat{\bbeta}_{\mbox{\tiny \rm  MPL}})}{S^{(0)}(Y_i^*, \hbeta)}\right\|\nonumber\\
   &&+\frac{1}{rn} \sumr \frac{\Delta_i^*}{\pi_i^*}\left\|\left\{\frac{S^{*(1)}(Y_i^*, \hbeta)}{S^{*(0)}(Y_i^*, \hbeta)}\right\}^{\otimes 2} - \left\{\frac{S^{(1)}(Y_i^*, \hbeta)}{S^{(0)}(Y_i^*, \hbeta)}\right\}^{\otimes 2}\right\|\nonumber\\
    &=&  \left\{ \frac{1}{rn} \sumr \frac{\Delta^*_i}{\pi_i^*} \right\}o_{P}(1)\nonumber\\
     &\leq& \max_{1\leq i \leq n}\left\{\frac{1}{n\pi_i}\right\}o_{P}(1)\nonumber\\
&=& o_{P}(1),
\end{eqnarray}
which is due to (\ref{SU7.5}) and the assumption \ref{assu4}.
 Thus, we have
\begin{align}\label{F713}
  \|\mathbf{\ddot{\ell}}^*(\hbeta) - \mathbf{\dot{V}}^*(\hbeta)\| = o_{P}(1).
\end{align}
By the triangle inequality, it is easy to derive that
\begin{align*}
  \|\mathbf{\ddot{\ell}}^*(\hbeta) - \mathbf{\Psi}\| &\leq \|\mathbf{\ddot{\ell}}^*(\hbeta) - \mathbf{\dot{V}}^*(\hbeta)\| + \|\mathbf{\dot{V}}^*(\hbeta) - \mathbf{\Psi}\|\\
                                                     &=o_{P}(1),
\end{align*}
where the last equality is owing to (\ref{F79}) and (\ref{F713}). This ends the
proof.

\vspace{1.5cm}
\noindent{\bf Proof of Theorem 1}.  First we establish the asymptotic normality of subsample-based estimator $\tbeta$ towards $\hbeta$ given
$\mathcal{D}_n$. As $ n \rightarrow \infty$ and $ r \rightarrow \infty$, it
follows from (\ref{A1}) and (\ref{A2}) that
$\dot{\ell}^*(\bbeta) - \dot{\ell}(\bbeta) \rightarrow 0$ in probability
conditional on $\mathcal{D}_n$. Because the parameter space $\Theta$ is compact,
the full data estimator $\hbeta$ is an unique solution to
$\dot{\ell}(\bbeta) = 0$ \cite[]{Andersen1982}. From Theorem 5.9 and its remark
of \cite{vander-1998}, conditional on $\mathcal{D}_n$ in probability, as
$n \rightarrow \infty$ and $r \rightarrow \infty$, we can obtain the following
conclusion:
\begin{align}\label{A2.9}
  \|\tbeta - \hbeta \| = o_{P|\mathcal{D}_n} (1).
\end{align}
i.e., for any $\epsilon > 0$, we have
$\lim_{r\rightarrow \infty} P(\|\tbeta - \hbeta \| >
\epsilon|\mathcal{D}_n) = 0$. According to \cite{XLletter-2008}, a random
sequence converges to zero in conditional probability also indicates that it
converges to zero in unconditional probability. For notational simplicity,
throughout the proofs we will use $o_{P}(1)$ instead of
$o_{P|\mathcal{D}_n}(1)$. In other word, we have
$\|\tbeta - \hbeta \| = o_{P}(1)$.

By the Taylor expansion, as $r \rightarrow \infty$ we can derive that
\begin{align}\label{TO1}
  0 = \dot{\ell}_j^*(\tbeta)= {\dot{\ell}_j}^*(\hbeta) + \frac{\partial{\dot{\ell}_j}^*(\hbeta)}{\partial \bbeta^\prime}(\tbeta - \hbeta) + R_j,
\end{align}
where $\dot{\ell}_j^*({\bbeta})$ is the  partial derivative of $\ell^*(\bbeta)$ with respect to $\beta_j$, and
\begin{eqnarray*}
R_j = (\tbeta - \hbeta)^\prime \int_0^1\int_0^1 \frac{\partial^2\dot{\ell}_j^* \{\hbeta + uv(\tbeta - \hbeta)\}}{\partial \bbeta \partial \bbeta^\prime}vdudv(\tbeta - \hbeta).
\end{eqnarray*}
Due to the assumptions \ref{assu3}-\ref{assu4}, some direct calculations lead to the following conclusion:
\begin{eqnarray*}
\sup_{\bbeta \in \Theta} \left\|\frac{\partial^2\dot{\ell}_j^* (\bbeta)}{\partial \bbeta \partial \bbeta^\prime}\right\| &\leq& \frac{K}{nr}\sumr \frac{\Delta_i^*}{\pi_i^*}\\
&\leq& \max_{1\leq i \leq r}\left\{\frac{1}{n \pi_i^*}\right\} \frac{K}{r} \sumr\Delta_i^*\\
&=& O_{P|\mathcal{D}_n} (1),
\end{eqnarray*}
where $K$ is a positive constant.
Therefore, $R_j = O_{P|\mathcal{D}_n} (\|\tbeta - \hbeta\|^2)$.
Based on (\ref{A2.8}), we know
$\mathbf{\ddot{\ell}}^*(\hbeta) = \mathbf{\Psi} + O_{P|\mathcal{D}_n} (r^{-1/2})
= O_{P|\mathcal{D}_n} (1)$. In view of (\ref{A2.7}) and (\ref{TO1}), we get
\begin{align}\label{M-715}
  \tbeta - \hbeta &= -\{\mathbf{\ddot{\ell}}^*(\hbeta)\}^{-1} \{\mathbf{\dot{\ell}}^*(\hbeta)+ O_{P|\mathcal{D}_n} (\|\tbeta - \hbeta\|^2) \}\nonumber\\
   &= O_{P|\mathcal{D}_n} (r^{-1/2}) +  o_{P|\mathcal{D}_n} (\|\tbeta - \hbeta\|)\nonumber\\
    &=  O_{P|\mathcal{D}_n} (r^{-1/2}).
\end{align}

Subsequently, we need to prove the asymptotic normality of $\tbeta$
towards $\hbeta$ given $\mathcal{D}_n$. Recall that
\begin{eqnarray*}
  \mathbf{U}^*(\hbeta) = -\frac{1}{nr}\sumr \frac{1}{\pi_i^*}\int_0^\tau \{\X_i^* - \bar{\X}(t,\hbeta)\}dM_i^*(t,\hbeta) = \sumr\xi_i^*,
\end{eqnarray*}
where
\begin{align*}
  \xi_i^* = -\frac{1}{nr\pi_i^*}\int_0^\tau \{\X_i^* - \bar{\X}(t,\hbeta)\}dM_i^*(t,\hbeta), ~~i=1,\cdots,r.
\end{align*}
Given $\mathcal{D}_n$, $\xi_1^*,\cdots,\xi_r^*$ are independent and identically
distributed random variables with
\begin{align*}
  E(\xi_i^* | \mathcal{D}_n) &=  -\frac{1}{nr}\sumn \int_0^\tau\{\X_i - \bar{\X}(t, \hbeta)\}dM_i(t,\hbeta)\\
                             & = \dot{\ell}(\hbeta)=0,
\end{align*}
and
\begin{align*}
  Var(\xi_i^* | \mathcal{D}_n) &= E\left(\frac{1}{n^2r^2\pi_i^{*2}}\left[ \int_0^\tau\left\{\X_i^* - \bar{\X}(t, \hbeta)\right\}dM_i^*(t,\hbeta)\right]^{\otimes 2}~ \vline ~\mathcal{D}_n\right)\\
                               &= \frac{1}{n^2r^2} \sumn \frac{1}{\pi_i} \left[\int_0^\tau\left\{\X_i - \bar{\X}(t, \hbeta)\right\}dM_i(t, \hbeta)\right]^{\otimes 2}.
\end{align*}
For every $\epsilon > 0$, we have
\begin{align*}
  &E\left(\sumr \| \xi_i^*\|^2 I( \|\xi_i^*\| > \epsilon)     ~\vline~ \mathcal{D}_n\right)\\
  & \leq  \frac{1}{\epsilon} \sumr E (\|\xi_i^*\|^3 | \mathcal{D}_n)\\
  &= \frac{1}{ r^{2}\epsilon}  \left\{\frac{1}{n^3} \sumn \frac{1}{\pi_i^2} \left\| \int_0^\tau \{\X_i - \bar{\X}(t, \hbeta)\}dM_i(t,\hbeta) \right\|^3\right\}\\
  &\leq \frac{1}{ r^{2}\epsilon} \max_{1\leq i \leq n}\left\{\frac{1}{n^2\pi_i^2}\right\} \left\{\onen \sumn \left\| \int_0^\tau \{\X_i - \bar{\X}(t, \hbeta)\}dM_i(t,\hbeta) \right\|^3\right\}\\
  &=o_P(1),~~{\rm as}~ r \rightarrow \infty.
\end{align*}
Here the last equality is from the assumption \ref{assu4}, and
\begin{align*}
  \onen \sumn \left\| \int_0^\tau \{\X_i - \bar{\X}(t, \hbeta)\}dM_i(t,\hbeta) \right\|^3=O_{P|\mathcal{D}_n} (1),
\end{align*}
which is due to the boundedness of $\X_i$'s in $\mathcal{D}_n$ and the
assumptions \ref{assu1} and \ref{assu3}.  Therefore, the Lindeberg-Feller conditions are
satisfied in probability. By the Lindeberg-Feller central limit theorem
(Proposition 2.27 of \cite{vander-1998}), as $r \rightarrow \infty$ and
conditional on $\mathcal{D}_n$, we get
\begin{align}\label{AR8.13}
  \mathbf{\Gamma}^{-1/2}\mathbf{U}^*(\hbeta) \stackrel{d}{\longrightarrow} \mathcal{N}(\mathbf{0},\mathbf{I}),
\end{align}
where
\begin{align}\label{Ga-720}
  \mathbf{\Gamma} = \frac{1}{n^2r} \sumn \frac{1}{\pi_i} \left[\int_0^\tau\left\{\X_i - \bar{\X}(t, \hbeta)\right\}dM_i(t, \hbeta)\right]^{\otimes 2} = O_{P|\mathcal{D}_n} (r^{-1}).
\end{align}

Conditional on $\mathcal{D}_n$, it follows from Theorem 2.7 of
\cite{vander-1998}, together with (\ref{A2}) and (\ref{AR8.13}) that as
$r\rightarrow \infty$,
\begin{eqnarray}\label{LL-720}
\mathbf{\Gamma}^{-1/2}\dot{\ell}^*(\hbeta) &=& \mathbf{\Gamma}^{-1/2}\mathbf{U}^*(\hbeta) + o_P(1)\nonumber\\
&\stackrel{d}{\longrightarrow}& \mathcal{N}(\mathbf{0},\mathbf{I}).
\end{eqnarray}
Based on (\ref{A2.8}) and (\ref{M-715}), we can deduce
the following conclusion:
\begin{align}\label{A2.15}
  \tbeta - \hbeta &=-\{{\ddot{\ell}}^*(\hbeta)\}^{-1} \mathbf{\dot{\ell}}^*(\hbeta) + O_{P|\mathcal{D}_n}(r^{-1}).
\end{align}
It follows from the assumption \ref{assu2} and (\ref{Ga-720}) that
\begin{eqnarray}\label{SI-724}
\mathbf{\Sigma} = \mathbf{\Psi}^{-1}\mathbf{\Gamma}\mathbf{\Psi}^{-1}=O_{P|\mathcal{D}_n}(r^{-1}).
\end{eqnarray}
By (\ref{SI-724}), it can be deduced that
\begin{eqnarray}\label{SI-725}
\{{\ddot{\ell}}^*(\hbeta)\}^{-1} -  \mathbf{\Psi}^{-1} = -\mathbf{\Psi}^{-1}\{{\ddot{\ell}}^*(\hbeta) - \mathbf{\Psi}\}\mathbf{\Psi}^{-1} =  o_P(1).
\end{eqnarray}
From (\ref{A2.15}), (\ref{SI-724}), (\ref{SI-725}) and Lemma \ref{lem3}, we get
\begin{align}\label{GS-726}
  \mathbf{\Sigma}^{-1/2}(\tbeta - \hbeta) &= -\mathbf{\Sigma}^{-1/2}\{{\ddot{\ell}}^*(\hbeta)\}^{-1} \mathbf{\dot{\ell}}^*(\hbeta) + O_{P|\mathcal{D}_n}(r^{-1/2})\nonumber\\
   &= -\mathbf{\Sigma}^{-1/2}\mathbf{\Psi}^{-1}\mathbf{\dot{\ell}}^*(\hbeta)- \mathbf{\Sigma}^{-1/2}[\{{\ddot{\ell}}^*(\hbeta)\}^{-1} -  \mathbf{\Psi}^{-1}]\mathbf{\dot{\ell}}^*(\hbeta) + O_{P|\mathcal{D}_n}(r^{-1/2})\nonumber\\
   &=-\mathbf{\Sigma}^{-1/2}\mathbf{\Psi}^{-1}\mathbf{\Gamma}^{1/2}\mathbf{\Gamma}^{-1/2}\mathbf{\dot{\ell}}^*(\hbeta) +o_P(1).
\end{align}

Furthermore, we observe that
\begin{align}\label{D17}
  \mathbf{\Sigma}^{-1/2}\mathbf{\Psi}^{-1} \mathbf{\Gamma}^{1/2}(\mathbf{\Sigma}^{-1/2}\mathbf{\Psi}^{-1} \mathbf{\Gamma}^{1/2})^\prime = \mathbf{\Sigma}^{-1/2}\mathbf{\Psi}^{-1} \mathbf{\Gamma}^{1/2}\mathbf{\Gamma}^{1/2}\mathbf{\Psi}^{-1}\mathbf{\Sigma}^{-1/2} = \mathbf{I}.
\end{align}
By (\ref{LL-720}), (\ref{GS-726}), (\ref{D17}) and the Slutsky's theorem,
conditional on $\mathcal{D}_n$, as $n \rightarrow \infty$ and
$r \rightarrow \infty$,
\begin{align*}
  \mathbf{\Sigma}^{-1/2}(\tbeta - \hbeta) \stackrel{d}{\longrightarrow} \mathcal{N}(0,\mathbf{I}).
\end{align*}
That is to say, for any $\mathbf{x} \in \mathbb{R}^p$ we have
$ P\{\mathbf{\Sigma}^{-1/2}(\tbeta - \hbeta) \leq
\mathbf{x}|\mathcal{D}_n\}\rightarrow \Phi(\mathbf{x}) $ in probability, where
$\Phi(\mathbf{x})$ is the cumulative distribution function of the standard
multivariate normal distribution.\\

\vspace{1cm}
\noindent{\bf {Proof of Theorem 2}}.  Note that
\begin{align*}
  tr(\mathbf\Gamma)&=tr\bigg(\frac{1}{n^2r} \sumn \frac{1}{\pi_i} \left[\int_0^\tau\left\{\X_i - \bar{\X}(t, \hbeta)\right\}dM_i(t, \hbeta)\right]^{\otimes 2}\bigg)\\
                   &= \frac{1}{rn^{2}}\sumn \frac{1}{\pi_i} \left\|\int_0^\tau\left\{\X_i - \bar{\X}(t, \hbeta)\right\}dM_i(t,\hbeta)\right\|^2\\
                   &= \frac{1}{rn^{2}} \sumn\pi_{i} \sumn \frac{1}{\pi_i}\left\|\int_0^\tau\left\{\X_i - \bar{\X}(t, \hbeta)\right\}dM_i(t,\hbeta)\right\|^2\\
                   &\geq  \frac{1}{rn^{2}} \left\{\sumn  \left\|\int_0^\tau\left\{\X_i - \bar{\X}(t, \hbeta)\right\}dM_i(t,\hbeta)\right\|\right\}^2,
\end{align*}
where the last inequality is from the Cauchy-Schwarz inequality, and its
equality holds if and only if
$\pi_{i}=\varsigma\|\int_0^\tau \{\X_i - \bar{\X}(t,
\hat{\bbeta}_{\mbox{\tiny \rm MPL}})\}dM_i(t,\hbeta)\| $ for some
$\varsigma > 0$.  Due to $\sumn \pi_i= 1$, we know
$ \varsigma = \{\sum_{j=1}^n \|\int_0^\tau \{\X_j - \bar{\X}(t,
\hat{\bbeta}_{\mbox{\tiny \rm MPL}})\}dM_j(t,\hbeta)\|\}^{-1}.  $ Therefore, the
optimal subsampling probabilities are
\begin{align*}
  \pi_{i}^{Lopt} =  \frac{\|\int_0^\tau\{\X_i - \bar{\X}(t, \hbeta)\}dM_i(t,\hbeta)\|}{\sum_{j=1}^n \|\int_0^\tau\{\X_j - \bar{\X}(t, \hbeta)\}dM_j(t,\hbeta)\|},~  \ \textrm{$i=1,\cdots,n$}.
\end{align*}
This completes the proof.

\vspace{1cm}
\begin{lemma}\label{lem4}
Under the assumptions \ref{assu1}-\ref{assu3}, as $n\rightarrow \infty$ and $r_0\rightarrow \infty$, conditional on $\mathcal{D}_n$ we have $\hat{\Lambda}_0^{\mbox{\tiny\rm  UNIF}}(t,\tbeta_0) = {\Lambda}_0(t) + O_{P|D_n} (r_0^{-1/2})$, i.e., for any $\epsilon>0$, with probability approaching one, there exists a finite $\Delta_\epsilon$ and $r_\epsilon$, such that
\begin{align}\label{L3-rate}
P\left(\left|\hat{\Lambda}_0^{\mbox{\tiny\rm  UNIF}}(t,\tbeta_0) - {\Lambda}_0(t)\right| \geq r_0^{-1/2}\Delta_\epsilon~\vline~\mathcal{D}_n\right) < \epsilon,
\end{align}
for all $r_0 \geq r_\epsilon$, where $\hat{\Lambda}_0^{\mbox{\tiny\rm  UNIF}}(t,\bbeta)$ is a uniform subsample Breslow-type estimator defined in (\ref{pi_Brew}).
\end{lemma}

\noindent{{\bf Proof}}. For any $t\in[0,\tau]$ and $\bbeta \in \Theta$,
conditional on $\mathcal{D}_n$ we need to prove the following two expressions:
\begin{align}\label{S-721}
  \frac{1}{r_0}\sum_{i=1}^{r_0} I(Y_i^{0*}\geq t)\exp(\bbeta^\prime \X_i^{0*}) = \onen \sumn I(Y_i^{}\geq t)\exp(\bbeta^\prime \X_i^{}) + O_{P|\mathcal{D}_n} (r_0^{-1/2}),
\end{align}
and
\begin{align}\label{S-722}
  \frac{1}{r_0} \sum_{i=1}^{r_0} \frac{\Delta_i^{0*}I(Y_i^{0*}\leq t)}{n^{-1}\sum_{j=1}^n I(Y_j^{} \geq Y_i^{0*})\exp(\bbeta^\prime \X_j^{})} = \hat{\Lambda}_0(t,\bbeta) + O_{P|\mathcal{D}_n} (r_0^{-1/2}),
\end{align}
where $\hat{\Lambda}_0(t,\bbeta)$ is the full data Breslow estimator given in
(\ref{Bre1}), and
$ \mathcal{D}_{r_0}^* = \{(\X^{0*}_i, \Delta^{0*}_i,
Y^{0*}_i)\}_{i=1}^{r_0}$ is a uniform subsample from the full data
$\mathcal{D}_n$.

Conditional on $\mathcal{D}_n$, it is straightforward to clarify that
\begin{align*}
  E\left\{\frac{1}{r_0}\sum_{i=1}^{r_0} I(Y_i^{0*}\geq t)\exp(\bbeta^\prime \X_i^{0*}) ~\vline~ \mathcal{D}_n\right\} = \onen \sumn I(Y_i^{}\geq t)\exp(\bbeta^\prime \X_i^{}),
\end{align*}
and
\begin{eqnarray*}
  &&Var\left\{\frac{1}{r_0}\sum_{i=1}^{r_0} I(Y_i^{0*}\geq t)\exp(\bbeta^\prime \X_i^{0*})~\vline~\mathcal{D}_n\right\}\\
  &&= \frac{1}{nr_0} \sumn I(Y_i \geq t)\exp(2\bbeta^\prime \X_i) - \frac{1}{n^2r}\left\{\sumn I(Y_i \geq t)  \exp(\bbeta^\prime \X_i)\right\}^2\\
                                              &&\leq \frac{1}{r_0n}\sumn \exp(2\bbeta^\prime \X_i)\\
                                              &&= O_{P|\mathcal{D}_n} (r^{-1}),
\end{eqnarray*}
 where the last equality is from the assumption \ref{assu3}. The Markov's inequality leads to (\ref{S-721}).

 For $i=1,\cdots,r_0$, we denote
\begin{align*}
  \eta_i^{0*}= \frac{\Delta_i^{0*}I(Y_i^{0*}\leq t)}{n^{-1}\sum_{j=1}^n I(Y_j^{} \geq Y_i^{0*})\exp(\bbeta^\prime \X_j^{})}.
\end{align*}
Conditional on $\mathcal{D}_n$, we have
\begin{align*}
  E\left(\frac{1}{r_0}\sum_{i=1}^{r_0}\eta_i^{0*} ~\vline~ \mathcal{D}_n\right) &= \sumn \frac{\Delta_i I(Y_i \leq t)}{ \sum_{j=1}^n I(Y_j \geq Y_i) \exp(\bbeta^\prime \X_j)} \\ &=\hat{\Lambda}_0(t,\bbeta),
\end{align*}
and
\begin{align*}
  Var\left(\frac{1}{r_0}\sum_{i=1}^{r_0} \eta_i^{0*}~\vline~\mathcal{D}_n\right)&= E\left\{\frac{1}{r_0}\sum_{i=1}^{r_0} \eta_i^{0*} - \hat{\Lambda}_0(t,\bbeta) ~\vline~\mathcal{D}_n\right\}^2\\
   &= \frac{1}{r_0} E\left\{(\eta_i^{0*})^2 - 2\eta_i^{0*}\hat{\Lambda}_0(t,\bbeta) + \hat{\Lambda}^2_0(t,\bbeta)~\vline~\mathcal{D}_n\right\}\\
   &= \frac{1}{r_0}E\left\{(\eta_i^{0*})^2 ~\vline~\mathcal{D}_n\right\} - \frac{1}{r_0}\hat{\Lambda}^2_0(t,\bbeta)\\
   &\leq \frac{1}{r_0n}\left[\sumn \frac{\Delta_i I(Y_i \leq t)}{\{n^{-1}\sum_{j=1}^n I(Y_j \geq Y_i)\exp(\X_j^\prime \bbeta)\}^2} \right]\\
   &=O_{P|\mathcal{D}_n}(r_0^{-1}).
\end{align*}
Here the last equality is due to
\begin{align*}
  \onen\sumn \frac{\Delta_i I(Y_i \leq t)}{\{n^{-1}\sum_{j=1}^n I(Y_j \geq Y_i)\exp(\X_j^\prime \bbeta)\}^2}= O_P(1),
\end{align*}
which is from the assumption \ref{assu3}. As a result, the Markov's inequality ensures
that (\ref{S-722}) holds, i.e.,
$r_0^{-1}\sum_{i=1}^{r_0}\eta_i^{0*} = \hat{\Lambda}_0(t,\bbeta) +
O_{P|\mathcal{D}_n} (r_0^{-1/2})$.  In addition, some direct calculations yield
that
\begin{eqnarray}\label{S-723}
  \hat{\Lambda}_0^{\mbox{\tiny\rm  UNIF}}(t,\bbeta)
  &=& \frac{1}{r_0}\sum_{i=1}^{r_0} \left\{\frac{\Delta_i^{0*} I(Y_i^{0*} \leq t)}{r_0^{-1} \sum_{j=1}^{r_0} I(Y_j^{0*} \geq Y_i^{0*}) \exp(\bbeta^\prime \X_j^{0*})} - \eta_i^{0*} + \eta_i^{0*}\right\}\nonumber\\
  &=& \frac{1}{r_0}\sum_{i=1}^{r_0} \Delta_i^{0*} I(Y_i^{0*} \leq t)\Bigg\{\frac{1}{r_0^{-1} \sum_{j=1}^{r_0} I(Y_j^{0*} \geq Y_i^{0*}) \exp(\bbeta^\prime \X_j^{0*})} \nonumber\\
  &&- \frac{1}{n^{-1}\sum_{j=1}^n I(Y_j^{} \geq Y_i^{0*})\exp(\bbeta^\prime \X_j^{})}\Bigg\}
     + \frac{1}{r_0}\sum_{i=1}^{r_0}\eta_i^{0*}\nonumber\\
  &=& \left\{\frac{1}{r_0}\sum_{i=1}^{r_0}\Delta_i^{0*} I(Y_i^{0*} \leq t)\right\}  O_{P|\mathcal{D}_n}(r_0^{-1/2}) + \hat{\Lambda}_0(t,\bbeta) + O_{P|\mathcal{D}_n} (r_0^{-1/2}),
\end{eqnarray}
where the last equality is owing to (\ref{S-721}) and (\ref{S-722}).

Given $\mathcal{D}_n$, it can be deduced that
\begin{eqnarray*}
  E\left\{\frac{1}{r_0} \sum_{i=1}^{r_0}\Delta_i^{0*} I(Y_i^{0*} \leq t)~\vline~\mathcal{D}_n\right\} = \onen\sumn \Delta_i I(Y_i \leq t),
\end{eqnarray*}
and
\begin{align*}
  Var\left\{\frac{1}{r_0}\sum_{i=1}^{r_0}\Delta_i^{0*} I(Y_i^{0*} \leq t)~\vline~\mathcal{D}_n\right\}&=E\left\{\frac{1}{r_0}\sum_{i=1}^{r_0}\Delta_i^{0*} I(Y_i^{0*} \leq t) - \onen\sumn \Delta_i I(Y_i \leq t)~\vline~\mathcal{D}_n\right\}^2\\
   &= \frac{1}{r_0}\left[ \onen\sumn \Delta_i I(Y_i \leq t) - \left\{\onen\sumn \Delta_i I(Y_i \leq t)\right\}^2 \right]\\
   &= O_{P|\mathcal{D}_n}(r_0^{-1}).
\end{align*}
Hence, we get
\begin{align}\label{S-724}
  \frac{1}{r_0}\sum_{i=1}^{r_0}\Delta_i^{0*} I(Y_i^{0*} \leq t) &= \onen\sumn \Delta_i I(Y_i \leq t) + O_{P|\mathcal{D}_n}(r_0^{-1/2})\nonumber\\
                                                                        & = O_{P|\mathcal{D}_n}(1).
\end{align}
It follows from (\ref{S-723}) and (\ref{S-724}) that
\begin{align}\label{S-726}
  \hat{\Lambda}_0^{\mbox{\tiny\rm  UNIF}}(t,\bbeta)
  = \hat{\Lambda}_0(t,\bbeta) + O_{P|\mathcal{D}_n} (r_0^{-1/2}).
\end{align}
In addition, we investigate the distance between $\hat{\Lambda}_0(t,\tbeta_0)$ and
$\hat{\Lambda}_0(t,\hbeta)$:
\begin{eqnarray}\label{dis-730}
  &&|\hat{\Lambda}_0(t,\tbeta_0) - \hat{\Lambda}_0(t,\hbeta)|\nonumber\\
  &&= \left|\sumn \frac{\Delta_i I(Y_i \leq t)}{ \sum_{j=1}^n I(Y_j \geq Y_i) \exp(\tbeta_0^\prime \X_j)} -  \sumn \frac{\Delta_i I(Y_i \leq t)}{ \sum_{j=1}^n I(Y_j \geq Y_i) \exp(\hbeta^\prime \X_j)}\right|\nonumber\\
  &&= \left|\onen\sumn\Delta_i I(Y_i \leq t)\frac{n^{-1}\sum_{j=1}^n I(Y_j \geq Y_i) \{\exp(\tbeta_0^\prime \X_j)-\exp(\hbeta^\prime \X_j)\}}{\{n^{-1}\sum_{j=1}^n I(Y_j \geq Y_i) \exp(\tbeta_0^\prime \X_j)\}\{n^{-1}\sum_{j=1}^n I(Y_j \geq Y_i) \exp(\hbeta^\prime \X_j)\}}\right|\nonumber\\
  &&\leq \left|\onen\sumn\frac{n^{-1}\sum_{j=1}^n I(Y_j \geq Y_i) \exp(\xi^\prime \X_j)\|\X_j\|}{\{n^{-1}\sum_{j=1}^n I(Y_j \geq Y_i) \exp(\tbeta_0^\prime \X_j)\}\{n^{-1}\sum_{j=1}^n I(Y_j \geq Y_i) \exp(\hbeta^\prime \X_j)\}}\right|\|\tbeta_0-\hbeta\|\nonumber\\
  &&= O_P(r_0^{-1/2}),
\end{eqnarray}
where $\xi$ is on the segment between $\tbeta_0$ and $\hbeta$, and the last
equality is due to assumption \ref{assu3} together with
$\|\tbeta_0-\hbeta\| = O_{P|\mathcal{D}_n}(r_0^{-1/2})$.

Based on \cite{Andersen1982}, the convergence rate of full data Breslow
estimator $\hat{\Lambda}_0(t,\hbeta)$ to $\Lambda_0(t)$ is $O_P(n^{-1/2})$,
i.e. $\hat{\Lambda}_0(t,\hbeta) - \Lambda_0(t) = O_P(n^{-1/2})$. This together
with (\ref{S-726}) and (\ref{dis-730}) ensures that
\begin{eqnarray*}
  |\hat{\Lambda}_0^{\mbox{\tiny\rm  UNIF}}(t,\tbeta_0) - {\Lambda}_0(t)| &\leq& |\hat{\Lambda}_0^{\mbox{\tiny\rm  UNIF}}(t,\tbeta_0) - \hat{\Lambda}_0(t,\tbeta_0)| + |\hat{\Lambda}_0(t,\tbeta_0) - \hat{\Lambda}_0(t,\hbeta) | \\
                                                                       &&+|\hat{\Lambda}_0(t,\hbeta) - {\Lambda}_0(t)|\\
                                                                       &=& O_{P|\mathcal{D}_n} (r_0^{-1/2}) + O_{P|\mathcal{D}_n} (r_0^{-1/2}) + O_P(n^{-1/2})\\
                                                                       &=& O_{P|\mathcal{D}_n} (r_0^{-1/2}).
\end{eqnarray*}
Therefore, the convergence rate given in (\ref{L3-rate}) is established. This ends the proof.\\

\vspace{0.5cm}

\begin{lemma} 
 Suppose the assumptions \ref{assu1}-\ref{assu3} hold, as
  $r_0\rightarrow \infty$, $r\rightarrow \infty$, and $n\rightarrow \infty$,
  conditional on $\mathcal{D}_n$ and $\tbeta_0$, we have
  \begin{eqnarray}\label{Eq-727}
    \mathbf{U}_{\tbeta_0}^*(\bbeta) = \dot{\ell}(\bbeta) + O_{P|\mathcal{D}_n,\tbeta_0}(r^{-1/2}),
  \end{eqnarray}
  and
  \begin{eqnarray}\label{Eq-728}
    \dot{\ell}_{\tbeta_0}^*(\bbeta) = \mathbf{U}_{\tbeta_0}^*(\bbeta)  + o_{P|\mathcal{D}_n,\tbeta_0}(r^{-1/2}),
  \end{eqnarray}
  where $\dot{\ell}_{\tbeta_0}^*(\bbeta)$ is given in (\ref{SC-45}), and
$$\mathbf{U}_{\tbeta_0}^*(\bbeta) = -\frac{1}{nr}\sumr \frac{1}{\appsi} \int_0^\tau \{\X_i^* - \bar{\X}(t,\bbeta)\} dM^*_i(t,\bbeta)$$
with $\X_i^*$, $\appsi$ and
$M^*_i(t,\bbeta)$ being given in (\ref{SC-45}), $i=1,\cdots,r$.
\end{lemma}
\vspace{0.5cm}
\noindent{{\bf Proof}}. Given $\mathcal{D}_n$ and $\tbeta_0$, it is direct to
deduce the unbiasedness of $\mathbf{U}_{\tbeta_0}^*(\bbeta)$ towards the score
$\dot{\ell}(\bbeta)$, i.e.,
\begin{align*}
  E\{\mathbf{U}_{\tbeta_0}^*(\bbeta)|\mathcal{D}_n, \tbeta_0\} &= -\onen \sumn \int_0^\tau \{\X_i - \bar{\X}(t,\bbeta)\}dM_i(t,\bbeta)\\
                                                           &=\dot{\ell}(\bbeta).
\end{align*}
Denote $\mathbf{U}_{\tbeta_0,j}^*(\bbeta)$ as the $j$th component of
$\mathbf{U}_{\tbeta_0}^*(\bbeta)$ with $1\leq j \leq p$, then we get
\begin{align*}
  Var\{\mathbf{U}_{\tbeta_0,j}^*(\bbeta)|\mathcal{D}_n, \tbeta_0\}&\leq \frac{1}{n^2r}\sumn \frac{1}{\appm}\left[ \int_0^\tau \{\X_{ij} - \bar{\X}_j(t,\bbeta)\}dM_i(t,\bbeta) \right]^2\\
                                                              &\leq \frac{1}{nr\delta}\sumn \left\| \int_0^\tau \{\X_{i} - \bar{\X}(t,\bbeta)\}dM_i(t,\bbeta) \right\|^2\\
                                                              &= O_{P|\mathcal{D}_n, \tbeta_0}(r^{-1}),
\end{align*}
where the last equality is from the assumptions \ref{assu1} and \ref{assu3}.  This together
with the Markov's inequality can ensure that (\ref{Eq-727}) holds.

In addition, some direct calculations lead to the following expressions:
\begin{align}\label{S-729}
  \dot{\ell}_{\tbeta_0}^*(\bbeta)&= - \oner\sumr \int_0^\tau \frac{1}{n\appsi} \left\{\X_i^* - \bar{\X}^*_{\tbeta_0}(t,\bbeta)\right\}d N_i^*(t)\nonumber\\
                               &= - \oner\sumr \int_0^\tau \frac{1}{n\appsi} \left\{\X_i^* - \bar{\X}^*_{\tbeta_0}(t,\bbeta)\right\}d M_i^*(t,\bbeta)\nonumber\\
                               &= - \oner\sumr \int_0^\tau \frac{1}{n\appsi} \left\{\X_i^* - \bar{\X}(t,\bbeta)+ \bar{\X}(t,\bbeta) - \bar{\X}^*_{\tbeta_0}(t,\bbeta)\right\}d M_i^*(t,\bbeta)\nonumber\\
                               &=\mathbf{U}_{\tbeta_0}^*(\bbeta)+ \underbrace{\oner\sumr \int_0^\tau \frac{1}{n\appsi} \left\{\bar{\X}^*_{\tbeta_0}(t,\bbeta) - \bar{\X}(t,\bbeta)  \right\}d M_i^*(t,\bbeta)}_{\mathbf{R}_{\delta}^*(\bbeta)}.
\end{align}
For $k=$0, 1, and 2, we denote
\begin{align}\label{E-730}
  S_{\tbeta_0}^{*(k)} (t,\bbeta)= \oner\sumr \frac{1}{n\appsi} I(Y_i^* \geq t) \X_i^{*\otimes k}\exp(\bbeta^\prime \X_i^*).
\end{align}

For  a subsample $\mathcal{D}_r^* = \{Z_i^*\}_{i=1}^r$ with $Z_i^* = (\X_i^*, \Delta_i^*, Y_i^*, \appsi)$, we define a subsample empirical measure conditional on $\mathcal{D}_n$ and $\tbeta_0$,
\begin{eqnarray*}
\mathbb{P}_{r|\tbeta_0, \mathcal{D}_n} = \oner\sumr \delta_{Z_i^*},
\end{eqnarray*}
and
\begin{eqnarray*}
\mathbb{P}_{r|\tbeta_0, \mathcal{D}_n}f = \oner\sumr f(Z_i^*).
\end{eqnarray*}
Based on the conditional empirical measure $\mathbb{P}_{r|\tbeta_0, \mathcal{D}_n}$,
we can rewrite $S_{\tbeta_0}^{*(k)}(t,\bbeta)$ as
\begin{align*}
S_{\tbeta_0}^{*(k)}(t,\bbeta) = \mathbb{P}_{r|\tbeta_0,\mathcal{D}_n}[ \{n\pi_\delta^{{\rm app}*}\}^{-1}I(Y^*\geq t)\X^{*\otimes k}\exp(\bbeta^\prime \X^*)],
\end{align*}
where $k=0$, 1 and 2. For convenience, we denote $\mathbf{P}_{\tbeta_0, \mathcal{D}_n}$ as
taking expectation conditional on $\mathcal{D}_n$ and $\tbeta_0$. e.g.
\begin{eqnarray}\label{PD-7.71}
\mathbf{P}_{\tbeta_0,\mathcal{D}_n} f (Z^*) = E\left\{f(Z^*)~\vline ~\tbeta_0, \mathcal{D}_n\right\} = \sumn \appm f(Z_i).
\end{eqnarray}
By (\ref{PD-7.71}), we can deduce the following expressions:
\begin{eqnarray*}
\mathbf{P}_{\tbeta_0,\mathcal{D}_n} [\{n\pi_{\delta}^{{\rm app}*}\}^{-1}I(Y^*\geq t) \X^{*\otimes k} \exp(\bbeta^\prime \X^*)]&=&
E\left[\frac{1}{n\pi_{\delta}^{{\rm app}*}}I(Y^*\geq t) \X^{*\otimes k} \exp(\bbeta^\prime \X^*)~ \vline ~\tbeta_0,\mathcal{D}_n\right]\\
&=&\onen\sumn I(Y_i\geq t) \X_i^{\otimes k} \exp(\bbeta^\prime \X_i)\\
&=& S^{(k)}(t,\bbeta).
\end{eqnarray*}
Due to \cite{Kosorok-2008} and the assumption \ref{assu3}, we get $\{(n\pi^{{\rm app}*}_{\delta})^{-1}I(Y^*\geq t)\X^{*\otimes k}\exp(\bbeta^\prime \X^*): t\in [0,\tau], \bbeta \in \Theta\}$ and $\{(n\pi^{{\rm app}*}_{\delta})^{-1}N(t): t\in [0,\tau]\}$ are Donsker, where $k=$0, 1 and 2. Therefore, conditional on $\mathcal{D}_n$ and $\tbeta_0$ we have
\begin{align}\label{SU7.5b}
\|S_{\tbeta_0}^{*(k)}(t,\bbeta) - S^{(k)}(t, \bbeta)\| \stackrel{P}{\longrightarrow} 0~{\rm uniformly~towards}~t.
\end{align}
Because ${S}^{(0)}(t,\bbeta)$ is  bounded away from zero \cite[]{Andersen1982}, then conditional on $\mathcal{D}_n$ and $\tbeta_0$,
\begin{eqnarray*}
\sup_{t\in [0,\tau]}\left\|\frac{\mathbb{P}_{r|\tbeta_0,\mathcal{D}_n}\{(n\pi^{{\rm app}*}_{\delta})^{-1}I(Y^*\geq t)\X^{*}\exp(\bbeta^\prime \X^*)\}}{\mathbb{P}_{r|\tbeta_0,\mathcal{D}_n}\{(n\pi^{{\rm app}*}_{\delta})^{-1}I(Y^*\geq t)\exp(\bbeta^\prime \X^*)\}} - \frac{\mathbf{P}_{\tbeta_0,\mathcal{D}_n} \{(n\pi^{{\rm app}*}_{\delta})^{-1}I(Y^*\geq t)\X^{*} \exp(\bbeta^\prime \X^*)\}}{\mathbf{P}_{\tbeta_0,\mathcal{D}_n} \{(n\pi^{{\rm app}*}_{\delta})^{-1}I(Y^*\geq t) \exp(\bbeta^\prime \X^*)\}}\right\| \stackrel{P}{\longrightarrow}0.
\end{eqnarray*}
i.e., as $r\rightarrow \infty$,
\begin{align}\label{S8.5bs}
\|\bar{\X}_{\tbeta_0}^{*}(t, \bbeta) - \bar{\X}(t, \bbeta)\| \stackrel{P}{\longrightarrow} 0~{\rm uniformly~towards}~t.
\end{align}

 Recall that
$dM^*_i(t,\bbeta) = dN^*_i(t) - I(Y_i^* \geq t)\exp(\bbeta^\prime
\X_i^*)\lambda_0(t) dt$, some derivations result in the following
expressions:
\begin{eqnarray}\label{RST-734}
  \mathbf{R}_{\delta}^*(\bbeta)
  &=&\frac{1}{rn} \sum_{i=1}^r \frac{1}{\appsi} \int_0^\tau\{\bar{\mathbf{X}}_{\tbeta_0}^{*}(t, \bbeta) - \bar{\mathbf{X}}(t, \bbeta) \}dM_i^*(t,\bbeta)\\
  &=& \underbrace{\int_0^\tau\{\bar{\mathbf{X}}_{\tbeta_0}^{*}(t, \bbeta) - \bar{\mathbf{X}}(t, \bbeta)\}d\bar{N}_{r\delta}^*(t)}_{\mathbf{R}_{1\delta}^*(\bbeta)} - \underbrace{\int_0^\tau\{\bar{\mathbf{X}}_{\tbeta_0}^{*}(t, \bbeta) - \bar{\mathbf{X}}(t, \bbeta)\}d\bar{\Lambda}_{r\delta}^*(t)}_{\mathbf{R}_{2\delta}^*(\bbeta)},\nonumber
\end{eqnarray}
where $\bar{N}_{r\delta}^*(t) = \frac{1}{rn} \sum_{i=1}^r \frac{1}{\appsi}N_i^*(t)$ and
$\bar{\Lambda}_{r\delta}^*(t) = \frac{1}{rn} \sum_{i=1}^r \frac{1}{\appsi}\int_0^t I(Y_i^* \geq u)\exp(\bbeta^\prime
\X_i^*)\lambda_0(u)du $. Notice that $\bar{N}_{r\delta}^*(t)$ and $\bar{\Lambda}_{r\delta}^*(t)$ are two nondecreasing processes, due to (\ref{S8.5bs}) we have
\begin{eqnarray*}
\|\mathbf{R}_{1\delta}^*(\bbeta)\| &=& \left\|\int_0^\tau\{\bar{\mathbf{X}}_{\tbeta_0}^{*}(t, \bbeta) - \bar{\mathbf{X}}(t, \bbeta)\}d\bar{N}_{r\delta}^*(t)\right\|\\
&\leq& \int_0^\tau\|\bar{\mathbf{X}}_{\tbeta_0}^{*}(t, \bbeta) - \bar{\mathbf{X}}(t, \bbeta)\|d\bar{N}_{r\delta}^*(t)\\
&=&\bar{N}_{r\delta}^*(\tau)o_P(1),
\end{eqnarray*}
and
\begin{eqnarray*}
\|\mathbf{R}_{2\delta}^*(\bbeta)\| &=& \left\|\int_0^\tau\{\bar{\mathbf{X}}_{\tbeta_0}^{*}(t, \bbeta) - \bar{\mathbf{X}}(t, \bbeta)\}d\bar{\Lambda}_{r\delta}^*(t)\right\|\\
&\leq& \int_0^\tau\|\bar{\mathbf{X}}_{\tbeta_0}^{*}(t, \bbeta) - \bar{\mathbf{X}}(t, \bbeta)\|d\bar{\Lambda}_{r\delta}^*(t)\\
&=&\bar{\Lambda}_{r\delta}^*(\tau)o_P(1).
\end{eqnarray*}
Therefore,
\begin{eqnarray*}
\mathbf{R}_{\delta}^*(\bbeta) &=& \{\bar{N}_{r\delta}^*(\tau) - \bar{\Lambda}_{r\delta}^*(\tau)\}o_P(1)\\
&=&\left\{\frac{1}{rn} \sum_{i=1}^r \frac{1}{\appsi}M_i^*(\tau)\right\} o_P(1).
\end{eqnarray*}

Conditional on $\mathcal{D}_n$ and $\tbeta_0$, we get
\begin{align*}
  E\left\{\frac{1}{r}\sum_{i=1}^r \frac{1}{n\appsi} M_i^*(\tau) ~\vline~ \mathcal{D}_n, \tbeta_0\right\}
  =\frac{1}{n}\sum_{i=1}^n M_i(\tau) = o_P(1),
\end{align*}
and
\begin{align*}
  Var\left\{\frac{1}{r}\sum_{i=1}^r \frac{1}{n\appsi} M_i^*(\tau)~\vline~ \mathcal{D}_n, \tbeta_0\right\}
  &=\frac{1}{n^2r} \sum_{i=1}^n \frac{1}{\appm}M_i^2(\tau) - \frac{1}{r} \left\{\frac{1}{n}\sum_{i=1}^n M_i(\tau)\right\}^2\\
  &\leq \frac{1}{rn\delta} \sum_{i=1}^n M_i^2(\tau) + o_P(r^{-1})\\
  &= O_P(r^{-1}),
\end{align*}
where the last equality is from the assumptions \ref{assu1} and \ref{assu3}.  By the Markov's
inequality, we know
\begin{align}\label{RST-735}
  \frac{1}{r}\sum_{i=1}^r \frac{1}{n\appsi}M_i^*(\tau) = O_{P|\mathcal{D}_n,\tbeta_0}(r^{-1/2}).
\end{align}
Conditional on $\mathcal{D}_n$ and $\tbeta_0$,
due to (\ref{RST-734}) and (\ref{RST-735}) we get
$$\mathbf{R}_{\delta}^*(\bbeta) = o_P(1)O_{P|\mathcal{D}_n,\tbeta_0} (r^{-1/2})=o_{P|\mathcal{D}_n,\tbeta_0} (r^{-1/2}).$$
This together with (\ref{S-729}) leads to the conclusion given in (\ref{Eq-728}), which completes the proof.\\

\vspace{0.5cm}
\begin{lemma} 
 Under the assumptions \ref{assu1}-\ref{assu3}, as
  $r_0\rightarrow \infty$, $r\rightarrow \infty$ and $n\rightarrow \infty$,
  conditional on $\mathcal{D}_n$ and $\tbeta_0$, we have
  \begin{align}\label{Eq-732}
    \dot{\ell}^*_{\tbeta_0}(\hbeta) = O_{P|\mathcal{D}_n,\tbeta_0}(r^{-1/2}),
  \end{align}
  and
  \begin{align}\label{Eq-733}
    \ddot{\ell}^*_{\tbeta_0}(\hbeta) = \mathbf{\Psi} + o_P(1),
  \end{align}
  where $\mathbf{\Psi}$ is given in (\ref{Eq5}), and
  \begin{align*}
    \mathbf{\ddot{\ell}}_{\tbeta_0}^*(\hbeta) = \frac{1}{nr}\sum_{i=1}^{r} \frac{\Delta_i^*}{\appsi} \left[\frac{S_{\tbeta_0}^{*(2)}(Y_i^*,\hat{\bbeta}_{\mbox{\tiny \rm  MPL}})}{S_{\tbeta_0}^{*(0)}(Y_i^*, \hbeta)} -  \left\{\frac{S_{\tbeta_0}^{*(1)}(Y_i^*, \hbeta)}{S_{\tbeta_0}^{*(0)}(Y_i^*, \hbeta)}\right\}^{\otimes 2}  \right].
  \end{align*}
\end{lemma}
\vspace{0.5cm}
\noindent{{\bf Proof}}. Conditional on $\mathcal{D}_n$ and $\tbeta_0$, it follows
from (\ref{Eq-727}) and (\ref{Eq-728}) that
$\dot{\ell}^*_{\tbeta_0}(\hbeta) = \dot{\ell}(\hbeta) +
O_{P|\mathcal{D}_n,\tbeta_0}(r^{-1/2})$. Due to $\dot{\ell}(\hbeta) =0$, then we
get $\dot{\ell}^*_{\tbeta_0}(\hbeta) = O_{P|\mathcal{D}_n,\tbeta_0}(r^{-1/2})$.

To prove (\ref{Eq-733}), we introduce a term as
\begin{align*}
  \mathbf{V}^*_{\tbeta_0}(\hbeta)=  -\frac{1}{nr}\sumr \frac{1}{\appsi} \int_0^\tau \{\X_i^* - \bar{\X}(t,\bbeta)\} dN^*_i(t,\bbeta).
\end{align*}
Furthermore, some direct calculations lead to the following expression:
\begin{align*}
  \mathbf{\dot{V}}_{\tbeta_0}^*(\hbeta) =  \frac{1}{nr}\sum_{i=1}^{r} \frac{\Delta_i^*}{\appsi} \left[\frac{S^{(2)}(Y_i^*,\hat{\bbeta}_{\mbox{\tiny \rm  MPL}})}{S^{(0)}(Y_i^*, \hbeta)} -  \left\{\frac{S^{(1)}(Y_i^*, \hbeta)}{S^{(0)}(Y_i^*, \hbeta)}\right\}^{\otimes 2}  \right].
\end{align*}
Then, we get
\begin{align*}
  E\{\mathbf{\dot{V}}_{\tbeta_0}^*(\hbeta) |\mathcal{D}_n, \tbeta_0\} &= \onen\sumn\Delta_i\left[\frac{S^{(2)}(Y_i,\hat{\bbeta}_{\mbox{\tiny \rm  MPL}})}{S^{(0)}(Y_i, \hbeta)} -  \left\{\frac{S^{(1)}(Y_i, \hbeta)}{S^{(0)}(Y_i, \hbeta)}\right\}^{\otimes 2}  \right]\\
                                                                  &=\mathbf{\Psi}.
\end{align*}
Let $\mathbf{\dot{V}}_{\tbeta_0,j_1j_2}^*(\hbeta)$ be any component of
$\mathbf{\dot{V}}_{\tbeta_0}^*(\hbeta)$ with $1\leq j_1,j_2 \leq p$, we can deduce
that
\begin{align*}
  Var\{\mathbf{\dot{V}}_{\tbeta_0,j_1j_2}^*(\hbeta)|\mathcal{F}_n, \tbeta_0\} &= \frac{1}{rn^2}\sum_{i=1}^{n} \frac{\Delta_i}{\appm} \left[\frac{S_{j_1j_2}^{(2)}(Y_i,\hat{\bbeta}_{\mbox{\tiny \rm  MPL}})}{S^{(0)}(Y_i, \hbeta)} -  \left\{\frac{S^{(1)}(Y_i, \hbeta)}{S^{(0)}(Y_i, \hbeta)}\right\}_{j_1j_2}^{\otimes 2}\right]^2 - \oner\mathbf{\Psi}_{j_1j_2}\\
   &\leq \frac{1}{r\delta}\left[\onen\sum_{i=1}^{n} {\Delta_i} \left\|\frac{S_{}^{(2)}(Y_i,\hat{\bbeta}_{\mbox{\tiny \rm  MPL}})}{S^{(0)}(Y_i, \hbeta)} -  \left\{\frac{S^{(1)}(Y_i, \hbeta)}{S^{(0)}(Y_i, \hbeta)}\right\}_{}^{\otimes 2}\right\|^2\right] \\
                                                                          &= O_{P|\mathcal{D}_n,\tbeta_0}(r^{-1}),
\end{align*}
where the last equality is due to the assumptions \ref{assu1}-\ref{assu3}. Conditional
on $\mathcal{F}_n$ and $\tbeta_0$, the Markov's inequality implies
\begin{align}\label{V737}
  \mathbf{\dot{V}}_{\tbeta_0}^*(\hbeta) = \mathbf{\Psi} + O_{P|\mathcal{D}_n,\tbeta_0}(r^{1/2}).
\end{align}

In view of (\ref{SU7.5b}), we can derive that
\begin{eqnarray}\label{Eq-737}
  \| \mathbf{\ddot{\ell}}_{\tbeta_0}^*(\hbeta) - \mathbf{\dot{V}}_{\tbeta_0}^*(\hbeta)\|&\leq& \frac{1}{rn} \sumr \frac{\Delta_i^*}{\appsi}\left\|\frac{S_{\tbeta_0}^{*(2)}(Y_i^*,\hat{\bbeta}_{\mbox{\tiny \rm  MPL}})}{S_{\tbeta_0}^{*(0)}(Y_i^*, \hbeta)} - \frac{S^{(2)}(Y_i^*,\hat{\bbeta}_{\mbox{\tiny \rm  MPL}})}{S^{(0)}(Y_i^*, \hbeta)}\right\|\nonumber\\
   &&+\frac{1}{rn} \sumr \frac{\Delta_i^*}{\appsi}\left\|\left\{\frac{S_{\tbeta_0}^{*(1)}(Y_i^*, \hbeta)}{S_{\tbeta_0}^{*(0)}(Y_i^*, \hbeta)}\right\}^{\otimes 2} - \left\{\frac{S^{(1)}(Y_i^*, \hbeta)}{S^{(0)}(Y_i^*, \hbeta)}\right\}^{\otimes 2}\right\|\nonumber\\
   &=& \left\{\frac{1}{nr}\sum_{i=1}^{r} \frac{\Delta_i^*}{\appsi}\right\}o_{P}(1)\nonumber\\
    &\leq &\left\{\frac{1}{r\delta}\sum_{i=1}^r \Delta_i^*\right\}o_{P}(1)\nonumber\\
    &=&o_{P}(1).
\end{eqnarray}

Accordingly, it follows from the triangle inequality that
\begin{eqnarray*}
  \|\ddot{\ell}^*_{\tbeta_0}(\hbeta) - \mathbf{\Psi}\| &\leq& \|\ddot{\ell}^*_{\tbeta_0}(\hbeta) - \mathbf{\dot{V}}_{\tbeta_0}^*(\hbeta) \| + \|\mathbf{\dot{V}}_{\tbeta_0}^*(\hbeta) - \mathbf{\Psi}\|\\
                                                     &=&o_{P}(1),
\end{eqnarray*}
where the last equality is due to (\ref{V737}) and (\ref{Eq-737}). Hence, we
obtain
$\ddot{\ell}^*_{\tbeta_0}(\hbeta) = \mathbf{\Psi} +
o_{P}(1)$. This finishes the proof.

\vspace{1.5cm}

\noindent{{\bf Proof of Theorem 3}}. By (\ref{Eq-727}) and
(\ref{Eq-728}), we get
$\dot{\ell}_{\tbeta_0}^*(\bbeta) = \dot{\ell}_{}^*(\bbeta) + o_P(1) $ as
$r_0\rightarrow \infty$ and $r\rightarrow \infty$. We observe that the full data estimator $\hbeta$ is a
unique solution to $\dot{\ell}_{}^*(\bbeta) = 0$, and the two-step subsample estimator $\breve{\bbeta}$ satisfies
$\dot{\ell}_{\tbeta_0}^*(\breve{\bbeta}) = 0$.  Conditional on $\mathcal{D}_n$ and
$\tbeta_0$, we know from Theorem 5.9 and its remark of \cite{vander-1998} that
\begin{align}\label{Eq-740}
  \|\breve{\bbeta} - \hbeta\| = o_P(1).
\end{align}
By the Taylor's theorem, we obtain
\begin{align}\label{Eq-741}
  0=\dot{\ell}_{\tbeta_0,j}^*(\breve{\bbeta}) = \dot{\ell}_{\tbeta_0,j}^*({\hbeta}) + \frac{\partial \dot{\ell}_{\tbeta_0,j}^*({\hbeta})}{\partial \bbeta^\prime}(\breve{\bbeta} - \hbeta) + R_{\tbeta_0,j},
\end{align}
where $\dot{\ell}_{\tbeta_0,j}^*(\cdot)$ is the $j$th component of $\dot{\ell}_{\tbeta_0}^*(\cdot)$, and
\begin{eqnarray*}
R_{\tbeta_0,j} = (\breve{\bbeta} - \hbeta)^\prime \int_0^1\int_0^1 \frac{\partial^2\dot{\ell}_{\tbeta_0,j}^* \{\hbeta + uv(\breve{\bbeta} - \hbeta)\}}{\partial \bbeta \partial \bbeta^\prime}vdudv(\breve{\bbeta} - \hbeta).
\end{eqnarray*}
From the assumptions \ref{assu1} and \ref{assu3}, we get
\begin{eqnarray*}
\sup_{\bbeta \in \Theta} \left\|\frac{\partial^2\dot{\ell}_{\tbeta_0,j}^* (\bbeta)}{\partial \bbeta \partial \bbeta^\prime}\right\| &\leq& \frac{K}{nr}\sumr \frac{\Delta_i^*}{\appsi}\\
&\leq&  \frac{K}{r\delta} \sumr\Delta_i^*\\
&=& O_{P|\mathcal{D}_n} (1),
\end{eqnarray*}
where $K$ is a positive constant. Hence, $R_{\tbeta_0,j} = O_{P|\mathcal{D}_n,\tbeta_0}(\|\breve{\bbeta} - \hbeta\|^2).$

By (\ref{Eq-733}), the assumption \ref{assu2} and the continuous mapping theorem (Theorem 2.3 of \cite{vander-1998}),
conditional on $\mathcal{D}_n$ and $\tbeta_0$, as $r\rightarrow \infty$ we get
\begin{align}\label{Eq-742}
\{\ddot{\ell}_{\tbeta_0}^*(\hbeta)\}^{-1}
                                &= \{\mathbf{\Psi} +  o_P(1)\}^{-1}\nonumber\\
                                &=O_P(1).
\end{align}
Conditional on $\mathcal{D}_n$ and $\tbeta_0$, it follows from
(\ref{Eq-732}), (\ref{Eq-741}) and (\ref{Eq-742}) that
\begin{align}\label{Eq-743}
  \breve{\bbeta} - \hbeta &= - \{\ddot{\ell}_{\tbeta_0}^*(\hbeta)\}^{-1} \{\dot{\ell}_{\tbeta_0}^*({\hbeta})+ O_{P|\mathcal{D}_n,\tbeta_0}(\|\breve{\bbeta} - \hbeta\|^2)\}\\
 &= O_{P|\mathcal{D}_n,\tbeta_0}(r^{-1/2}) + o_{P|\mathcal{D}_n,\tbeta_0}(\|\breve{\bbeta} - \hbeta\|)\nonumber\\
 &= O_{P|\mathcal{D}_n,\tbeta_0}(r^{-1/2}).\nonumber
\end{align}
Therefore, $\breve{\bbeta} - \hbeta = o_P(1)$, i.e., $\breve{\bbeta}$ is
consistent to $\hbeta$ as $r\rightarrow \infty$.

We start to prove the asymptotic normality of the error term
$\breve{\bbeta} - \hbeta$ conditional on $\mathcal{D}_n$ and $\tbeta_0$. Recall
that
\begin{align*}
  \mathbf{U}_{\tbeta_0}^*(\hbeta) = \sumr \xi_i^{*\tbeta_0},
\end{align*}
where
\begin{align*}
  \xi_i^{*\tbeta_0} = -\frac{1}{nr\appsi}\int_0^\tau \{\X_i^* - \bar{\X}(t,\hbeta)\}dM_i^*(t,\hbeta), ~~i=1,\cdots,r.
\end{align*}
Conditional on $\mathcal{D}_n$ and $\tbeta_0$,
$\xi_1^{*\tbeta_0}, \cdots, \xi_r^{*\tbeta_0}$ are independent and identically
distributed random variables with
\begin{align*}
  E(\xi_i^{*\tbeta_0} | \mathcal{D}_n, \tbeta_0) &=  -\frac{1}{nr}\sumn \int_0^\tau\{\X_i - \bar{\X}(t, \hbeta)\}dM_i(t,\hbeta)\\
                                             & =0,
\end{align*}
and
\begin{align*}
  Var(\xi_i^{*\tbeta_0} | \mathcal{D}_n, \tbeta_0) &= E\left(\frac{1}{n^2r^2\{\appsi\}^2}\left[ \int_0^\tau\left\{\X_i^* - \bar{\X}(t, \hbeta)\right\}dM_i^*(t,\hbeta)\right]^{\otimes 2}~ \vline ~\mathcal{D}_n, \tbeta_0 \right)\\
                                               &= \frac{1}{n^2r^2} \sumn \frac{1}{\appm} \left[\int_0^\tau\left\{\X_i - \bar{\X}(t, \hbeta)\right\}dM_i(t, \hbeta)\right]^{\otimes 2}.
\end{align*}
For every $\epsilon > 0$, we can deduce that
\begin{align*}
  &E\left(\sumr \| \xi_i^{*\tbeta_0}\|^2 I( \|\xi_i^{*\tbeta_0}\| > \epsilon)     ~\vline~ \mathcal{D}_n, \tbeta_0\right)\\
  & \leq  \frac{1}{\epsilon} \sumr E (\|\xi_i^{*\tbeta_0}\|^3 | \mathcal{D}_n, \tbeta_0)\\
  &= \frac{1}{ r^{2}\epsilon}  \left\{\frac{1}{n^3} \sumn \frac{1}{(\appm)^2} \left\| \int_0^\tau \{\X_i - \bar{\X}(t, \hbeta)\}dM_i(t,\hbeta) \right\|^3\right\}\\
  &\leq \frac{1}{ \delta^2 r^{2}\epsilon}  \left\{\onen \sumn \left\| \int_0^\tau \{\X_i - \bar{\X}(t, \hbeta)\}dM_i(t,\hbeta) \right\|^3\right\}\\
  &=o_P(1),~~{\rm as}~ r \rightarrow \infty,
\end{align*}
where $\delta$ is a factor controlling the mixture proportion in (\ref{PSP1}), and
the last equality is from the assumptions \ref{assu1} and \ref{assu3}. Therefore, the
Lindeberg-Feller conditions are satisfied in probability. By the
Lindeberg-Feller central limit theorem (Proposition 2.27 of \cite{vander-1998}),
as $r_0 \rightarrow \infty$, $r \rightarrow \infty$, $ n\rightarrow \infty$,
conditional on $\mathcal{F}_n$ and $\tbeta_0$, we have
\begin{align}\label{N-744}
  \mathbf{\Gamma}_{\tbeta_0}^{-1/2} \mathbf{U}_{\tbeta_0}^*(\hbeta) \stackrel{d}{\longrightarrow} N(0,\mathbf{I}),
\end{align}
where
\begin{align*}
  \mathbf{\Gamma}_{\tbeta_0}  = \frac{1}{n^2r} \sumn \frac{1}{\appm} \left[\int_0^\tau\left\{\X_i - \bar{\X}(t, \hbeta)\right\}dM_i(t, \hbeta)\right]^{\otimes 2} = O_{P|\mathcal{D}_n, \tbeta_0}(r^{-1}).
\end{align*}

In addition, we need to consider the distance between $\mathbf{\Gamma}_{\tbeta_0}$
and $\mathbf{\Gamma}$. More specifically,
\begin{align}\label{Eq-745}
  \mathbf{\Gamma}_{\tbeta_0} - \mathbf{\Gamma} = \frac{1}{nr} \sumn  \left[\int_0^\tau\left\{\X_i - \bar{\X}(t, \hbeta)\right\}dM_i(t, \hbeta)\right]^{\otimes 2} \left\{\frac{1}{n\appm} - \frac{1}{n\pi_{\delta i}^{\rm{Lopt}}}\right\},
\end{align}
where $\appm$ and $\pi_{\delta i}^{\rm{Lopt}}$ are given in
(\ref{PSP1}) and (\ref{Pd-46}), respectively.  For notational convenience, we
denote
\begin{align*}
  \phi_i &=  \left\|\int_0^\tau\{\X_i - \bar{\X}^{0*}(t, {\tbeta_0})\}d\hat{M}_i(t,\tbeta_0)\right\|,
\end{align*}
and
\begin{align*}
  \psi_i&= \left\|\int_0^\tau\{\X_i - \bar{\X}(t, \hbeta)\}dM_i(t,\hbeta)\right\|.
\end{align*}
Then, we can rewrite the expressions of $\appm$ and
$\pi_{\delta i}^{\rm Lopt}$ with
\begin{eqnarray*}
\appm &=&
  (1-\delta)\frac{\phi_i}{\sum_{j=1}^n \phi_j} + \delta \onen,
\end{eqnarray*}
and
\begin{eqnarray*}
{\pi_{\delta i}^{\rm Lopt}} &=&
  (1-\delta)\frac{\psi_i}{\sum_{j=1}^n \psi_j} + \delta \onen,
\end{eqnarray*}
respectively. Note that
\begin{align*}
  \left|\frac{1}{n\appm} - \frac{1}{n\pi_{\delta i}^{\rm Lopt}}\right|&=\frac{1}{n\appm\pi_{\delta i}^{\rm Lopt}} |\appm - \pi_{\delta i}^{\rm Lopt}|\nonumber\\
                                                                                  &\leq \frac{n(1-\delta)}{\delta^2} \left|\frac{\phi_i}{\sum_{j=1}^n \phi_j} - \frac{\psi_i}{\sum_{j=1}^n \psi_j}\right|\nonumber\\
                                                                                  &= \frac{(1-\delta)}{\delta^2} \left|\frac{\phi_i n^{-1}\sum_{j=1}^n\psi_j - \psi_i n^{-1}\sum_{j=1}^n\phi_j}{(n^{-1}\sum_{j=1}^n\phi_j)(n^{-1}\sum_{j=1}^n\psi_j)}\right|.
\end{align*}
For any $t\in [0,\tau]$, we observe that
\begin{eqnarray}\label{L3-754}
  \bar{\X}^{0*}(t, {\tbeta_0}) - \bar{\X}(t, \hbeta) &=&\bar{\X}^{0*}(t, {\tbeta_0}) -\bar{\X}(t, \tbeta_0) + \bar{\X}(t, \tbeta_0) - \bar{\X}(t, \hbeta)\nonumber\\
                                                                   &=& \bar{\X}(t, \tbeta_0) - \bar{\X}(t, \hbeta) + O_{P|\mathcal{D}_n,\tbeta_0}(r^{-1/2}),
\end{eqnarray}
where the last equality is from (\ref{S8.5bs}).  In view of the fact that
$\tbeta_0 - \hbeta = O_{P|\mathcal{D}_n}(r_0^{-1/2})$, for any $t\in [0, \tau]$ we
obtain
\begin{eqnarray}\label{L2-754}
  \bar{\X}(t, \tbeta_0) - \bar{\X}(t, \hbeta) &=& \frac{S^{(1)}(t,\tbeta_0)}{S^{(0)}(t,\tbeta_0)} - \frac{S^{(1)}(t,\hbeta)}{S^{(0)}(t,\hbeta)}\nonumber\\
                                                            &=&\frac{1}{S^{(0)}(t,\tbeta_0)S^{(0)}(t,\hbeta)} \{S^{(0)}(t,\hbeta)S^{(1)}(t,\tbeta_0) - S^{(0)}(t,\tbeta_0)S^{(1)}(t,\hbeta)\}\nonumber\\
                                                            &=&\frac{1}{S^{(0)}(t,\tbeta_0)S^{(0)}(t,\hbeta)} \{S^{(0)}(t,\hbeta)S^{(1)}(t,\tbeta_0) - S^{(0)}(t,\tbeta_0)S^{(1)}(t,\tbeta_0)\nonumber\\
                                                            &&+ S^{(0)}(t,\tbeta_0)S^{(1)}(t,\tbeta_0) -  S^{(0)}(t,\tbeta_0)S^{(1)}(t,\hbeta)\}\nonumber\\
                                                            &=&\frac{1}{S^{(0)}(t,\tbeta_0)S^{(0)}(t,\hbeta)}\{S^{(1)}(t,\xi_1)^\prime (\hbeta-\tbeta_0)S^{(1)}(t,\tbeta_0)\nonumber\\
                                                            && -  S^{(0)}(t,\tbeta_0)S^{(2)}(t,\xi_2)(\hbeta-\tbeta_0)\}\nonumber\\
                                                            &=& O_{P|\mathcal{D}_n}(r_0^{-1/2}),
\end{eqnarray}
where $\xi_1$ and $\xi_2$ are on the segment between $\hbeta$ and $\tbeta_0$, and
the last equality is from the assumption \ref{assu3}. It follows from (\ref{L3-754}),
(\ref{L2-754}) and typically $r_0 < r$ that for any $t\in [0, \tau]$
\begin{align}\label{L6-755}
 \|\bar{\X}^{0*}(t, {\tbeta_0}) - \bar{\X}(t, \hbeta)\| = O_{P|\mathcal{D}_n,\tbeta_0}(r_0^{-1/2}).
\end{align}

Recall that
$\hat{M}_i(t,\tbeta_0) = N_i(t) - \int_0^tI(\tilde{T}_i \geq u)\exp(\tbeta_0^\prime
\X_i)d\hat{\Lambda}_0^{\mbox{\tiny\rm UNIF}}(u,\tbeta_0)$ and
${M}_i(t,\hbeta) = N_i(t) - \int_0^t I(\tilde{T}_i \geq u)\exp(\hbeta^\prime
\X_i)d{\Lambda}_0(u)$, it is straightforward to derive that
\begin{eqnarray}\label{L4-756}
  \hat{M}_i(t,\tbeta_0) - {M}_i(t,\hbeta)  &=&  \exp(\hbeta^\prime \X_i){\Lambda}_0(min\{t,Y_i\}) - \exp(\tbeta_0^\prime \X_i)\hat{\Lambda}_0^{\mbox{\tiny\rm  UNIF}}(min\{t,Y_i\},\tbeta_0) \nonumber\\
                                         &=& \exp(\hbeta^\prime \X_i){\Lambda}_0(min\{t,Y_i\}) - \exp(\tbeta_0^\prime \X_i){\Lambda}_0(min\{t,Y_i\})  \nonumber\\
                                         && +\exp(\tbeta_0^\prime \X_i){\Lambda}_0(min\{t,Y_i\}) - \exp(\tbeta_0^\prime \X_i)\hat{\Lambda}_0^{\mbox{\tiny\rm  UNIF}}(min\{t,Y_i\},\tbeta_0)\nonumber\\
                                         &=& \exp(\xi^\prime \X_i)\X_i^\prime(\tbeta_0 - \hbeta){\Lambda}_0(min\{t,Y_i\})\nonumber\\
                                         && - \exp(\tbeta_0^\prime \X_i)\{\hat{\Lambda}_0^{\mbox{\tiny\rm  UNIF}}(min\{t,Y_i\},\tbeta_0) - {\Lambda}_0(min\{t,Y_i\})\}
                                            \nonumber\\
                                         &=&O_{P|\mathcal{D}_n}(r_0^{-1/2}),
\end{eqnarray}
where $\xi$ is between $\tbeta_0$ and $\hbeta$, the last equality is due to
$\tbeta_0 - \hbeta = O_{P|\mathcal{D}_n}(r_0^{-1/2})$, the assumption \ref{assu3} and
Lemma \ref{lem4}.

Furthermore, some direct calculations lead to the following expressions:
\begin{eqnarray*}
  &&\int_0^\tau\{\X_i - \bar{\X}^{0*}(t, {\tbeta_0})\}d\hat{M}_i(t,\tbeta_0)\\
  &&= \int_0^\tau\{\X_i - \bar{\X}(t, \hbeta) + \bar{\X}(t, \hbeta) - \bar{\X}^{0*}(t, {\tbeta_0})\}\{d\hat{M}_i(t,\tbeta_0) - dM_i(t,\hbeta) + dM_i(t,\hbeta)\}\\
  &&=\int_0^\tau\{\X_i - \bar{\X}(t, \hbeta)\}dM_i(t,\hbeta) + \mathbf{R}_1 + \mathbf{R}_2+ \mathbf{R}_3,
\end{eqnarray*}
where
\begin{eqnarray*}
  \mathbf{R}_1&=& \int_0^\tau\{\X_i - \bar{\X}(t, \hbeta)\}\{d\hat{M}_i(t,\tbeta_0) - dM_i(t,\hbeta)\},\\
  \mathbf{R}_2&=&\int_0^\tau\{\bar{\X}(t, \hbeta) - \bar{\X}^{0*}(t, {\tbeta_0})\}\{d\hat{M}_i(t,\tbeta_0) - dM_i(t,\hbeta)\},\\
  \mathbf{R}_3&=&\int_0^\tau\{\bar{\X}(t, \hbeta) - \bar{\X}^{0*}(t, {\tbeta_0})\}dM_i(t,\hbeta).\\
\end{eqnarray*}
Combining the boundedness of $\X_i$'s in $\mathcal{D}_n$, the
assumptions \ref{assu1}-\ref{assu3}, (\ref{L6-755}), (\ref{L4-756}) and Lemma \ref{Lem0}, we can deduce that $\mathbf{R}_1 = o_P(1)$ as
$r_0 \rightarrow \infty$. In a similar way, we have $\mathbf{R}_2 = o_P(1)$ and
$\mathbf{R}_3 = o_P(1)$. For $i=1,\cdots,n$, we know
\begin{eqnarray}\label{EB-757}
  \phi_i = \psi_i + o_P(1),~ {\rm as}~ r_0 \rightarrow \infty,
\end{eqnarray}
indicating that conditional on $\mathcal{D}_n$ and $\tbeta_0$,
\begin{eqnarray}\label{LL-744}
  \onen\sumn \phi_i = \onen\sumn \psi_i + o_P(1).
\end{eqnarray}
Moreover, both (\ref{EB-757}) and (\ref{LL-744}) lead to
\begin{eqnarray*}
  \left|\frac{\phi_i}{n}\sum_{j=1}^n\psi_j - \frac{\psi_i}{n}\sum_{j=1}^n\phi_j\right| &\leq& |\phi_i - \psi_i|\onen\sum_{j=1}^n\psi_j + \psi_i \left|\onen\sum_{j=1}^n\psi_j - \onen\sum_{j=1}^n\phi_j\right|\\
                                                                                       &=& o_P(1).
\end{eqnarray*}
Therefore, as $r_0\rightarrow \infty$ and $r\rightarrow \infty$, we get the
following conclusion:
\begin{eqnarray}\label{LL-S44}
  \left|\frac{1}{n\appm} - \frac{1}{n\pi_{\delta i}^{\rm Lopt}}\right| = o_P(1).
\end{eqnarray}

Combining the assumptions \ref{assu1}-\ref{assu3}, (\ref{Eq-745}) and (\ref{LL-S44}),
conditional on $\mathcal{D}_n$ and $\tbeta_0$ we have
\begin{eqnarray}\label{GaM726}
\mathbf{\Gamma}_{\tbeta_0} - \mathbf{\Gamma}= o_{P|\mathcal{D}_n, \tbeta_0}(r^{-1}).
\end{eqnarray}
Based on
(\ref{Eq-728}), (\ref{Eq-742}), (\ref{Eq-743}), the Slutsky's theorem, and
Theorem 2.7 of \cite{vander-1998}, we can derive that
\begin{align*}
\mathbf{\Gamma}_{\tbeta_0}^{-1/2} \dot{\ell}_{\tbeta_0}^*({\hbeta}) &= \mathbf{\Gamma}_{\tbeta_0}^{-1/2} \mathbf{U}_{\tbeta_0}^*(\hbeta) + o_P(1)\nonumber\\
&\stackrel{d}{\longrightarrow} N(0,\mathbf{I}).
\end{align*}
From (\ref{Eq-742}) and (\ref{Eq-743}),
\begin{align}\label{BE723}
  \breve{\bbeta} - \hbeta &= - \{\ddot{\ell}_{\tbeta_0}^*(\hbeta)\}^{-1} \dot{\ell}_{\tbeta_0}^*({\hbeta})+ O_{P|\mathcal{D}_n,\tbeta_0}(r^{-1}).
\end{align}
Due to the assumption \ref{assu2} and (\ref{Eq-733}), we can derive that
\begin{eqnarray}\label{BE764}
\{{\ddot{\ell}}^*(\hbeta)\}^{-1} -  \mathbf{\Psi}^{-1} = -\mathbf{\Psi}^{-1}\{{\ddot{\ell}}^*(\hbeta) - \mathbf{\Psi}\}\mathbf{\Psi}^{-1} =  o_P(1).
\end{eqnarray}
In view of (\ref{BE723}) and (\ref{BE764}), we have
\begin{align}\label{Eq-748}
  &\bSigma^{-1/2}(\breve{\bbeta} - \hbeta) = -\bSigma^{-1/2}\{\ddot{\ell}_{\tbeta_0}^*(\hbeta)\}^{-1} \dot{\ell}_{\tbeta_0}^*({\hbeta}) + O_{P|\mathcal{D}_n,\tbeta_0}(r^{-1/2})\nonumber\\
  &=  -\bSigma^{-1/2}\mathbf{\Psi}^{-1} \dot{\ell}_{\tbeta_0}^*({\hbeta}) -\bSigma^{-1/2}[\{{\ddot{\ell}}^*(\hbeta)\}^{-1} -  \mathbf{\Psi}^{-1}]\dot{\ell}_{\tbeta_0}^*({\hbeta})+O_{P|\mathcal{D}_n}(r^{-1/2})\nonumber\\
  &=  -\bSigma^{-1/2}\mathbf{\Psi}^{-1} \mathbf{\Gamma}_{\tbeta_0}^{1/2}\mathbf{\Gamma}_{\tbeta_0}^{-1/2}\dot{\ell}_{\tbeta_0}^*({\hbeta}) + o_P(1).
\end{align}
From (\ref{GaM726}), as $r_0\rightarrow \infty$ and $r\rightarrow \infty$,
\begin{align*}
  (\bSigma^{-1/2}\mathbf{\Psi}^{-1}\mathbf{\Gamma}_{\tbeta_0}^{1/2})(\bSigma^{-1/2}\mathbf{\Psi}^{-1}\mathbf{\Gamma}_{\tbeta_0}^{1/2})^\prime
  &=\bSigma^{-1/2}\mathbf{\Psi}^{-1}\mathbf{\Gamma}_{\tbeta_0}\mathbf{\Psi}^{-1}\bSigma^{-1/2}\\
  &=\bSigma^{-1/2}\mathbf{\Psi}^{-1}\mathbf{\Gamma}_{}\mathbf{\Psi}^{-1}\bSigma^{-1/2} + o_P(1)\\
  &= \mathbf{I} + o_P(1).
\end{align*}
Conditional on $\mathcal{D}_n$ and $\tbeta_0$, the Slutsky's theorem, together
with (\ref{N-744}) and (\ref{Eq-748}) ensures that as $r_0\rightarrow \infty$
and $r\rightarrow \infty$,
$$\bSigma^{-1/2}(\breve{\bbeta} - \hbeta) \stackrel{d}{\rightarrow} N(0,\mathbf{I}).$$
This ends the proof.
\vspace{0.2cm}

\vspace{1.5cm}

{\noindent{{\bf Proof of Proposition 1}}}. Conditional on $\mathcal{D}_n$ and $\tbeta_0$, it is direct to derive that
\begin{align*}
\|\breve{\bbeta}- \bbeta_0\| &\leq \|\breve{\bbeta} - \hbeta\| + \|\hbeta - \bbeta_0\|\\
&=O_{P|\mathcal{D}_n,\tbeta_0} (r^{-1/2}) + O_P(n^{-1/2}),
\end{align*}
where the equality is due to Eq. (\ref{Eq-743}). i.e., $\|\breve{\bbeta}- \bbeta_0\| = O_{P|\mathcal{D}_n,\tbeta_0} (r^{-1/2})$. It follows from Proposition 2 of \cite{IEEE-Poisson} that
\begin{align*}
\|\breve{\bbeta}- \bbeta_0\| = O_P(r^{-1/2}).
\end{align*}

Next, we prove the asymptotic normality of $\breve{\bbeta}$ with respect to the true parameter. Note that
\begin{align*}
r^{1/2}(\breve{\bbeta}- \bbeta_0)&= r^{1/2}(\breve{\bbeta}- \hbeta) + r^{1/2}(\hbeta-\bbeta_0)\\
&=r^{1/2}(\breve{\bbeta}- \hbeta) + o_P(1),
\end{align*}
where the last equality is due to $\hbeta-\bbeta_0= o_P(n^{-1/2})$ and the assumption $r=o(n)$. Hence, conditional on $\mathcal{D}_n$ and $\tbeta_0$, the asymptotic distribution of $\breve{\bbeta}- \bbeta_0$ is the same as that of $\breve{\bbeta}- \hbeta$. That is to say, conditional on $\mathcal{D}_n$ and $\tbeta_0$ we have
    \begin{align}\label{S-Normmality}
    \bSigma^{-1/2}(\breve{\bbeta}- \bbeta_0)
    \stackrel{d}{\longrightarrow} N(0,\mathbf{I}),
  \end{align}
  where $\bSigma$ is given in Theorem~\ref{Th3}. Based on Proposition 2 of \cite{IEEE-Poisson}, the asymptotic normality in (\ref{S-Normmality}) also holds without
   conditioning on $\mathcal{D}_n$ and $\tbeta_0$. This completes the proof.

\bibliographystyle{natbib}
\bibliography{reference}

\end{document}